\documentclass[12pt,preprint]{aastex}

\slugcomment{ApJ in press (July 2009 - 1 v699 issue)}
\shorttitle{Dynamics of Ionized Gas}
\shortauthors{Zhao et al.}
\begin{document}
\title{Dynamics of Ionized Gas at the Galactic Center: \\
Very Large Array Observations of the Three-Dimensional Velocity Field and Location 
of the Ionized Streams in 
Sagittarius A West}    
\author{Jun-Hui Zhao}
\affil{Harvard-Smithsonian CfA, 60 Garden St, Cambridge, MA 02138}
\email{jzhao@cfa.harvard.edu}
\author{Mark R. Morris}
\affil{Department of Physics and Astronomy, UCLA, 405 Hilgard Avenue, Los Angeles, CA 90095}
\author{W. M. Goss}
\affil{NRAO, P. O. Box O, Socorro, NM 87801}
\and 
\author{Tao An}
\affil{Shanghai Astronomical Observatory, Chinese Academy of Sciences, Shanghai 200030, China}   
\begin{abstract}
We present new results based on  high-resolution observations of Sgr A West 
at the Galactic center with the Very Large Array (VLA) at 1.3 cm. By combining 
recent observations with those made at earlier epochs with the VLA at 
wavelengths of 1.3 and 3.6 cm, we measured proper motions for 71 compact 
HII components in the central 80\arcsec~(3 pc, assuming $D=8$ pc). Using 
VLA archival data for the H92$\alpha$ radio recombination line, we also 
investigated radial velocities in the LSR velocity range from +200 to --415 
\kms. Combining proper motion and radial velocity measurements, we have 
determined the three-dimensional velocity distribution in Sgr A West. We 
find that the three ionized streams (Northern Arm, Eastern Arm, and Western 
Arc) in the central 3 pc can be modeled with three bundles of Keplerian 
orbits around Sgr A*. Assuming that each of the observed streams of ionized 
gas follows a single orbit, we determined  the five orbital parameters 
($a$, $e$, $\Omega$, $\omega$, $i$) for each of them using least-square 
fitting to the loci of the streams. The degeneracy in the orbital solutions 
for both the direction of flow and the two mirror images can be further
resolved using the information obtained from the velocity measurements.
Our results confirm earlier results on the streams in the Western Arc 
and the Northern Arm to be in Keplerian orbits, suggesting that the 
stream in the Eastern Arm is also consistent with an elliptical orbit.  
All three are confined within the central three pc. Both the Northern 
and Eastern Arm streams have high eccentricities ($e=0.83\pm0.10$ and 
$0.82\pm0.05$, respectively), while the Western Arc stream is nearly 
circular, with $e=0.2\pm0.15$. All three streams orbit around Sgr A* 
in a counterclockwise sense (viewed from the Earth) and have orbital periods 
in the range 4-8$\times10^4$ yr. To verify the fit, the distributions of 
radial and transverse velocity vectors in Sgr A West were also computed 
using the Keplerian model and they show good agreement with both the 
proper motion and radial velocity data. In addition, the computed orbits 
suggest that the Northern and Eastern Arm streams may collide in the 
``Bar'' region (a few arcsec south of Sgr A*) and that most of the 
orbiting ionized gas in the ``Bar'' region is located behind Sgr A*.
We also report an ionized nebula associated with IRS 8, including a bow 
shock in radio continuum emission which shows  excellent agreement with 
near IR observations. From the H92$\alpha$ line data, we find evidence 
of substantial interaction between the IRS 8 nebula and the Northern 
Arm stream occurring in the bow-shock region. Other new morphological 
features revealed in our high-resolution image include: 1) a helical 
structure in the Northern Arm, suggesting that MHD plays an important 
role in the motion of the ionized gas, in addition to the dynamics 
determined by the central gravitational field and 2) a linear feature in 
the IRS 16 region, suggesting that the compressed edge of the Northern Arm
may result from the collective winds and radiation pressure from the 
high mass stars in the IRS16 cluster.  
\end{abstract}

\keywords{Galaxy:center -- HII regions -- ISM: individual (Sagittarius A West)
-- ISM: kinematics and dynamics -- radio lines: ISM}
\section{Introduction}
Proper motion measurements of infrared (IR) stars in the central
1\arcsec~of the Galaxy suggest the existence of a supermassive 
black hole (SMBH) with a mass of several million solar masses 
(M$_\odot$) at the position of the compact radio source Sgr A* 
\citep{scho02,ghez03,ghez05}. A recent determination of the mass  
(4.2$\times$10$^6$ M$_\odot$) has been reported by \citet{ghez08} and
\citet{gill09}. The radio source size, 37$^{+16}_{-10}$ $\mu$as, 
was determined with recent very long baseline interferometry (VLBI) 
observations at 1.3 mm, suggesting that the intrinsic shape of Sgr A* 
is not symmetric about the SMBH, and that it arises from the surrounding 
accretion flow \citep{doel08}. 
Accretion onto Sgr A* must also be fed at various times by the 
numerous gas streams that have been observed in the surrounding 
HII region, Sgr A West, notably the ionized ``Mini-spiral'' 
arms discovered by  \citet{lo83} and \citet{eker83} with the Very 
Large Array (VLA). The VLA and IR observations have revealed
that Sgr A West consists of high velocity bulk flows toward, 
or around, Sgr A*. The radial velocity fields of the ionized flows were 
imaged based on  IR observations of the [Ne II] 12.8 $\mu$m 
fine-structure line \citep{lacy80,sera85,sera88}, the IR recombination 
lines Br-$\alpha$ and/or Br-$\gamma$ \citep{geba87,herb93,paum04} and  
VLA observations of the H76$\alpha$ line \citep{schw89} and the H92$\alpha$ 
line \citep{rob93,rob96}. \cite{schw89} found a nearly complete 
ionized ring from the H76$\alpha$ line data and the kinematics of 
this ionized ring could be modeled with a circular orbit. \citet{lacy91} 
interpreted the kinematics combining the Northern Arm and Western Arc as a 
one-armed spiral in a Keplerian disk with nearly circular orbits.
Later, Roberts and Goss (1993) used VLA observations of the H92$\alpha$ 
line to divide the ionized ``ring'' into three separate components: 
the Western Arc, which is indeed in circular motion, the Northern Arm, 
and the Bar. 
The H92$\alpha$ line observations were subsequently
used by \cite{liszt03} to argue that the Bar and the
Eastern Arm together represent a single stream of gas orbiting
about the Galactic center.
Recently, Paumard et al. (2004) used observations of the 
Br-$\gamma$ line to suggest that the motions of the ionized stream 
in the Northern Arm could be fitted with a family of elliptical, Keplerian 
orbits. 

Using high-resolution radio observations,  the motions of the HII components 
on the sky plane can be elucidated. Large proper motions of the HII 
components were inferred from data acquired during the period 1991--1997 
\citep{yusef98,zhao98,zhao99}. A noticeable fraction of the HII 
components shows large motions that cannot be explained purely in terms 
of the bounded orbital motions occurring in reaction to the central 
gravitational field. The large peculiar motions might be explained by 
stellar winds or stellar wind interactions, possible explosive events, 
or debris ejected from a partially disrupted stellar envelope of a star 
due to the tidal force near the SMBH. These explanations need to be 
assessed with the improved observations presented here.

In this paper, we present new results on the dynamics of the ionized gas in the 
central 3 pc of the Galaxy based on  VLA measurements of three-dimensional 
velocities, along with numerical modeling. Section 2 describes observations and data 
reductions. Section 3 shows the measurements of proper motions and radial velocities. 
Section 4 presents the results on the three-dimensional velocities of individual sources 
and regions. Section 5 describes the dynamical models for the Mini-spiral arms in Sgr A West. 
Section 6 discusses an MHD effect on the detailed structure of the ionized streams. 
Section 7 summarizes the results and conclusions. 

\section{Observations and data reductions}

\subsection{Data at 1.3 cm}
The new observations of Sgr A* and its vicinity were made at 1.3 cm on 
2004 December 23 and 2005 April 26 with the VLA in its A and B configurations,
respectively, in order to better determine the proper motions of the HII 
components near Sgr A* with a long time span. Table 1 summarizes the observations 
and data that are used in this paper. At 1.3 cm, all the observations were made 
in the continuum mode with a total bandwidth of 100 MHz. The flux 
density scales for each of the data sets were determined using the primary 
calibrators 3C 286 and 3C 48 following the standard VLA calibration procedure 
in the Astronomical Image Processing System (AIPS). The complex gains were 
calibrated using the quasi-stellar object (QSO) NRAO 530. In order to remove the 
time variability of Sgr A*, the flux density of the point source was measured in time bins 
of 0.5 hr. Then the time variation in flux density from Sgr A* was accounted 
for using the technique described in \citet{zhao91} so that these variations 
do not lead to uncorrected sidelobe emission that could affect the proper 
motion determinations. In addition, archive data taken by different groups 
using different pointing positions slightly offset from Sgr A* are in B1950 
or J2000 coordinate systems. We therefore recalculated the coordinates 
in each of the data sets with UVFIX in AIPS, converting all coordinate 
frames to the J2000 equinox and shifting the phase centers to the position  
of Sgr A*. The errors owing to differential aberration caused by the on-line 
program were automatically corrected with the off-line program UVFIX. 
Further corrections for the baseline-based residual errors were applied 
using the gains determined from the model of NRAO 530. Then, the data from 
the observations in the consecutive A and B configurations were combined 
to form separate epoch data sets for the proper motion measurements. The remaining 
residual errors in the complex gains of each data set were further corrected 
in both phase and amplitude using the self-calibration technique. Weighting 
by the number of visibilities in each combined data set, we determined effective 
dates of the three combined data sets to be MJD 53419 (2005 February 18), MJD 51422 
(1999 September 01) and MJD 48485 (1991 August 17), spanning 13.5 years. Hereafter, we 
refer to these epochs' data sets as 2005K, 1999K, and 1991K, respectively. 

A dirty image at 1.3 cm  was constructed by applying an FFT of the 2005K data 
with  robustness weighting of R=0. The dirty image was cleaned with the 
Clark--Steer hybrid algorithm and the clean component image was convolved with 
a Gaussian beam and the resultant FWHM beam is 0.\arcsec2$\times$0.\arcsec1~
(P.A.=11\arcdeg). The rms noise of the ``cleaned'' image (Fig. 1) is 20 $\mu$Jy 
and the peak flux density is $>$1 Jy beam$^{-1}$. The dynamical range of the 
image is $>$50,000:1. With higher system temperatures at the earlier epochs,
the images of 1999K and 1991K have higher rms fluctuations of 40 and 
50 $\mu$Jy, respectively. The delay beam in the image at 1.3 cm is 
92\arcsec$\times$46\arcsec~(P.A.=11\arcdeg) and the source offsets from 
the delay center are all less than 16\arcsec~in the measurements. Thus,
the effects due to the delay beam can be neglected.
 
\subsection{Data at 3.6 cm}
At 3.6 cm, a total of nine observations was acquired from the VLA archive. All 
the observations were made in the spectral line mode for the H92$\alpha$ line.
The flux density scale was determined using 3C 286 and 3C 48. The calibration 
for complex gains utilized NRAO 530 following the AIPS procedure. In addition, 
we also corrected for the bandpass shape using 3C 84 and NRAO 530. We corrected
flux density variations of Sgr A* using the method discussed by \citet{zhao91}.
  
The on-line effect of differential aberration was corrected with the same 
procedure as used for the data at 1.3 cm. The phase centers of all the data 
sets were shifted to the same common position for Sgr A*. We combined 
two data sets taken in the BnA configuration on 2 consecutive days to 
form four epochs' data sets (2002X,1999X,1993X and 1990X). The effective 
dates of the four data sets --- MJD 52415.5 (2002 May 22), MJD 51468.5 (1999 October 18), 
MJD 49030.5 (1993 February 13) and MJD 48075.5 (1990 July 04)--- span nearly 12 years. 

Combining all the data at 3.6 cm, we constructed a 8192$\times$8192 dirty
image with robustness weighting R=0 and  the multi-frequency synthesis 
technique, which avoids the bandwidth smearing effect. Thus, the frequency 
dependence of the $uv$ coordinate was used to provide better $uv$ coverage and 
to minimize  frequency smearing. The cleaned image shown in Fig. 2 was 
achieved with the Clark--Steer hybrid deconvolution algorithm, convolving 
with the synthesized beam FWHM = 0.\arcsec70$\times$0.\arcsec57 (57\arcdeg). 
The rms noise is 20 $\mu$Jy and the peak flux density is 0.8 Jy beam$^{-1}$, 
giving a dynamical range of 40,000:1.
 
\section{Measurements of three-dimensional velocities}

\subsection{Proper motion}
Both the $K$-band and $X$-band  data sets were used for proper motion 
measurements. In order to eliminate  confusion from the extended, diffusion 
emission in Sgr A West, we filtered out the short baseline data in the 
imaging process. We constructed images at each epoch with the visibility 
data at  projected baseline lengths $>$ 100 $k\lambda$ at both 1.3 and 
3.6 cm. The cleaned images were convolved to the common synthesized beams 
of $0.^{\prime\prime}2\times0.^{\prime\prime}1$ (P.A.=0\arcdeg) and
$0.^{\prime\prime}6\times0.^{\prime\prime}5$ (P.A.=0\arcdeg) at 1.3 and 3.6 cm, 
respectively. Then, we fitted each of the compact HII components with a single 
Gaussian and a slope that removes the continuum baseline. The peak position 
offsets ${\rm \Delta\alpha,~\Delta\delta}$ from Sgr A* were determined in 
the J2000 coordinate system along with 1$\sigma$ errors in the least-square 
(LSQ) fitting. All source positions were determined as offsets from Sgr A*.
The motions of the HII components with respect to Sgr A* can be determined 
from the difference in the position offsets between two epochs. In order 
to minimize the systematic errors, we made the best fit to all the measurements 
($\Delta\alpha_i, \Delta\delta_i$) at each epoch ($t_i$) using a linear regression.
The slopes $\mu_\alpha$ and $\mu_\delta$ corresponding to proper motions in RA 
and Dec, along with the position offsets ($\Delta\alpha_0$, $\Delta\delta_0$) 
at reference time $t_0$, can be determined from the linear fitting. Fig. 3 
shows the linear fitting to the position offsets and the determination of 
the proper motions. Table 2 summarizes the measurements of the proper motions. 
Columns 3 and 4 in Table 2 list the peak position offsets at epochs 2005K 
and 2002X for 1.3 and 3.6 cm, respectively. The deconvolved sizes of the major 
($\theta_{maj}$) and minor ($\theta_{min}$) axes and the position angle (P.A.) 
of the sources, along with their 1 $\sigma$ errors, are given in column 5. 
The measurements of the proper motions 
($\mu_\alpha$, $\mu_\delta$) are given in  columns 6 and 7. The proper 
motion vectors determined from both the 1.3 and 3.6 cm data are overlaid 
on the continuum image of Sgr A West at 3.6 cm (Fig. 2).
 
\subsection{Radial velocity}   
\subsubsection{Construction of the H92$\alpha$ line cube}
The H92$\alpha$ line ($\nu_0=8309.393$ MHz) observations were centered 
at two radial velocities (see Table 1), 0 km s$^{-1}$ at the early epochs 
(before 1992) and $-200$ km s$^{-1}$ at the epochs after 1993. The observations 
can be calibrated to form two line cubes at 0 km s$^{-1}$ and $-200$ km s$^{-1}$,
respectively.
Both data cubes have 31 channels, with a channel width of 0.3906 MHz, 
corresponding to 14.1 km s$^{-1}$ in velocity resolution. Excluding the 
first channel due to poor bandpass signals in the correlator configuration, 
each of the cubes covers a velocity range of 423 km s$^{-1}$. Due to the 
broad velocity distribution of the H92$\alpha$ line in the range from --350 
to +350 km s$^{-1}$ in the Galactic center region, none of the observations 
covers the full extent of the line emission. This limitation makes it 
difficult to construct a continuum-free line cube. The final line cubes
were made using the following technique: (1) for the $-200$ 
km~s$^{-1}$ data cube, we used  channels 29 and 30 at velocities of 
$-383$ and $-397$ km~s$^{-1}$ to represent the line-free channels
and concatenate them to the other end of the $-200$ km s$^{-1}$ cube; 
(2) using UVLIN, the continuum emission was linearly interpolated and 
the fitted continuum emission was subtracted from the $uv$ data across 
the line cube to produce continuum-free line data. For the 0 km s$^{-1}$ 
cube, there are essentially no line-free channels. However, the two 
line cubes overlap in the velocity range 0 to $-200$ km s$^{-1}$. We 
averaged the four overlapping channels at velocities ranging from $-169$ 
to $-212$ km~s$^{-1}$. In the same velocity range, the line emission 
can be determined from the $-200$ km~s$^{-1}$ cube. A line-free channel 
was created by subtracting the line emission from the averaged channel. 
Then, the line-free channel was attached to both ends of the 0 km~s$^{-1}$ 
cube and the continuum level was interpolated linearly across the line 
cube. Again the fitted continuum  was subtracted from the visibility data 
to form the line (continuum-free) cube. We note that this technique works 
only if the change ($\Delta S$) of the continuum flux density ($S$) over 
the  velocity range ($\Delta V=550$ km~s$^{-1}$ or $\Delta \nu=15.3$ MHz)
is less than the noise-to-signal ratio, 
namely, 
$\displaystyle {\left|{\Delta S \over S}\right| \approx \left|\alpha 
{\Delta \nu \over \nu}\right| < {\sigma_{\rm ch} \over S},} 
$ where
$S \propto \nu^{-\alpha}$ and 1 $\sigma_{\rm ch}$ noise in each channel.
For the optically thin free-free emission from Sgr A West, $\alpha\sim 0.1$ and 
${\left|\displaystyle{\Delta S \over S}\right|\approx 0.002 \alpha}=0.0002$. 
The uncertainties introduced from the continuum subtraction is much less than  
the thermal noise.

At this point, we regridded the visibility data by interpolating the velocity
so that the velocities at the adjacent channels between the
0 km s$^{-1}$ and $-200$ km s$^{-1}$ cubes are matched. The final 
velocity separation between channels is then resampled to 15~km s$^{-1}$.

Furthermore, the two image line cubes were constructed from the 0 km
s$^{-1}$ and $-200$ km s$^{-1}$ image line data cubes  with robustness 
weighting ($R=0$). Using the robustness parameters of 0 and 2, we achieved 
the synthesized beams of $1.^{\prime\prime}0\times0.^{\prime\prime}95$
(19\arcdeg) and 1$.^{\prime\prime}0\times0.^{\prime\prime}83$
(3\arcdeg) for the 0 km s$^{-1}$ and --200 km s$^{-1}$ image 
line cubes, respectively. The two image cubes were combined to form the final 
high-resolution line image cube by convolving the synthesized image to 
a circular beam of 1.25$^{\prime\prime}$. We also convolved the 
line image cube to a large circular beam (2$^{\prime\prime}$) to detect  
a low-surface brightness structure at larger scales. The rms fluctuations
over the cube were not uniform with values with 0.15 mJy 
beam$^{-1}$ for the channels from the $-200$ km~s$^{-1}$ data cube and 0.27 mJy  
beam$^{-1}$ for the 0 km~s$^{-1}$ data cube.

Fig. 4 shows the channel images covering the velocity range  --415 
to 200 \kms. The current data are in good agreement with the data of
\citet{rob93} except for the ``Bar'', the Mini-cavity and the IRS 6 
region where the high negative-velocity range ($V_{\rm LSR} <-212$ km~s$^{-1}$) 
was not covered by these authors. The high negative-velocity components were 
imaged separately and discussed by 
\citet{rob96}.

\subsubsection{Radial velocity in the three mini-spiral arms}

Fig. 5a shows the image of the integrated H92$\alpha$ line intensity. This  
clearly delineates the three well-known kinematic features of Sgr A West: 
the Northern Arm, Eastern Arm and Western Arc. The H92$\alpha$ line spectra 
at the positions marked by the circles in each of the streams (Fig. 5b) 
are shown in Fig. 5c. We determined the kinematic parameters (radial velocity 
and line-width) from the spectra at these locations using Gaussian fitting. 
Table 3 summarizes the results. Columns 1, 4 and 7 are the identification 
of the locations in the three arms. Columns 2, 5 and 8 are the RA and Dec 
offsets from Sgr A*. Columns 3, 6 and 9 list the LSR velocity and FWHM 
line-width of the velocity components from the LSQ fitting.
 
\section{Results}
\subsection{Three-dimensional velocities of the winds from supergiants IRS 7 and IRS 8} 
Fig. 6 shows the finding chart for the IR sources in the radio image of 
Sgr A West.

{\bf IRS 7.} IRS 7 is the brightest near-IR source at the Galactic center, associated 
with an M supergiant \citep{woll82,lebo82}. In our high-angular-resolution 
map at 1.3 cm (Fig. 7a), the head of IRS 7 was resolved into several bright 
components in a shape consistent with a bow shock and a tailed structure 
$\sim$2\arcsec~(0.1 pc) north of the head. The shape of the bow shock is 
in good agreement with that observed with lower resolution by \cite{yusef92} 
at 2 cm. The bridge between the head of the bow shock and the emission knot 
in the tail is resolved out at our  resolution of 0.1\arcsec$\times$0.2\arcsec.
Fig. 7c shows the detailed structure of the radio emission from the bow shock. 

We overlay the IRS 7-SiO maser position of \citet{reid07}, and star positions 
from several IR measurements on the radio images for comparison in Fig. 7a. 
The IRS 7-SiO maser appears to be surrounded by radio continuum emission in the 
high-resolution image at 1.3 cm, and agrees with the centroid position of the 
radio emission in the low-resolution image at 3.6 cm (contours in Fig. 7a) within 
an offset $\le 0.1\arcsec$. The centroid position at 3.6 cm at epoch 2002X is 
in good agreement with that determined by \citet{yusef89} at 2 cm at an early 
epoch, after correction for proper motion. The positions of the IR star determined 
in the IR measurements by a number of IR groups \citep{blum96,genz00,vieh06} 
show  offsets ($\sim$0.\arcsec1-0.\arcsec2) from the radio positions (Fig. 7a). The 
IRS 7-SiO maser, believed to be located close to the M supergiant ($<10$ mas), 
has been used for the position registration in the coordinate frames at IR and 
radio wavelengths \citep{ment97, reid03, reid07}. Therefore, we use the position 
of the IRS7-SiO maser measured at 2006.23 \citep{reid07} as the reference position 
for the star of IRS 7, that is, the IR positions used in this paper have been 
corrected for the offsets between the position of the IRS 7 star from various 
published catalogs and that of the IRS 7-SiO maser at 2006.23.

The proper motion of IRS 7, $\mu_\alpha=-1.5\pm1.2$ mas y$^{-1}$
and $\mu_\delta=-5.4\pm0.6$ mas y$^{-1}$ at the epoch 2002X,
was determined from the four epochs' data at 3.6 cm, 
giving ${ V_x = -56\pm44}$ \kms~and  ${V_y = -205\pm21}$ \kms,
agreeing well with the proper motion of the IR star from \cite{genz00}. 
The radial velocity determined from the H92$\alpha$ line
is $V_{\rm LSR} = -123\pm7$ \kms~(Fig. 7b), which is in good agreement with
that derived from [NeII] emission line \citep{sera91}. 
The total velocity of the ionized gas,  
${V=\sqrt{V_x^2+V_y^2+V^2_{\rm LSR}}}= 246\pm20$ \kms,
agrees with the results measured from the SiO maser 
within 3$\sigma$ \citep{reid07}. Based on the cometary morphology,
\citet{yusef91}  proposed that the radio structure (Fig. 7c) is 
caused by the ram pressure of a nuclear wind or by radiative pressure
arising from the IRS 16 cluster of stars surrounding Sgr A*.
The current observations agree with the models discussed by 
\citet{yusef91} and \citet{yusef92}.

{\bf IRS 8.} IRS 8 was detected at 2.2 $\mu$m by \citet{beck75}, 
with an estimated luminosity  $\sim1\times10^{5}~L_\sun$. The bow-shock 
structure of radio continuum emission in the high-resolution image of 
IRS 8 at 1.3 cm (Fig. 8) is in excellent agreement with 
the IR structure and the location of the IRS 8 star  at $\Delta\alpha=$1.\arcsec2~and
$\Delta\delta=$29.\arcsec3~\citep{geba04}. The position of 
the radio continuum peak agrees with that of the IR apex in the 
bow-shock. The proper motion of the radio peak is measured to
be $\mu_\alpha=1.8\pm0.3$ mas y$^{-1}$, 
$\mu_\delta=-1.1\pm0.5$ mas y$^{-1}$, or $V_{\rm t} =\sqrt{ V_x^2 +V_y^2}= 
80\pm14$ \kms~(P.A.$=121\arcdeg\pm24\arcdeg$).

In the lower resolution ($\theta_{\rm FWHM}=2^{\prime\prime}$) channel image 
in H92$\alpha$, a symmetrical line emission has been detected at $-10$ km s$^{-1}$ 
in the IRS 8 region. We integrated the velocity from $-25$ to 5 km s$^{-1}$ 
to show the integrated-line-flux image (contours in Fig. 9a). We refer to this 
structure, which has a size of 5\arcsec~(EW) by 6\arcsec~(NS) (blue), as the 
``IRS 8 nebula'' since its radial velocity is consistent with that of the IRS 8 
star \citep{geba06}. In Fig. 9a, the IRS 8 nebula is compared to the pseudo-color 
image of the integrated-line emission from the Northern Arm in the velocity 
range 35--170 km s$^{-1}$. The intensity-weighted velocity images are also 
shown in Fig. 9b, for the negative-velocity gas in the IRS 8 nebula. Fig. 9c 
shows the intensity-weighted velocity distribution of the redshifted gas in 
the Northern Arm. A high-spatial-resolution (1.\arcsec25) spectrum in Fig. 9d 
was created from region 1 with the dominant line emission at $-18\pm4$ \kms~from 
the IRS 8 nebula and much weaker line emission from the Northern Arm. A spectrum 
of the line emission at $87\pm7$ \kms~from region 3 in the Northern Arm is shown 
in Fig. 9e. In Fig. 9f, the spectrum (black) from region 2 is compared to the 
spectrum (red) from the position of the radio continuum peak in the bow shock. 
The H92$\alpha$ spectra in the bow-shock region show both a broad, redshifted 
feature at $V_{\rm LSR}=80\pm28$ km s$^{-1}$ with $\Delta V_{\rm FWHM}=137\pm53$ 
km s$^{-1}$ and a somewhat narrower feature at $V_{\rm LSR}=-14\pm 5$ km s$^{-1}$ 
with $\Delta V_{\rm FWHM}=52\pm17$ km s$^{-1}$. The H92$\alpha$ line spectra 
in the shocked region agree with the spectrum of [Ne{\rm II}] at IRS 8 in a 
larger area \citep{geba04}. Furthermore, a comparison of H92$\alpha$ line 
spectra shows that line emission from region 2 (in the tail of the IRS 8 
bow shock) is a factor of 1.5 stronger than that from the position at the 
continuum peak (the head of the IRS 8 bow shock). We derived line-to-continuum 
ratios (L/C) of 5$\pm1$\% and  2.5$\pm0.5$\% from the tail and head of the 
bow shock, respectively. The significant difference in L/C between the head 
and the tail suggests that the local thermodynamic equilibroum (LTE) electron 
temperature in the head 
($T_e\sim1.3\times10^4$ K) is higher than that in the tail ($T_e\sim7.0\times10^3$ 
K), if the gas is optically thin. The suggested higher electron temperature 
in the head of the bow shock may result from a strong shock interaction 
of the IRS 8 nebula with gas in the Northern Arm. Fig. 9a also shows that 
at the position of region 2, both the positive-velocity emission from 
the Northern Arm (color) and the negative velocity emission from the IRS 8 nebula
(contours) are enhanced, possible evidence that the ionized gas in both 
the velocity components may be compressed due to their collision. If the 
interaction can be verified with improved data, we would conclude that 
both the IRS 8 star and the nebula are located close to the Northern Arm 
and approach the Mini-spiral arm from the far side with respect to Sgr~A* 
considering their opposite radial velocities.
 
Fig. 9g shows a spectrum toward IRS 8 nebula, integrated over 
a region of size 6\arcsec$\times$5\arcsec~(P.A.=0\arcdeg) centered at an offset 
from Sgr A* of $\Delta\alpha=-$1.\arcsec19, $\Delta\delta=$26.\arcsec3. 
The IR star is located $\sim$4\arcsec~NE of the nebula center. We fit two 
Gaussian components to the integrated spectrum, giving $S_{\rm L}=34\pm5$ mJy, 
$V_{\rm LSR}=-12$ \kms, $\Delta V_{\rm FWHM}=46\pm20$ \kms~from the IRS 8 
nebula and $S_{\rm L}=26\pm4$ mJy, $V_{\rm LSR}=75\pm24$ \kms, 
$\Delta V_{\rm FWHM}=122\pm50$ \kms~from the Northern Arm stream.
The two components are clearly separated in the integrated spectrum.  
For the IRS 8 nebula (the negative-velocity component), 
the emission measure (EM$\approx 8\times10^6$ cm$^{-6}$ pc) and 
the mean electron density ($n_e\approx 6\times10^3$ cm$^{-3}$)
are inferred assuming that the H92$\alpha$ line is optically thin. 
Based on the derived excitation parameter ($U\approx 35$ cm$^{-2}$ pc),  the 
rate of ionizing photons ($N_{Lyc}\approx1.5\times10^{49}$ 
phot. s$^{-1}$) is inferred on the assumption that only 10\% of 
the Lyman continuum flux contributes to ionizing the IRS 8 nebula owing 
to the fact that the ionizing star is located on the edge of the nebula. 
Thus, the ionizing source is equivalent to an O5.5-6 zero-age 
main sequence (ZAMS) star \citep{pana73}, which agrees well 
with the spectral type inferred from the IR spectrum \citep{geba06}.
The location of the IRS 8 star with respect to the edge of the nebula,
displaced by 4\arcsec~($\sim0.15$ pc) from the center, is striking. This 
displacement suggests that the star runs away from the progenital ionized 
cloud (as delineated by the blue contours in Fig. 9a) into a dense medium, 
the Northern Arm (as scaled with  color in the same figure). The shape 
and the location of the observed bow shock with respect to the IR star 
also agree with this scenario. The velocity of the star is probably 
similar to the bow-shock velocity, that is, $V_{\rm bs}\sim\Delta V_{\rm FWHM}=46$ 
\kms~with respect to the nebular center which is at a velocity of $\sim$0 \kms. 
The observed transverse velocity $V_{\rm t} =80\pm16$ \kms~(P.A.=121\arcdeg$\pm$25\arcdeg)
from our proper motion measurement of the apex corresponds to 
the velocity of the medium flow in the Northern Arm shocked by the stellar wind
from IRS 8. The time 
for the star to move from the center to the edge of the nebula would be  
$>3\times10^3$ y if the star has the same velocity as that of the bow 
shock ($V_{\rm bs}$).

\subsection{Motions of the sources in the Northern Arm}
{\bf IRS 5SE1.} A bright radio source, X7 (S$\sim$7 mJy at 3.6 cm), 
can be identified in the IRS 5 field. Within the errors, this source 
agrees with the mid-IR position of IRS 5SE1 of 
\cite{vieh06} (we note that the IR position reported by \cite{perg08} 
is offset by more than 1\arcsec~from X7). The radio emission  X7 is resolved 
at 1.3 cm (Fig. 10a \& Fig. 22) showing a triangular structure 
0.\arcsec5~in size pointing toward Sgr A*. The radio morphology of IRS 5SE1 
agrees with the structure of the dust tail observed in the $L\arcmin$ band by \cite{perg08}.
The proper motion data of X7 at 3.6 cm suggest that the source  moves 
toward the SW with a transverse velocity of $V_{\rm t}=313\pm31$ 
km s$^{-1}$ (P.A.$=-148\arcdeg\pm7\arcdeg$),  nearly perpendicular
to the proper motion vector of IRS 5SE1 determined from 2.18 $\mu$m 
astrometry \citep{perg08}. 
The H92$\alpha$ line spectrum shows a spatially extended, broad spectral feature 
at $V_{\rm LSR}=72\pm16$ \kms~with $\Delta V_{\rm FWHM}= 116\pm41$ \kms~(Fig. 3c). 
The radial velocity is consistent with the velocity of the Northern Arm but 
the transverse velocity is 2 times large. Based on the radio observations, 
IRS 5SE1 appears to run into the Northern Arm. 

{\bf IRS 10W.} IRS 10W is located in the Northern Arm. A shell-like radio 
continuum source 
(K8 in Fig. 10b) is revealed in the high-resolution image at 1.3 cm, 
suggesting an association with a bow shock. We note that the orientation 
of the radio bow shock appears to be very different (by $\sim$120\arcdeg) from
that determined from the near-IR observations \citep{tann05}. The apex of 
the radio bow-shock emission appears to be located SW with respect to  
the mid-IR source IRS 10W \citep{vieh06}. Proper motion measurements of 
the radio bow shock imply a southward 
motion of $V_{\rm t}=73\pm30$ km s$^{-1}$ (P.A.$=171\arcdeg\pm20\arcdeg$). 
The H92$\alpha$ line spectrum at K8  shows a prominent line emission feature
at $V_{\rm LSR}=67\pm2$ km s$^{-1}$ with  $\Delta V_{\rm FWHM}=49\pm4$ km s$^{-1}$
(Fig. 3a),
suggesting that this radio source is associated with the Northern Arm.

{\bf IRS 21.} In the high-resolution 1.3-cm radio image, the core of the ionized 
gas surrounding IRS 21 is resolved with a deconvolved size of  
0.\arcsec24$\times$0.\arcsec15~(P.A.=98\arcdeg) and a total flux density 
of 9.0$\pm$0.5 mJy (Fig. 10c); the source is slightly resolved at 3.6 cm with 
a total flux density of 10$\pm$1 mJy.  The measurements are consistent 
with a free--free origin of  the radio  emission from the externally
ionized winds of the IRS 21 star. The morphology of the ionized gas appears to be in 
good agreement with the dust emission structure observed in the IR bands 
\citep{tann02}. The proper motion measurements suggest that the radio source (K36) 
of IRS 21  moves SW with a transverse velocity of 
$V_{\rm t}=114\pm17$ \kms~(P.A.$=-127\arcdeg\pm15\arcdeg$). The H92$\alpha$ line 
spectrum (K36 in Fig. 3c) shows a radial velocity of 
$V_{\rm LSR}=-92\pm4$ km s$^{-1}$ with $\Delta V_{\rm FWHM}= 97\pm10$ \kms.  
The moving direction of the radio source K36 appears to be similar to that 
of the Northern Arm gas at this location 
($V_{\rm z}=-130$ \kms, $V_{\rm t}=401$ \kms with P.A.$=-132$\arcdeg) 
as predicted from the Keplerian model (see Section 5) but the magnitude of total velocity 
appears to be significantly less than (by a factor of $\sim$3) the model values. 
This discrepancy suggests that the radio source K36 of 
IRS 21 might be in an orbit significantly different from that of the Northern Arm.

{\bf CXOGC J174540.029003.} This  X-ray transient was discovered by the Chandra X-ray 
observatory \citep{muno05, porq05}, while a double radio source was found to be 
associated with the X-ray source \citep{bow05}. The double radio source (at the 2005 
epoch during the burst) is shown in Fig. 10f. The SW component was present in all three 
epochs with  varying flux densities of 1.1$\pm0.1$, 1.1$\pm$0.1 and 3.4$\pm$0.1 mJy  
in 1991K, 1999K and 2005K, respectively. The position of the SW component agrees with 
the X-ray source within the uncertainties, suggesting that this component is associated 
with that of the X-ray source, presumably a low-mass X-ray binary. The NE component was 
present only in the  2005K epoch with flux density 3.5$\pm$0.1 mJy and was not detected 
at the 3$\sigma$ level of 0.11 and 0.14 mJy beam$^{-1}$ in the 1999K and 1991K images, 
respectively; this suggests that the NE component was a hot spot associated with the 
head of the jet ejected from the persistent, radio-variable star. The SW component shows 
a large proper motion, with the transverse velocity of $V_{\rm t}= 468\pm60$ 
\kms~(P.A.$=-35\arcdeg\pm6\arcdeg$) toward the NW (Fig. 10). The radial velocity of the 
gas in this direction, as determined from the H92$\alpha$ line data, is 
$V_{\rm LSR}=-182\pm19$ \kms. Both the value of radial velocity and the vector of a 
transverse velocity are consistent with the velocity predicted in the numerical model 
for the Northern Arm stream (see Figs. 21b and 21c and Section 5). Of course, there are no 
reasons that the X-ray transient CXOGC J174540.029003 should share the motion of the 
Northern Arm. The agreement between the transverse velocity of the radio source 
associated with the X-ray transient and the kinematics of the gas in the Northern 
Arm might be just a coincidence. Also, the X-ray transient is shown to be located within 
0.\arcsec15 (1$\sigma$) of IRS 33SW (marked with a plus in Fig. 10f) which was detected 
in the mid-IR \citep{vieh06}. The positional error ($\pm$0.\arcsec15) in the mid-IR 
appears to be large. The position coincidence between the X-ray transient and the 
mid-IR source deserves to be further investigated. 
 
Assuming that the NE component was ejected from the  X-ray source located at  K40,
a proper motion of $\mu_\alpha>89$ mas y$^{-1}$ and $\mu_\delta>58$ mas y$^{-1}$ is 
inferred, that is, the velocity of the NE component with respect to K40 is $>$4000 \kms. 
The jet velocity ejected from the radio transient is about an order of magnitude greater 
than its orbital velocity of $\sim$500 \kms~in this region, as observed and predicted 
from the orbital model for the Northern Arm (see Fig. 21c).

\subsection{The sources in the Eastern Arm}
Fig. 11 shows the radio components in the Eastern Arm.  One of the striking radio features 
in the Eastern Arm is the arc at $\Delta \alpha=$10.\arcsec6~ and $\Delta \delta=-$6.\arcsec0  
with a width of 0.\arcsec2 and  a length of at least 3\arcsec~(Fig. 11a). The same morphology  
is observed at both 1.3 and 3.6 cm (Figs. 1, 2 and 6). The IRS 28/IRS 4 star is located close 
to the location where this radio arc is bent. The structure is resolved at 1.3 cm and proper 
motion measurements at this wavelength give a large uncertainty; therefore, we use the 3.6 cm 
data for our reported proper motion measurements. The proper motion derived from the radio 
component X18 near the center of the arc indicates that it moves westward at 
$V_{\rm t}=89\pm12$ \kms~(P.A.$=-102\arcdeg\pm56\arcdeg$). At this location, the radial 
velocity is $V_{\rm LSR}=154\pm4$ \kms.
These gas motions contrast with those of the star IRS 28: 
$V_{\rm t}=215$ \kms~(P.A.$=159\arcdeg\pm4\arcdeg$) 
and $V_{\rm LSR}=-55$ \kms~determined from the observations of the SiO maser \citep{reid07}. 
Thus, the ionized gas and the evolved star are probably unrelated. On the other hand, an He I 
star (He I 17) in the IRS 4 region shows a broad line profile ($\Delta V_{\rm FWHM}\sim 500$ 
\kms)~centered at $V_{\rm LSR}\sim130$ \kms~\citep{krab91}, which agrees with that of X18.
We raise the possibility that the high-velocity wind from He I 17 has shocked the ambient
gas in the Eastern Arm, and has thereby produced the arc of radio emission. 

The component K66 appears to be associated with the mid IR source VISIR 12 
\citep{vieh06}, showing an arc-like structure with a NW proper motion,
$V_{\rm t} = 87\pm20$ \kms~(P.A.$=-29\arcdeg\pm13\arcdeg$) and $V_{\rm LSR}=163\pm4$ 
\kms~(Fig. 11c). The H92$\alpha$ line shows a radial velocity gradient of 5 km s$^{-1}$ per 
arcsec across this feature. There is no obvious velocity discontinuity. It is possible 
that K66 is a continuation of the arc feature described above.  

In the IRS 9 region there are two ``head-tail'' radio components 
(K57 and K59) located NE and SW of the IR star (IRS 9N) (Fig. 11b),
which is located at the southern rim of the  Eastern cavity 
\citep{zhao98}. Both radio components move generally SW, 
with $V_{\rm t}=101\pm27$ \kms~(P.A.$=-110\arcdeg\pm50\arcdeg$) and
186$\pm19$ \kms~(P.A.$=-168\arcdeg\pm4\arcdeg$) for K57 and  K59, 
respectively. The radial velocities of $V_{\rm LSR}=133\pm4$ 
\kms~and $145\pm4$ \kms~from the H92$\alpha$ observations are 
consistent with the values measured from other lines such as Br-$\gamma$
\citep{paum04}, but differ greatly from  that of the IRS 9-SiO maser: 
$-342$ \kms~\citep{reid07}, suggesting that IRS 9  is not 
associated with the Eastern Arm.

The radio complex (K48, K50, and K51) may well be associated with  
IRS 9NW (Fig. 11d). The mean proper motion of the complex is
$\overline \mu_\alpha=-4.25\pm0.38$ mas y$^{-1}$ and 
$\overline \mu_\delta=2.15\pm0.34$ mas y$^{-1}$, 
implying $\overline V_{\rm t}=181\pm14$ \kms~(P.A.$=-63\arcdeg\pm9\arcdeg$). 
The H92$\alpha$ line spectra (Fig. 3b) show multiple velocity components 
at $V_{\rm LSR}=-70\pm15$ \kms~and $98\pm92$ \kms~from K48,  $V_{\rm LSR}=-76\pm22$
\kms~and $91\pm15$ \kms~and $163\pm21$ \kms~from K50, and
$V_{\rm LSR}=98\pm20$ \kms~and $163\pm21$ \kms~from K51, suggesting 
the presence of more than one velocity stream along the line of sight. 
We also note that any redshifted components at V$_{\rm LSR} > 200$ 
\kms~would not be covered by the observations.

The Eastern Arm may in fact have several kinematic components. In the H92$\alpha$ 
line profiles along the portion of the Eastern Arm (see E1-E5 in Fig. 5c) 
that  \cite{paum04} termed the ``Ribbon'', one can clearly see the positive velocity
features around +70 to +110 \kms~that correspond to what Paumard et
al. (2004) called the ``Eastern Bridge'', a separate kinematic
component. In addition, a weak negative-velocity feature at $\sim-$100 km s$^{-1}$
is also present (see E7-E9 in Fig. 5c). Unfortunately, our H92$\alpha$ line data 
cube does not contain the high positive velocities ($>$ 220 \kms) that are found 
in the ``Tip'' feature, which may be related to the Eastern Arm (see  
Vollmer \& Duschl, 2000), 
although the two are quite distinct in the Br-$\gamma$ spectra of Paumard et al 
(2004). While Paumard et al. (2004) drew a distinction between the Eastern Arm
(their ``Ribbon'') and the Bar, we find that the quasi-continuity in
the position and velocity of the gas in the Eastern Arm (positions E1 to
E11) and the Bar (positions E13 to E17; see Figs. 5 and 20) reinforces
the possibility that these features can be unified as a single
dynamical entity, as was suggested by \cite{liszt03}. 
He advanced
a circular ring model, which connects the dynamics and placement
of the innermost parts of the Eastern Arm with those of the Bar, but
does not appear to reproduce the outer structure of the Eastern Arm.
Indeed, we show below in Section 5 that the radial
velocities and positions are consistent with motion around a single
Keplerian orbit.  However, the continuity is unclear at around
position E12, so the coherence of this feature remains to be
demonstrated.

\subsection{The sources in the Western Arc} 
The radio emission from the Western Arc (Fig. 6) is weak and only a few sources 
are compact and bright enough ($S>1.2$ mJy beam$^{-1}$) for proper motion 
measurements. Fig. 12 shows radio components with measurements of proper motion 
in the Western Arc. A compact radio component, X6, in the Northern part of the 
Western Arc shows a double structure (Fig. 12a). It shows a NE-ward motion with 
an implied transverse velocity of $V_{\rm t}=299\pm36$ \kms~(P.A.$=66\arcdeg\pm12\arcdeg$), 
and the radial velocity from the H92$\alpha$ line is $V_{\rm LSR}=31\pm11$ \kms~at 
this location. Somewhat to the east of the Western Arc, the compact radio component 
X24 is identified with the IR star MP-9.0-14.4 (Fig. 12b). The proper motion 
measurements suggest that X24 moves SW-ward with $V_{\rm t}=244\pm43$ 
\kms~(P.A.$=-147\arcdeg\pm9\arcdeg$) and $V_{\rm LSR}=-18\pm8$ \kms. 
In the Western Arc (Fig. 12c), the filament X23 shows a NW-ward motion with 
$V_{\rm t}=537\pm74$ \kms~(P.A.$=-25\arcdeg\pm6\arcdeg$) and 
$V_{\rm LSR}=-44\pm4$ \kms. Located 5\arcsec~NE to X23, X21 shows nearly W-ward 
motion with  $V_{\rm t}=187\pm44$ \kms~(P.A.$=-72\arcdeg\pm26\arcdeg$). 
The H92$\alpha$ line of X21 shows a double profile with peaks at 
$V_{\rm LSR}=-51\pm5$ \kms~and $-110\pm7$ \kms. The weaker component, 
X19, shows NE-ward motion with $V_{\rm t}=244\pm63$ \kms~(P.A.$=64\arcdeg\pm23\arcdeg$) 
and $V_{\rm LSR}=-30\pm4$ \kms. We note that the proper motions of the 
three components (X19, X21 and X23) appear to be inconsistent with each 
other, which could result from a substantial contribution of non-orbital
motions. If non-orbital motions are random, a vector average could help 
reduce the effect of non-orbital motion in the orbital fitting program 
discussed in Section 5. The mean proper motion in this region obtained 
by vector averaging the three components  is $\overline \mu_\alpha=-3.61\pm0.74$ 
mas yr$^{-1}$ and $\overline \mu_\delta=2.52\pm0.49$ mas y$^{-1}$, 
corresponding to $\overline V_{\rm t}=167\pm25$ \kms~(P.Ai.$=-55\arcdeg\pm13\arcdeg$) 
NW-ward.
 
\subsection{The sources near Sgr A*}
Fig 13 shows the radio images at 1.3 cm (color for the epoch 2005K) and 3.6 cm 
(contours for the epoch 2002X ) along with the positions of the IR stars in 
the IRS 16 region. The most prominent feature in both the 1.3 and 3.6 cm 
images is the linear feature ($\sim4$\arcsec~in length, $<0.$\arcsec5~in width), 
located at the NW edge of the Northern Arm. In the middle of the linear feature, 
there is an elongated  radio component, K20. The radio source K20 may be 
associated with the HeI star (HeI-N2), which shows a P-Cygni HeI profile
 \citep{paum01}. The radial velocity of the star determined from IR line 
observations, $-97^{-64}_{+56}$ \kms, is similar to that of the Northern Arm 
gas in the same direction, as determined from the H92$\alpha$ line. However, 
this coincidence may just be fortuitous. On the other hand, K20 appears to be 
the compressed, ionized edge of the Northern Arm, which was also observed clearly 
at 3.8 $\mu$m with the VLT \citep{muzi07}. This compression plausibly results from
the collective winds and radiation pressure from all of the high mass stars in the 
IRS 16 cluster.  

In addition, we searched the literature and only
found that MP+2.37--0.29 \citep{genz00} is close to HeI-N2, with 
proper motion measurement. The magnitude of the star velocity agrees 
well with that of K20 but the P.A. of the transverse
velocity between the two objects differs more than 100\arcdeg.
K20 shows a SW-ward proper motion of $V_{\rm t}=354\pm19$ 
\kms~(P.A.$=-136\arcdeg\pm4\arcdeg$), which is consistent with this compressional 
front being caused by the IRS 16 stars. Two Gaussian components were fitted to the 
broad spectrum of the H92$\alpha$ line at K20, the dominant one being at 
${V_{\rm LSR}=-41\pm11}$ \kms~with ${\Delta V_{\rm FWHM}} = 137\pm37$
\kms. The total velocity is then $V=356\pm20$ \kms. Considering the fact that 
the gas in the Northern Arm stream is located behind Sgr A* based on the LSQ 
fitting and numerical calculations in Section 5.5, the velocity vector of K20 suggests 
that the source moves toward its periapse in an orbit that carries it around 
Sgr A*. If the proper motion of the linear feature indicates the track of the 
gas motion governed by the gravity of the SMBH, we would expect the velocity of 
K20 to increase as it approaches Sgr A*. A comparison between its motion and 
what the dynamical models (Section 5) predict for the transverse velocity would 
be helpful in understanding the accretion process occuring 
in the immediate vicinity around Sgr A*.  

\subsection{Mini-cavity and IRS 13}

Fig. 14a shows the locations of a dozen  IR stars in 
the Mini-cavity region, including the IRS 13 and IRS 2 complexes 
to the West and the IRS 33 complex at the Eastern edge of the cavity. 
The proper motions 
of the compact continuum components are shown (Fig. 14b), 
suggesting that the bulk motion of the ionized gas in this region 
is toward the West. The source $\varepsilon$  \citep{yusef91} (K22), 
the closest component to Sgr A* (offset 1.\arcsec6), has a proper motion of 
$\mu_\alpha=-8.8\pm0.5$  mas y$^{-1}$ and $\mu_\delta=-1.4\pm1.5$ 
mas y$^{-1}$ or $V_{\rm t} = 338\pm21$  \kms~(P.A.$=-99\arcdeg\pm61\arcdeg$) 
to the SW, almost directly away from Sgr A*. We note that this is consistent
with the model of \cite{ward92}, who postulated that source $\varepsilon$ 
has resulted from the confluence of gravitationally focused winds from the IRS 16 stars.

The mean proper motion of the components K25, K28, K33, K40, and K45 in the 
IRS 33 region (NE rim of the ``Mini-cavity'') is 
$\overline \mu_\alpha=-9.3\pm0.2$ mas y$^{-1}$ 
and $\overline \mu_\delta=1.4\pm0.3$ mas y$^{-1}$ 
($\overline V_{\rm t}=355\pm8$ \kms, P.A.$=-81\arcdeg\pm12\arcdeg$), 
while for components K32 (source $\zeta$), K41 and K42 (source $\eta$) \citep{yusef90}
in the middle of the ``Mini-cavity'',
$\overline \mu_\alpha=-5.5\pm0.2$ mas y$^{-1}$ and $\overline \mu_\delta=1.7\pm0.5$ 
mas y$^{-1}$ ($\overline V_{\rm t}=218\pm9$ \kms, 
P.A.$=-73\arcdeg\pm16\arcdeg$). The components K47, K49, K53 and K54 
around the SW rim show $\overline \mu_\alpha=-5.1\pm0.2$ mas y$^{-1}$ and 
$\overline \mu_\delta=-1.2\pm0.4$ mas y$^{-1}$ ($\overline V_{\rm t}=199\pm8$ \kms,
P.A.$=-103\arcdeg\pm19\arcdeg$). The magnitude of the transverse velocity 
determined from proper motion measurements appears to be comparable to that 
of the radial velocity observed from the H92$\alpha$ line. 

The most surprising region is the IRS 13 complex (Fig. 15), with proper motion 
vectors pointing in many directions, suggesting that the local stellar winds 
might be playing a significant role in the motion of the ionized gas 
\citep{zhao98,muzi08}. However, averaging over the proper motion vectors 
from the HII components (K21, K23, K24, K26, K27, K29, K31, and K35) in 
IRS 13, the resultant vector is to the NW, that is, $\overline \mu_\alpha=-2.91\pm0.04$
mas y$^{-1}$ and $\overline \mu_\delta=2.50\pm0.07$ mas y$^{-1}$,
corresponding to $V_{\rm t}=145\pm2$ \kms~(P.A.$=-49\arcdeg\pm1\arcdeg$).

On the other hand, the observations of the H92$\alpha$ line show 
that in the ``Bar'' region, a bright emission feature south of 
Sgr A* \citep{eker83}, there are two distinct velocity components, 
one associated with the Northern Arm  and the other related to 
the ionized stream in the Eastern Arm \citep{rob93}. These components
are also shown in the Br-$\gamma$ line observations \citep{paum04}
although the authors did not regard the bar emission as part of Eastern Arm.
Fig. 16 shows the spectra at the HII components in the ``Mini-cavity'' 
region covering the velocity range from --415 to 200 \kms.
Around the NE rim, the positive velocity associated with the 
Eastern Arm stream is prominent while the negative-velocity
component associated with the Northern Arm stream is weak with 
broad-line profile ($\Delta V_{\rm FWHM} > 100$ \kms). In the Mini-cavity, 
the spectra close to Sgr A* are dominated by components with positive 
radial velocity. For example, the source $\varepsilon$ (K22) has a broad 
component at $V_{\rm LSR} =37\pm7$ \kms~with $\Delta V_{\rm FWHM}=63\pm16$
\kms~and a narrower ($\Delta V_{\rm FWHM}=22\pm15$ \kms)
feature at a high positive velocity ($V_{\rm LSR}=165\pm5$ \kms).
Of course, possible lines at $V_{\rm LSR} > 200$ \kms~may
have been missed due to the limited velocity coverage.
The negative-velocity component at $-275$ \kms~becomes 
strong about 5\arcsec~to the SW of Sgr A* and peaks at the Southern rim 
(K47) of the Mini-cavity. Along the Western rim, the negative-velocity 
gas shows a trend similar to that in the Mini-cavity, {\it i.e.} 
it appears to be strong, broad and localized at the SW rim (K43) and
it becomes weak as it moves closer to Sgr A*. In comparison with the 
negative-velocity component, the positive-velocity ($\sim$+50 \kms) 
component becomes stronger and broader as it gets closer to Sgr A*. 
The H92$\alpha$ line observations 
also suggest that the Mini-cavity was created by strong interactions between 
various flows. \citet{morr87} proposed that the Mini-cavity might be due to the  
winds from massive stars in this region. Other  interactions, such 
as the interaction between the  Eastern and Northern Arm streams 
or the interactions between outflows from Sgr A* and the streams might also
cause high temperature and therefore a depression in the radio free--free
emission. The high LTE temperature ($T_e^*$) in the Bar reported by
\cite{rob93} might be owing to such an interaction.
 
\subsection{The IRS 6 and IRS 34 region}
Fig. 17 shows the radio sources in the IRS 6/IRS 34 region,
which is located 6\arcsec~NW of Sgr A* (Fig. 6). In Fig. 18,
the H92$\alpha$ line shows a noticeable gradient in radial velocity 
from $V_{\rm LSR}=-102\pm10$ \kms~at (K19) 
to $V_{\rm LSR}=-184\pm5$ \kms~at (K11). 
This velocity gradient was also detected by \citet{rob96}. The velocity 
gradient observed in the H92$\alpha$ line agrees well with that observed 
in the Br-$\gamma$ line (see Fig.A.2b of Paumard et al. 2004). The H92$\alpha$ 
line spectrum at K19 (Fig. 3a) contains at least 
three Gaussian components. Averaging the vectors from the HII components 
(K10, K11, K13, K14, K17 and K19) in the IRS 6 region, we derive the mean 
proper motion of $\overline \mu_\alpha=-3.4\pm0.2$ and $\overline \mu_\delta=1.4\pm0.2$ 
mas y$^{-1}$, suggesting a bulk motion toward the NW 
($V_{\rm t}=139\pm8$ \kms, P.A.$=-68\arcdeg\pm8\arcdeg$).

High velocity of the Bullet (X12) was reported early based on
proper motion measurements at 2 and 1.3 cm \citep{yusef98,zhao98}. 
At 1.3 cm, the structure is resolved  to at least three knots with 
varying peak intensities in time, which results in large uncertainties in proper
motion measurements.  Instead, we use the low-resolution data at 3.6 cm
to determine the proper motion of the centroid position of the source. 
We confirm that the source moves NW but the magnitude of the transverse
velocity is reduced by a factor of $\sim$3: $V_{\rm t}=319\pm37$ 
\kms~(P.A.$=-53\arcdeg\pm12\arcdeg$). The revised transverse velocity appears 
to be consistent with the velocity of the ionized flow in the Eastern 
Arm at this location. However the radial velocities of the
two spectral features, $V_{\rm LSR}=-25\pm9$ \kms~and 
$V_{\rm LSR}=70\pm13$ \kms (Fig. 3c),
deviates significantly from the value at this location predicted from the Keplerian model.

\subsection{The overall three-dimensional velocity field}

We imaged the peak radial velocity values derived from the H92$\alpha$ line in 
the central 80\arcsec$\times$80\arcsec~(Fig. 18). The mean transverse velocities  
along the $x$ and $y$ axes  of the radio knots in the three ionized arms 
were determined by averaging the proper motions of several individual 
radio knots given in Table 2. The mean values of the transverse velocities 
with respect to Sgr A* ($\overline{V}_{x}$, $\overline{V}_{y}$), 
the mean radial velocity with respect to the LSR ($\overline{V}_{\rm LSR}$) and 
the mean position offsets from Sgr A* (${\rm \overline{\Delta \alpha}}$,
${\rm \overline{\Delta \delta}}$) were derived and are summarized in Table 4.
Column 1 lists the name of the region in which the three-dimensional velocities 
were determined. The radio IDs in column 2 indicate the radio knots that were 
included in the averaging. Column 3 gives the IRS names of the sources in the corresponding 
regions. Fig. 18 shows the images of the vectors of the transverse velocities
superimposed on the image of peak radial velocities in the  ionized arms
based on the VLA  observations of the H92$\alpha$ line. The proper motion 
vectors determined from the VLA observations suggest that all three streams 
appear to move around Sgr A* in the sense of counterclockwise rotation, as viewed  
from  the Earth.

\section{Dynamic models for the ionized streams}
\subsection{Three bundles of Keplerian orbits}

In the case of elliptical motion, adopting the distance ($D=8$ kpc), there are  
six orbital parameters ($a$, $e$, $\Omega$, $\omega$, $i$, $M_{\rm dyn}$)
to completely describe an orbit projected onto the plane of the sky. Here $a$ 
is the size of the semi-major, $e$ is eccentricity, $M_{\rm dyn}$ is 
the dynamical mass, $\Omega$ is the longitude of the line of nodes, 
$\omega$ is the angle of periapsis with respect to the line of nodes, 
and $i$ is the inclination angle. Based on the assumptions that the positions 
determined in Section 3, which follow the locus of the stream in each of the arms,
are located along a single, Keplerian orbit, and that the central mass 
is fixed at the location of Sgr A*, the orbital parameters ($a$, $e$, 
$\Omega$, $\omega$, $i$) for the orbital geometry can be determined with 
LSQ fitting to those positions. The dynamical mass 
($M_{\rm dyn}$), on the other hand, can be subsequently determined by
fitting the radial velocities in each of the streams. The detailed procedure 
for determining  the orbital parameters is developed and described 
in the Appendix.

We apply this technique to Sgr A West. Fig. 19 shows the fits (solid ellipses) 
to the observed loci (dots) for each of the three streams. For the loci along 
each stream, four sets of solutions for the orbital parameters ($a$, $e$, 
$\Omega$, $\omega$, $i$) can be derived from the LSQ fitting of the quadratic 
equation (A1). The first two geometrical parameters ($a$, $e$) are the same  
for all the solutions. The four combinations of the three angular parameters 
($\Omega$, $\omega$, $i$) correspond to the two mirror images about the sky plane,
and in each of these, the direction of flow can be clockwise or counterclockwise.
The proper motions can be used to decide between clockwise and counterclockwise, 
and the sign of the radial velocities determines which of the mirror images is 
correct.

Fig. 18 shows the  measured vectors of the transverse velocities distributed 
in each of the three streams, showing that the ionized streams are in 
counterclockwise rotation around the dynamical center. The remaining degeneracy of the 
two mirror images can be solved by the radial velocity determined from the 
H92$\alpha$ line. Thus, we can unambiguously determine the orbital parameters 
for the three major streams in Sgr A West.

There is no principle dictating that the gas arrayed along any given
stream must all be on the same Keplerian orbit. Indeed, gas in a tidally sheared 
cloud, which may describe these streams, will in general be different orbits, 
depending on where it was located in the original cloud. However,
except under unusual circumstances, such as a radially infalling
cloud, the orbits assumed by the different parcels of gas within a shearing 
cloud will be rather similar, and the ensemble of orbits will be a family
of similar orbits. We therefore adopt the simplification of a single
orbit in order to characterize the orbit family. The assumptions underlying
this procedure are somewhat vindicated by our success in
fitting single orbits to the streams.

Table 5 summarizes the LSQ solutions of the orbital parameters  
for the three streams from the best fits. Both the Northern Arm and 
Eastern Arm streams appear to be in Keplerian motion on high-eccentricity
elliptical orbits of $e=0.83\pm0.10$ and  $0.82\pm0.05$, respectively.
Our best fit for the Western Arc stream indicates that it is in a nearly circular 
orbit ($e=0.20\pm0.15$), consistent with what has previously been suggested 
\citep{rob93,lacy91}. Constrained by  proper motion measurements which show that
all  three streams rotate in a counterclockwise sense  around Sgr A*, the 
inclination angles must be greater than 90\arcdeg. From the best-fitted 
solutions above, we can unambiguously determine the inclination angles to be  
139\arcdeg$\pm10$\arcdeg, 122\arcdeg$\pm5$\arcdeg~and 117\arcdeg$\pm3$\arcdeg~for 
the Northern Arm, Eastern Arm and Western Arc, respectively.

\subsection{Dynamical mass in Sgr A West}

The above analysis does not yield information on the mass.
In order to determine the dynamical mass, velocity information is required.  
If the radial velocity of Sgr A* with respect to the LSR, 
the transverse velocity of Sgr A* with respect to that of the 
dynamical center and the position offsets of Sgr A* from the 
dynamical center can be neglected, then  $V_z=V_{\rm LSR}$ 
and the location of Sgr A* defines  the dynamic center. 
For a given total velocity $V=\sqrt{V_x^2+V_y^2+V_z^2}$
at position ($x,y,z$), the dynamical equation relating $V$ to the mass
($M_{\rm dyn}$) is 

\begin{equation} 
V={M_{\rm dyn}^{1/2}\left(\frac{2}{r} 
-\frac{1}{a}\right)^{1/2}},
\end{equation}
\noindent where $r$ is the radial distance from  the dynamical center, 
which can be expressed as a function of the
projected coordinates ($x,y$) in  Equation (B2).

In principle, the dynamical mass in Sgr A West can be derived by 
fitting the total velocity to Equation (1). Since only a few measurements 
of proper motion in the outer region of Sgr A West are available, the fit 
of the measured total velocities leaves a sizeable uncertainty in the 
determination of the dynamical mass. On the other hand, our radial velocity 
measurements from each of the three orbits cover a large fraction of the 
complete orbits of each flow, so we can use the radial velocities to 
constrain the dynamical mass with Equation (B1). With the best-fit 
orbital parameters ($a$, $e$, $\Omega$, $\omega$, $i$), we solve Equation 
(B1) using our orbit models for three values of the central
mass, that is,  $M_{\rm dyn} =1.2,~4.2,~10\times10^6$ M$_\odot$,
and compare the predicted radial velocities to those measured at 
each position along the orbits. Fig. 20 shows the results of fitting in 
each of the three streams. In general, the model with a mass of  
$M_{\rm dyn}=4.2\times 10^{6}$ M$_\odot$ fits the data, consistent 
with the prevailing view  that the mass of Sgr A*  determined by 
fitting the orbits of the IR stars \citep{ghez08,ghez03,scho02} 
dominates the central dynamics. However, not surprisingly, there are 
some deviations between the model and data (see Fig. 20). The deviations 
of the radial velocity in the inner region  can be attributed  to some local 
non-gravitational effects such as winds, MHD flows, magnetic pressure 
gradients, and possibly interactions between streams. 
Deviations at larger radii might result from the increasingly 
important contribution of stars to the enclosed mass as 
the radius increases, thus causing deviations from pure 
Keplerian orbits. Of course, many of these apparent deviations could 
simply result from the failure of the basic assumption in this analysis: 
that each of the three streams follows a unique Keplerian orbit.
In any case, the total dynamical mass in the central 3 pc appears 
to be $\sim1\times10^7$ $M_\odot$.

\subsection{Model calculation and three-dimensional location of the three streams}
The fitted parameters are summarized in Table 5 for the three streams in
Sgr A West. Using these parameters, we have modeled
these streams as the  model of three bundles of 
Keplerian orbits. In addition to the best-fitted orbital parameters 
(Table 5), we also used  the  dynamical mass of 
$M_{\rm dyn}=4.2\times10^6$ $M_{\odot}$ and a dispersion of 
$\Delta a =\pm 0.25a$ in the semi-major axis
to roughly match the width of the orbits in
each of the streams. The ranges of the true anomaly
192\arcdeg -- 460\arcdeg, 195\arcdeg -- 430\arcdeg, and 
130\arcdeg -- 420\arcdeg, were used to calculate the three 
streams for~the Northern~Arm,
the Eastern~Arm, and the Western~Arc, respectively.

The comparison between the observed kinematics of Sgr A West and 
the calculation based on the best-fitted orbital model is shown in Fig. 21.
The radial velocities ($V_{z}$) of the streams are plotted in 
Fig. 21b. The vectors of the transverse velocity ($V_{ x},V_{y}$) 
are illustrated in Fig. 21c. The three-dimensional velocity field calculated from 
our model appears to be in good agreement with the observations.
In other words, the three bundles of  elliptical orbits along with
the orbital parameters ($a$, $e$, $\Omega$, $\omega$, $i$)
derived from the LSQ fitting are consistent with the model 
calculations.

The locations of the three streams along the  line of sight are shown in 
Fig. 21d, suggesting that most of the orbiting ionized gas in the ``Bar'' 
region is located behind Sgr A*. The highly inhomogeneous distribution of 
the ionized gas allows  the low-frequency ($\nu\sim0.3$ GHz) 
radio emission from Sgr A*, which is subject to strong free--free absorption,
to pass through the gap in the ionized streams, as was suggested from 
the VLA and Giant Metrewave Radio Telescope (GMRT) observations \citep{ant05}.
 
The total velocity and kinetic energy per unit mass were also calculated and
are shown in Figs. 21e and 21f, respectively. The figures show that 
the specific kinetic energy of both the Northern and Eastern Arms reaches a maximum in
the ``Bar'' region. The high specific kinetic energy density there emphasizes the possibility 
that the Mini-cavity was created by strong shocks resulting from the collision 
of the two streams.

\section{A magnetically determined morphology in the Northern Arm?}
The dynamical model discussed above is based on a spherical gravitational 
potential centered on  Sgr A*.  However, the magnetic field in the Northern 
Arm might be strong enough to cause observable deviations from the kinematics 
determined by gravity alone. Polarization measurements at 12.5 $\mu$m of 
thermal dust emission arising in the Northern Arm indicate that the dust 
grains in this structure are extremely well aligned by a uniform magnetic field 
oriented along the long axis of the arm \citep{aitk98,glas03}. 
The uniformity  of the field direction is attributed to shear in the direction of the flow 
\citep{aitk91}, which is inevitable, given that the finite breadth of the 
Northern Arm gives rise to a significant differential gravitational 
force across the arm. The magnetic field strength estimated from dynamical 
considerations and from the application of the method of \citet{chan53} 
is $\gtrsim$ 2 mG. 

Our high-resolution image at 1.3 cm (Fig. 22) reveals a  suggestive 
sinuous structure in the Northern Arm that  could have a magnetic origin.  
Although we do not know the line-of-sight structure of this feature, 
the morphology is consistent with a projected helix having 1.5--2 well-defined 
wavelengths between declinations (J2000) of --29\arcdeg~00\arcmin~17.\arcsec5 and 
--29\arcdeg~00\arcmin~22\arcsec. To illustrate this, we show model-projected 
helices in Fig. 22 for comparison with the observed feature. 
The velocity ($V_x=-18$ \kms, 
$V_y=-223$ \kms, $V_z=70$ \kms) at this location ($\Delta x =7$\arcsec, 
$\Delta y =10$\arcsec) is obtained from the orbital model. The tilt (the 
angle of the velocity vector with respect to the line of sight) 
of this Northern Arm fragment is arccos$(Vz/V)\sim$72\arcdeg,
which is reasonably consistent with the best-fit projection angles of the model
helices.  A helix is a natural morphology for a twisted magnetic field frozen 
into a plasma; the twist could result, for example, from torques exerted on the 
plasma by reconnection between the Arm field and the ambient field at the surface of 
the Northern Arm. This process is important in the Earth's magnetotail, and results
from the ``draping'' of the solar wind field around the Earth's magnetosphere.  
It is possible that a similar process occurs in the Northern Arm where the 
head of the Northern Arm plows through the ambient Galactic center field.  

A sinuous morphology in the Northern Arm had previously been noted by \cite{yusef89}
in a lower-quality 2-cm VLA image (see, e.g., Fig.  1 of
\cite{yusef90}) and this structure is evident in Fig. 22 of
this paper.  However, in that case, the wavelength was suggested 
to be 0.4 pc, much larger than the 0.12 pc wavelength of the feature 
considered here. \cite{yusef93} interpreted this longer wavelength waviness in 
terms of Rayleigh--Taylor or Kelvin--Helmholtz instabilities caused by winds from the young
stars in the IRS 16 cluster. Perhaps there is a homologous structure at multiple 
scales in the Northern Arm, or perhaps there is an evolution toward 
increasing wavelength from N to S along the arm as the gas accelerates 
toward the black hole. In order to investigate this possibility, we 
considered a model helix in which the wavelength of the helical turns 
increases southward along the Northern Arm as Sgr A* is approached. This is illustrated by
the model helix superimposed on the right side of Fig. 22. Although the observed structure
becomes more diffuse to the south, there is a good correspondence between this model
and the observations.

Observations of the mid-IR polarization vectors and 
of the velocity field with spatial resolution comparable to that in Fig. 22 
would provide a test of the hypothesis that a strong magnetic field imposes 
a helical geometry on the otherwise laminar flow of the Northern Arm.  
If the pitch angle of $\sim$15\arcdeg~reflects the ratio of 
velocities perpendicular to and along the arm, $\sim$0.25, then velocity perturbations
as large as 60 km s$^{-1}$ are predicted, using a total flow velocity of ~250 km s$^{-1}$
at this location.  In radial velocity only, the perturbation may be only a few tens of
km s$^{-1}$. In addition the mid-IR polarization vectors should everywhere be 
tangent to the helix. 

\section{Summary and Conclusion}

Using the VLA, we have observed the Galactic center with 
an angular resolution of 0.\arcsec1 at 1.3 cm. Along with the data 
obtained at previous epochs at 1.3 and 3.6 cm, we measured the proper 
motions of 71 compact HII components in Sgr A West. Using VLA archival data, 
we constructed an H92$\alpha$ line image cube with angular and velocity 
resolutions of 1.\arcsec25 and 15 \kms, respectively, covering
the velocity range from +200 \kms~to --415 \kms. 

With the observed proper motion and  radial velocity measurements 
from the ionized gas, we have investigated the three-dimensional velocity distributions 
in the central 3 pc. The loci and motions of the three well-known 
ionized streams in Sgr A West can be modeled with Keplerian motions in 
three bundles of elliptical orbits. We developed a procedure using a 
least-square fit of the loci of the ionized streams at the 
Galactic center. The orbital parameters ($a$, $e$, $\Omega$, $\omega$, $i$)
were determined confirming that the three ionized streams corresponding to
the three radio continuum features (Northern Arm, Eastern Arm and Western Arc) 
are confined within the central 3 pc. The dynamics appear to be 
dominated by the SMBH at Sgr A*, and the total dynamical mass within a 
radius of 1.5 pc is $\lesssim1\times10^7$ M$_\odot$.
With the Keplerian model and the fitted orbital parameters, 
we numerically calculated the kinematics and spatial distribution of the 
ionized gas. The calculated three-dimensional velocities are in good agreement with the 
observed velocity distribution, based on the observations
of  proper motions and radial velocities. This numerical model 
confirms that three bundles of elliptical orbits with the orbital parameters 
($a$, $e$, $\Omega$, $\omega$, $i$) derived from our LSQ fitting and an assumed mass
for the SMBH
of 4.2$\times10^6$ M$_\odot$ at the dynamical center are  
a good approach to modeling the dynamics of the ionized gas 
in the central 3 pc.

With the determined orbital parameters for the ionized 
streams, our numerical calculations suggest that the Northern and 
Eastern streams might have collided in the ``Bar'' region located 
5\arcsec~SW of Sgr A*. Such a substantial interaction between the two 
streams could  have  contributed to the formation  of 
the Mini-cavity.

In the IRS 16 region located 2\arcsec~E of Sgr A*, the 
high-resolution observations reveal a linear radio feature (K20).
The collective stellar winds and radiation 
from the IRS 16 cluster are likely responsible for compressing the
edge of the Northern Arm at this location, and causing it to be 
particularly bright.
 
In the IRS 8 region, a radio bow-shock structure is observed,
similar to that revealed by IR observations. A shell-like nebula
is revealed by the H92$\alpha$ line image showing
a narrow feature ($\Delta V_{\rm FWHM} =52\pm17$ \kms) at 
$V_{\rm LSR}=-12$ \kms. The H92$\alpha$ line observations show that 
the NE part of this nebula physically interacts with the Northern Arm.   
   
Finally, from the high-resolution VLA observations, a helical structure
is revealed in the Northern Arm, suggesting that, in addition to the 
predominantly Keplerian dynamics attributable to the central mass, MHD effects
play a role in the production of the detailed structure in the orbiting ionized 
gas. The predictions from the MHD model are testable with future high-resolution 
observations of the mid-IR polarization vectors and the velocity field.
\acknowledgments
The VLA is operated by the National Radio Astronomy 
Observatory (NRAO). The NRAO is a facility of the National Science Foundation 
operated under cooperative agreement by Associated Universities, Inc.
M.R.M. has been partially supported in this work by NSF grant AST-0406816 to UCLA. 
T.A. has been supported by the NSFC (10503008). 
\appendix

\section{LSQ fit of an ellipse to a projected orbit} 
In a Cartesian coordinate system with +$x$-axis (East), +$y$-axis (North), 
+$z$-axis (pointing away from the Earth), and the origin at the dynamical center, 
the coordinates giving the locus of an elliptical orbit projected on the 
plane of the sky must satisfy
\begin{equation}
{
\frac{(yT_{11} - xT_{21})^2}{(b {\rm cos} i )^2} + \frac{(xT_{22}+a e {\rm cos} i
-y T_{12})^2}{(a {\rm cos} i)^2}=1
}
\end{equation}

\noindent where $T_{ij}$ are the elements of the coordinate transformation
matrix that is used to project the orbit plane onto  the sky plane.
$T_{ij}$ are functions of the orbital angles
($\Omega$, $\omega$, $i$), that is,
\begin{displaymath}
T_{11} = +{\rm cos} \omega {\rm cos} \Omega - {\rm sin} \omega {\rm sin} \Omega
{\rm cos} i
\end{displaymath}
\begin{displaymath}
T_{12} = -{\rm sin} \omega {\rm cos} \Omega - {\rm cos} \omega {\rm sin} \Omega
{\rm cos} i
\end{displaymath}
\begin{displaymath}
T_{21} = +{\rm cos} \omega {\rm sin} \Omega + {\rm sin} \omega {\rm cos} \Omega
{\rm cos} i
\end{displaymath}
\begin{displaymath}
T_{22} = +{\rm sin} \omega {\rm sin} \Omega + {\rm cos} \omega {\rm cos} \Omega
{\rm cos} i.
\end{displaymath}

The other parameters required to describe the orbit geometry
are the eccentricity $e$, semi-major axis $a$, and semi-minor axis 
$b=a\sqrt{1-e^2}$. Using Equation (A1) as the input constraint equation for
fitting an orbit, we utilized the LSQ subroutine LSQGEN of \citet{bran99},  
which is based on an algorithm of orthogonal transforms. The values
of the orbital parameters are derived by this program along with their 
1$\sigma$ errors computed from the square root of the covariance matrix 
under the assumption that the measurement errors are small.
The assumption of small measurement errors in the 
determination of the loci ($\sim\pm$1.\arcsec25, 
the radius of the circles in Fig. 5a) is valid in this paper.

\section{Dynamical mass constrained by radial velocities}

Under the assumption of Keplerian motion, the dynamical mass
of a system can be constrained by the observed radial velocities
at the projected location ($x,y$) and the five orbital parameters 
($a$, $e$, $\Omega$, $\omega$, $i$) determined from the LSQ fit of 
the orbit geometry as discussed above,
\begin{equation}
V_z = \frac{M^{1/2}_{\rm dyn}}{\sqrt{a(1-e^2)}} \left[T_{32}\left(e+
{\rm cos}\nu\right)
-T_{31}
{\rm sin }\nu \right],
\end{equation}
where $\nu$ is the angle of the true anomaly:
\begin{displaymath}
{\rm cos} \nu = \frac{a (1-e^2)}{er} - \frac{1}{e},
\end{displaymath}
where $r$ is the distance of the object with respect to the dynamical center:
\begin{equation}
r= \sqrt{(x {\rm cos} \Omega + y {\rm sin} \Omega)^2 + \frac{1}{{\rm cos}^2 i}
(y {\rm cos} \Omega - x {\rm sin} \Omega)^2},
\end{equation}
and the corresponding elements of the transformation matrix are
\begin{displaymath}
T_{31}={\rm sin}\omega{\rm sin}i,
\end{displaymath}
\begin{displaymath}
T_{32}={\rm cos}\omega{\rm sin}i.
\end{displaymath}

\clearpage
\begin{figure}[ht]
\includegraphics[width=44pc,angle=-90]{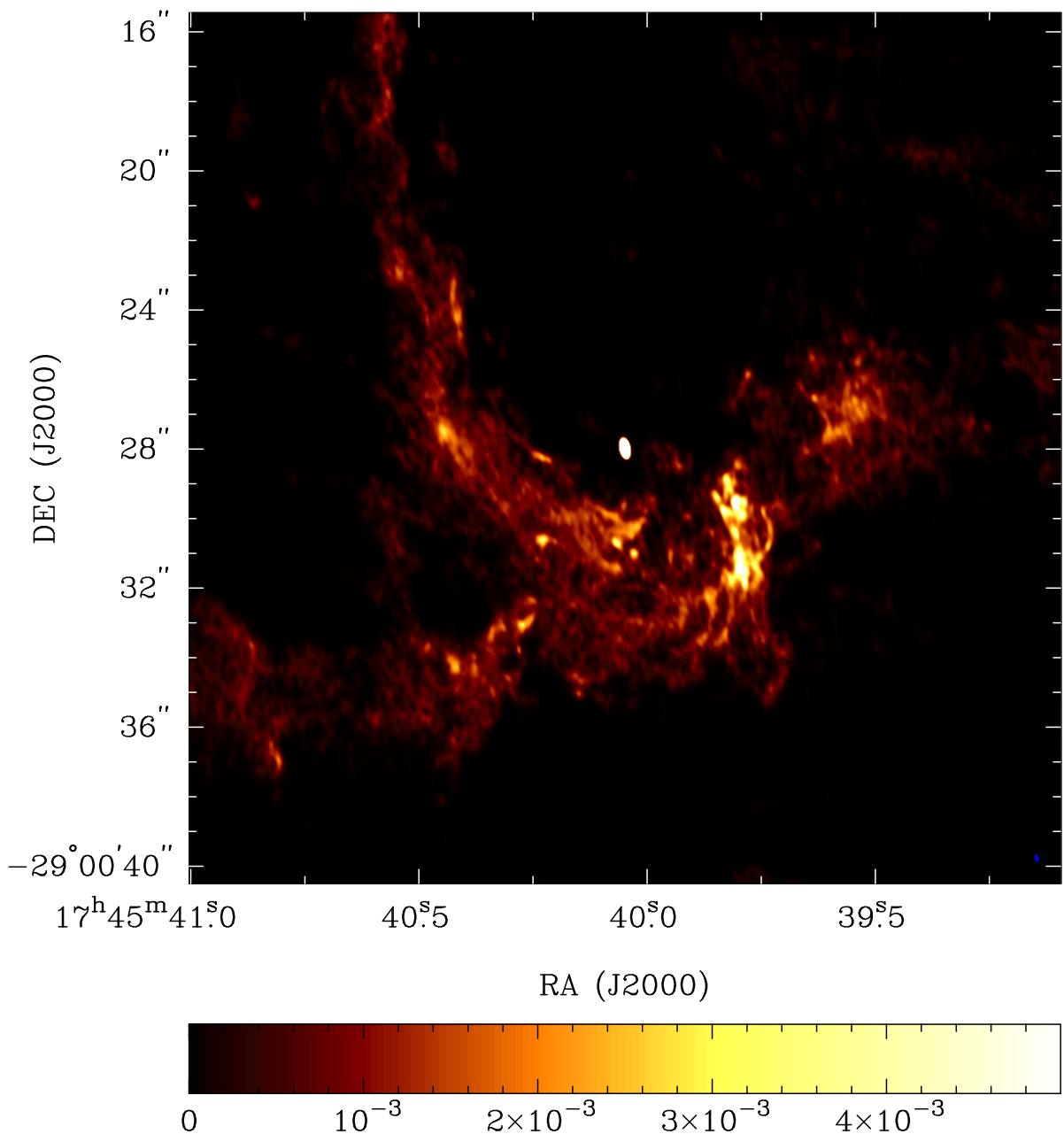}
\caption{
The VLA continuum image of the central 25\arcsec~(833$\times$833 pixels)
of Sgr A West at 1.3 cm, made with the epoch 2005K  data with
FWHM=0.2\arcsec$\times$0.1\arcsec (P.A.=11\arcdeg) (Right-bottom).
The size of the original image is 8192$\times$8192 pixels, constructed
with all the visibility data (not filtered) sampled with the VLA in A and
B configurations. The color wedge (bottom) shows the intensity of
the continuum in units of Jy beam$^{-1}$. The rms noise is
20 $\mu$Jy. 
\label{fig1}
}
\end{figure}
\clearpage
\begin{figure}[ht]
\centering
\includegraphics[width=40pc]{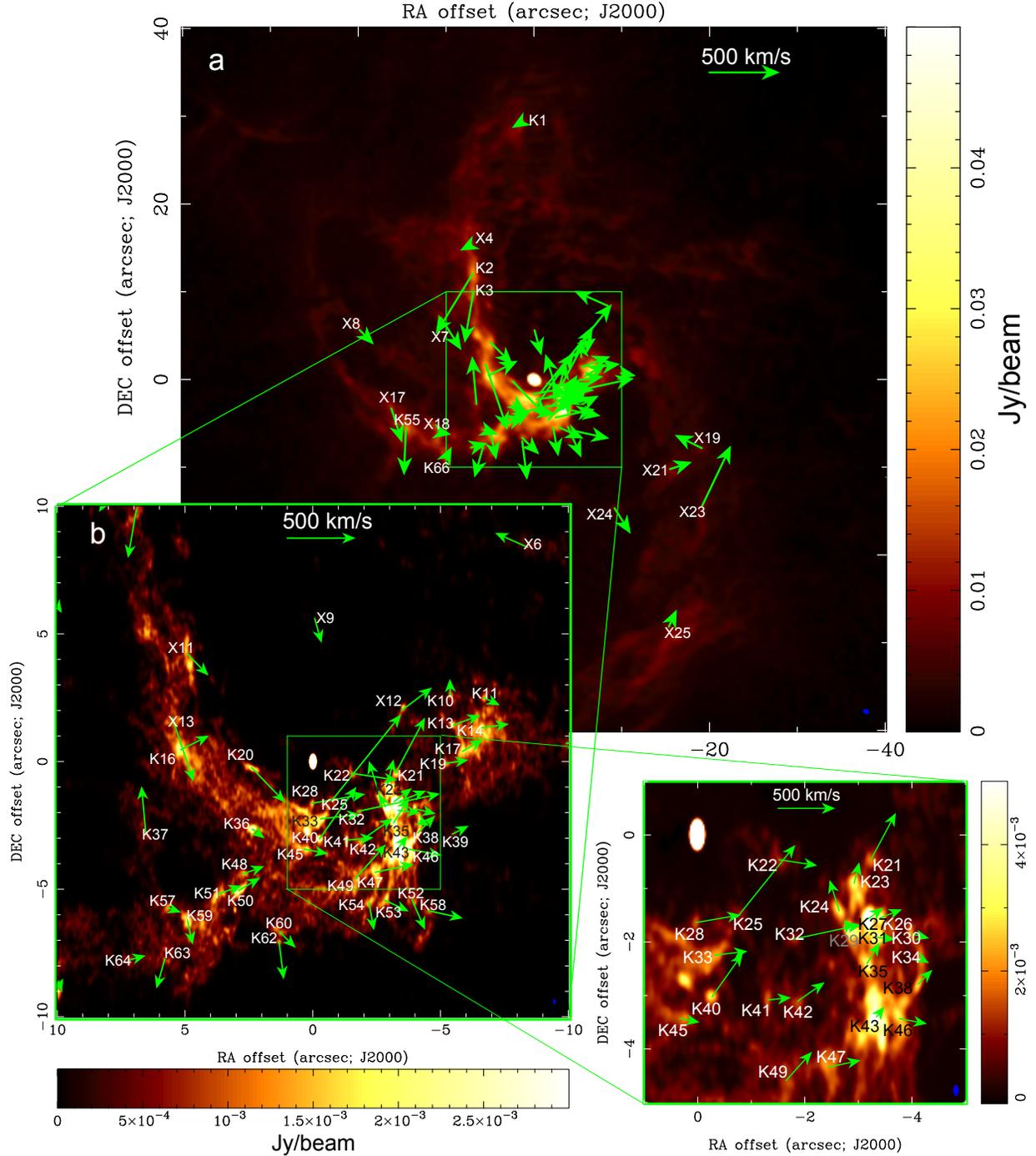}
\caption{
a. The VLA continuum image of Sgr A West at 3.6 cm (background) made from combining 
all the data listed in Table 1, with all the baselines that are not filtered. The 
FWHM of the beam (right-bottom corner) is 0.70\arcsec$\times$0.57\arcsec~(P.A.=57\arcdeg). 
The color wedge to the right shows the intensity of the continuum emission in units 
of Jy beam$^{-1}$. Proper motion vectors of the 71 HII knots (Table 2) in Sgr A 
West are overlaid on the radio continuum image.  
b. The inset (foreground) shows the proper motion
vectors in the inner 20\arcsec$\times$20\arcsec~region,
overlaid on the continuum image at 1.3 cm (Fig. 1).
c. The inset at left-bottom shows 
the inner 6\arcsec$\times$6\arcsec~region.
The velocity scaling vectors  for proper motions
are shown on each of the images.  
\label{fig2}}
\end{figure}
\clearpage
\begin{figure}[ht]
\figurenum{3a}
\centering
\includegraphics[width=45pc,angle=-90]{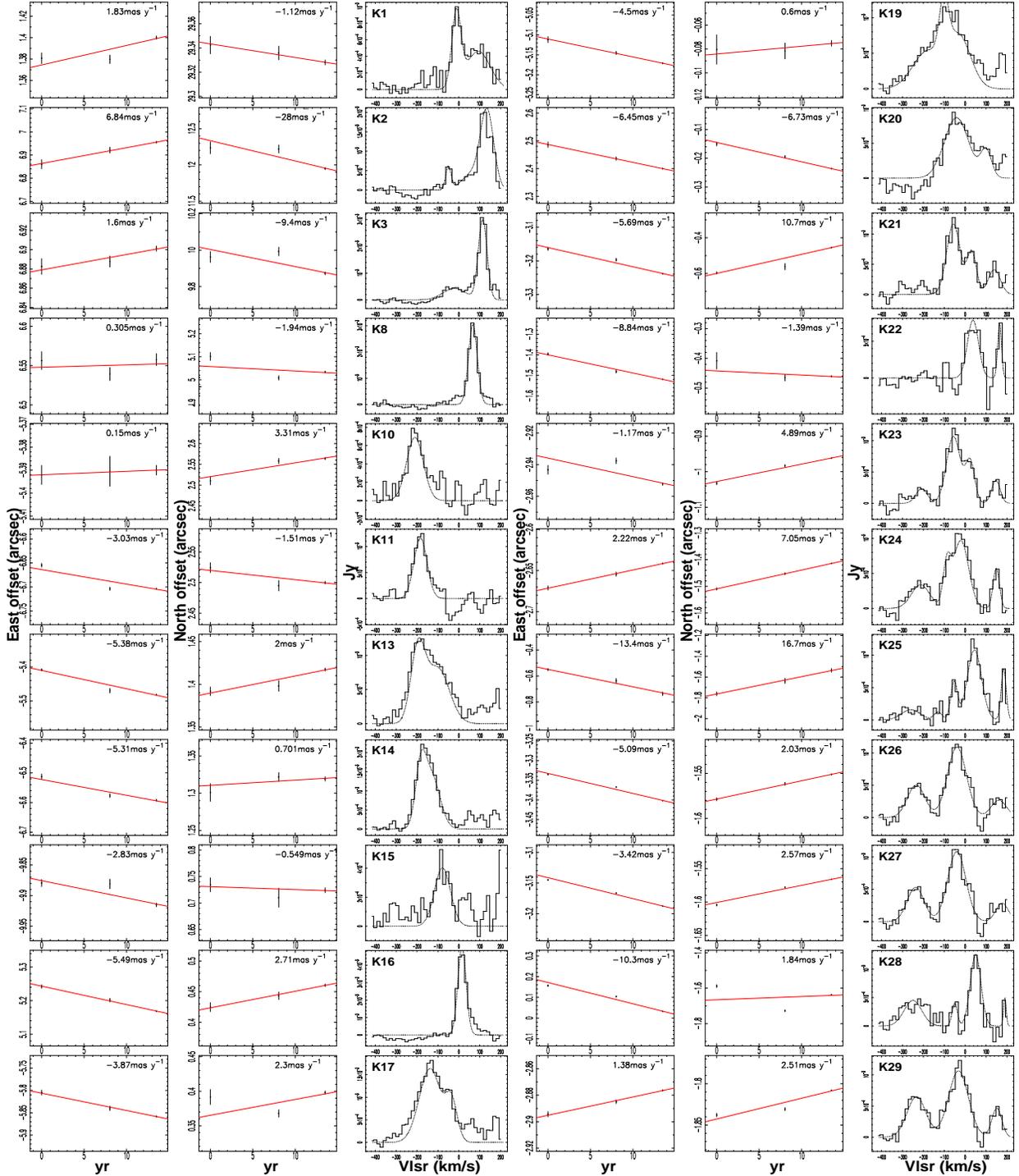}
\caption{Plots of the eastward (col. 1 \& 4) and
northward (col. 2 \& 5) position offsets from Sgr A* vs. time,
and the H92$\alpha$ line spectra (col. 3 \& 6) in LSR for 
70 HII components. The component IDs are labelled
(top-left) on each  H92$\alpha$ line spectrum panel.
The uncertainty (1$\sigma$) in position offsets
is marked with a vertical bar. The linear least-square fitting
(red) to the position offsets weighted by $\sigma^2$
was used to determine the slope (proper motions) of the 
displacements. The determined  proper motion are labelled on each of
the corresponding panels. The least-square fitting to the
spectra with multiple Gaussian (dashed curves) was
used to determine the radial velocities
of each component. The results are summarized in Table 2.
\label{fig3a}}
\end{figure}
\begin{figure}[ht]
\figurenum{3b}
\centering
\includegraphics[width=53pc,angle=-90]{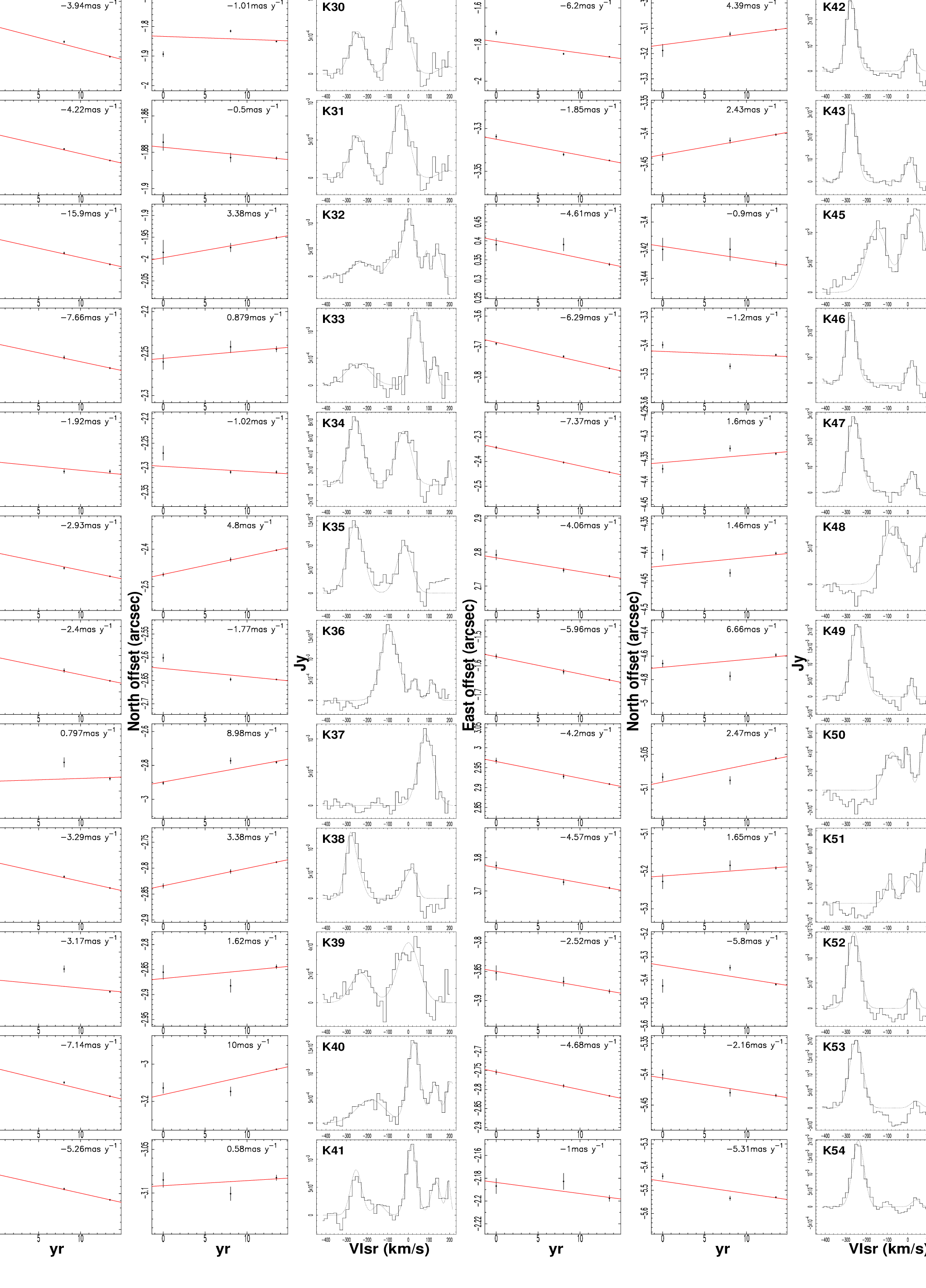}
\caption{Continued: Proper motions and radial velocities
\label{fig3b}}
\end{figure}
\clearpage
\begin{figure}[ht]
\figurenum{3c}
\centering
\includegraphics[width=53pc,angle=-90]{f3c.eps}
\caption{Continued: Proper motions and radial velocities
\label{fig3c}}
\end{figure}
\clearpage
\begin{figure}[ht]
\figurenum{4}
\centering
\includegraphics[width=40pc,angle=-90]{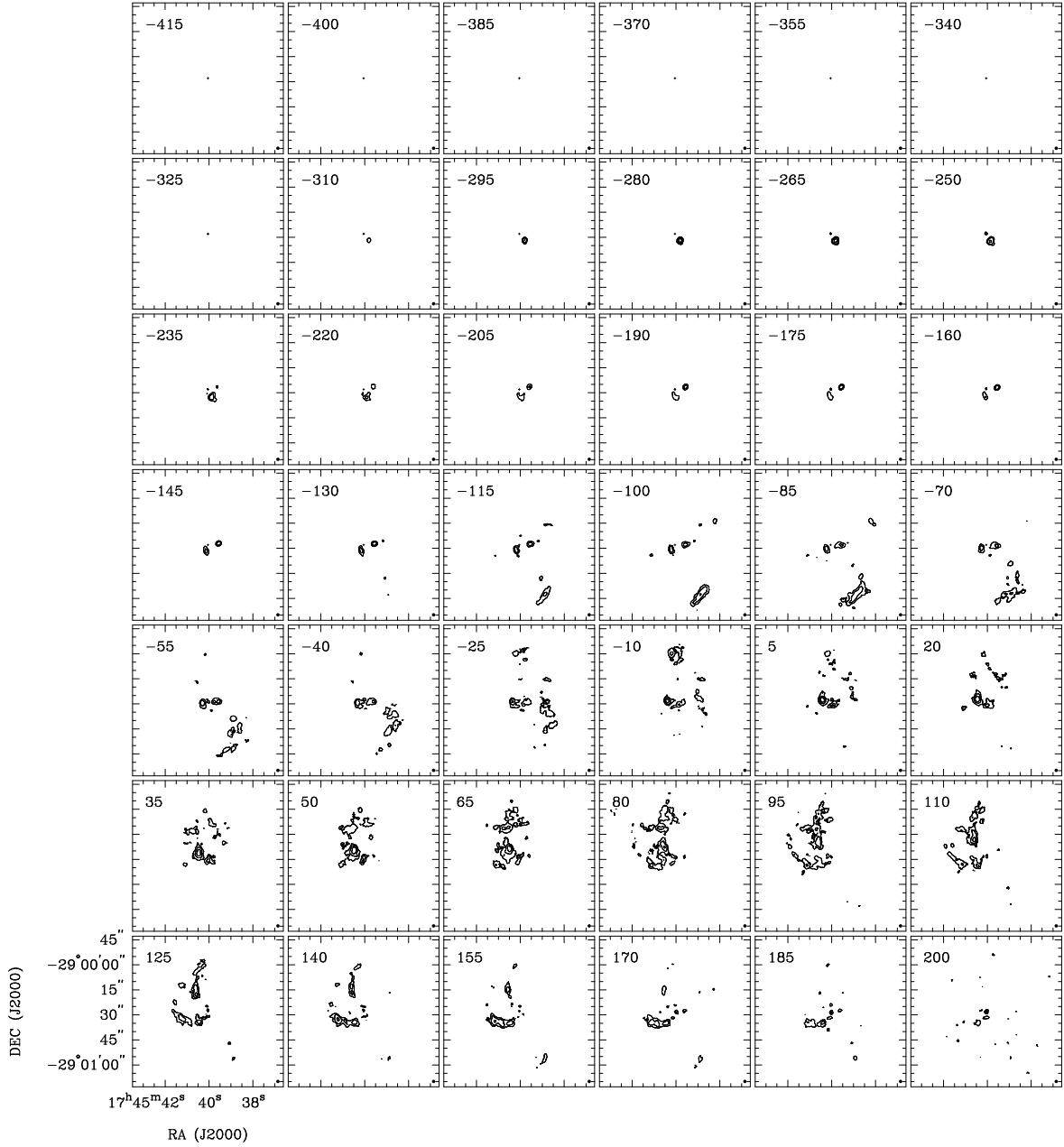}
\caption{
Channel images of H92$\alpha$ line convolved with
a circular beam of 2$^{\prime\prime}$. The contours are
0.27 mJy beam$^{-1}\times$(--5,5,10,20,40,80,160).
The LSR velocities (--415 to +200 \kms)
are labelled on each of the channel
images. The rms is in the range 0.15-0.27 mJy beam$^{-1}$. \label{fig4}}
\end{figure}
\clearpage
\begin{figure}[ht]
\figurenum{5a,b}
\centering
\includegraphics[scale=0.5,angle=90]{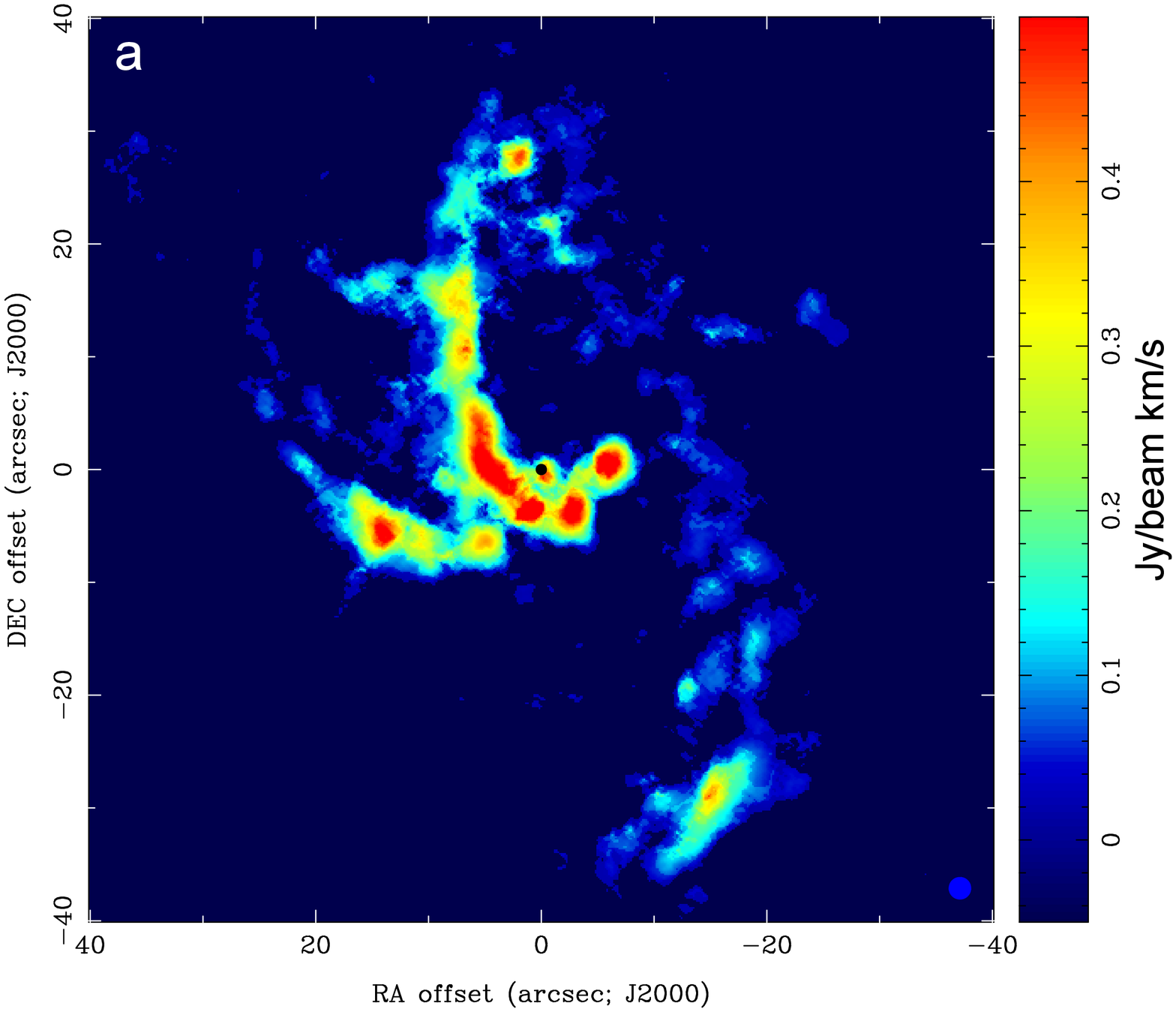}
\vfil
\includegraphics[scale=0.5]{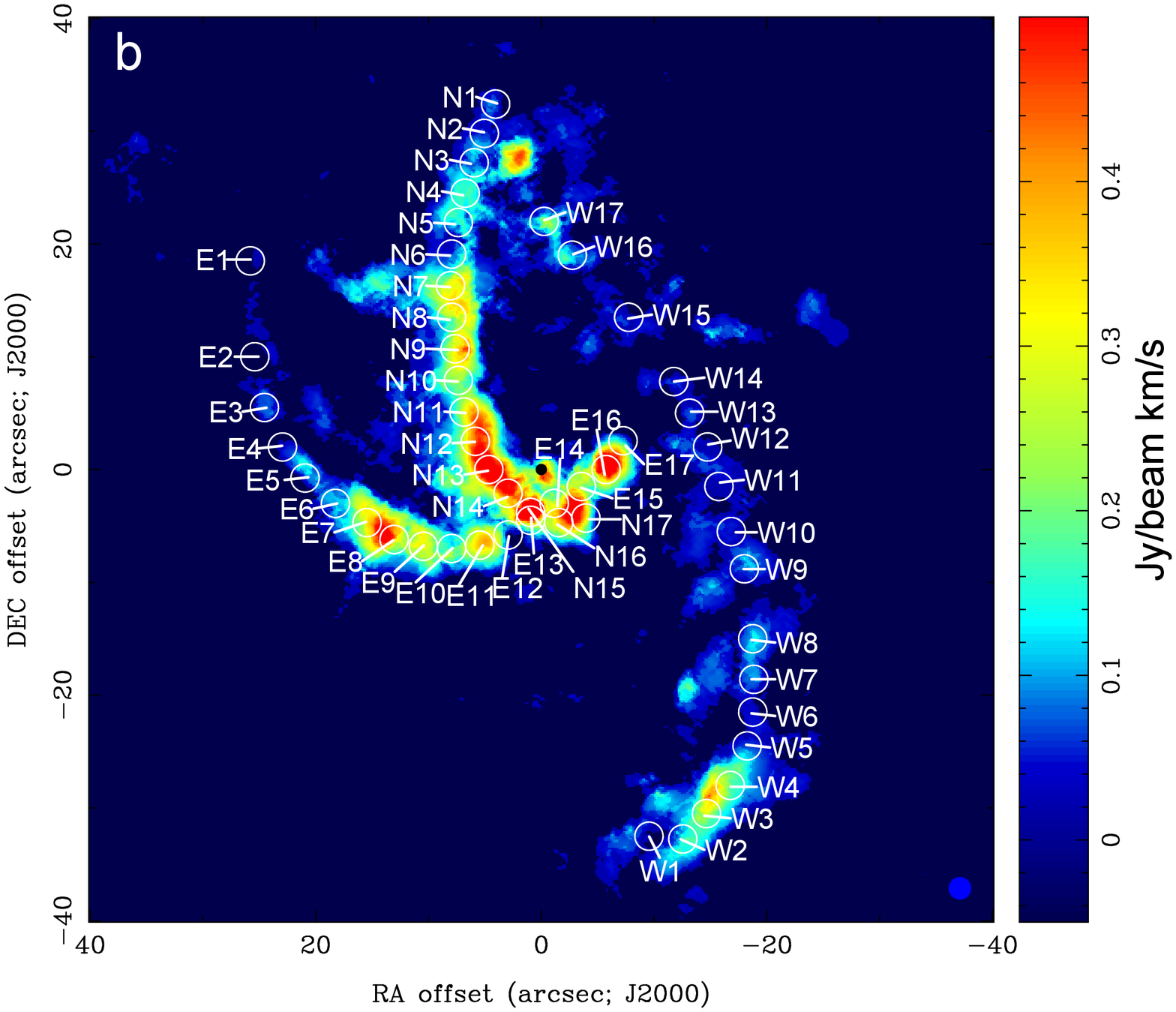}
\caption{The integrated H92$\alpha$ line intensity image 
scaled (in color). The FWHM beam (right-bottom) is 2\arcsec. 
The solid dot in the images marks the position of Sgr A*.
The locations of the spectra shown in Fig. 5c are marked with
the white circles in Fig. 5b and their positions are tabulated in
Table 3. 
\label{fig5a,b}}
\end{figure}
\clearpage
\begin{figure}[ht]
\figurenum{5c}
\centering
\includegraphics[angle=-90,scale=0.75]{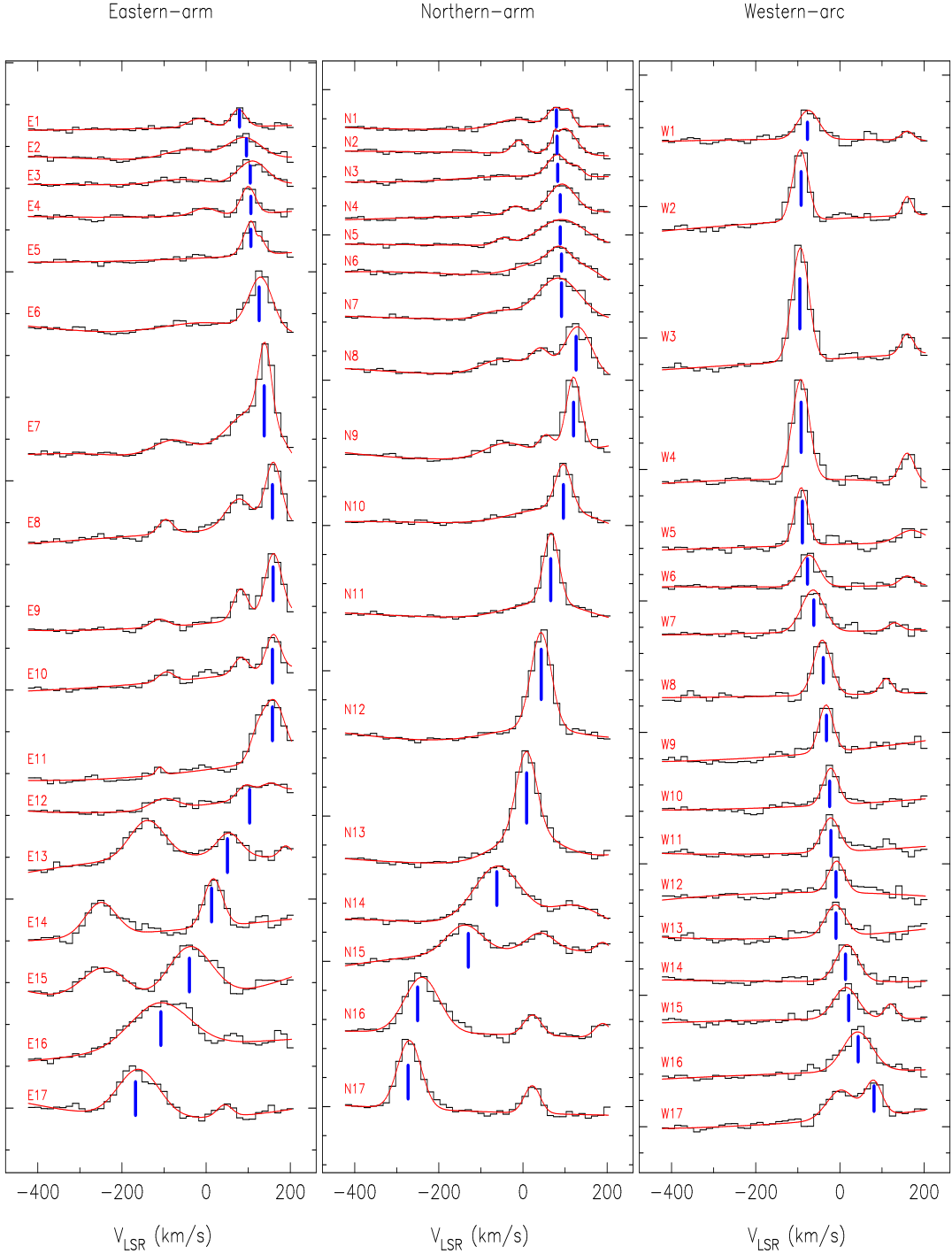}
\caption{
The spectra from a sequence of locations (marked in Fig. 5a) in each of the 
arms. The red curves are the LSQ fits with multiple-gaussians. The blue bars 
indicate the observed peak velocity of the kinematic components for the 
corresponding streams. The measurements of radial velocities are summarized
in Table 3.
\label{fig5c}}
\end{figure}
\clearpage
\begin{figure}[ht]
\figurenum{6}
\centering
\includegraphics[width=37pc]{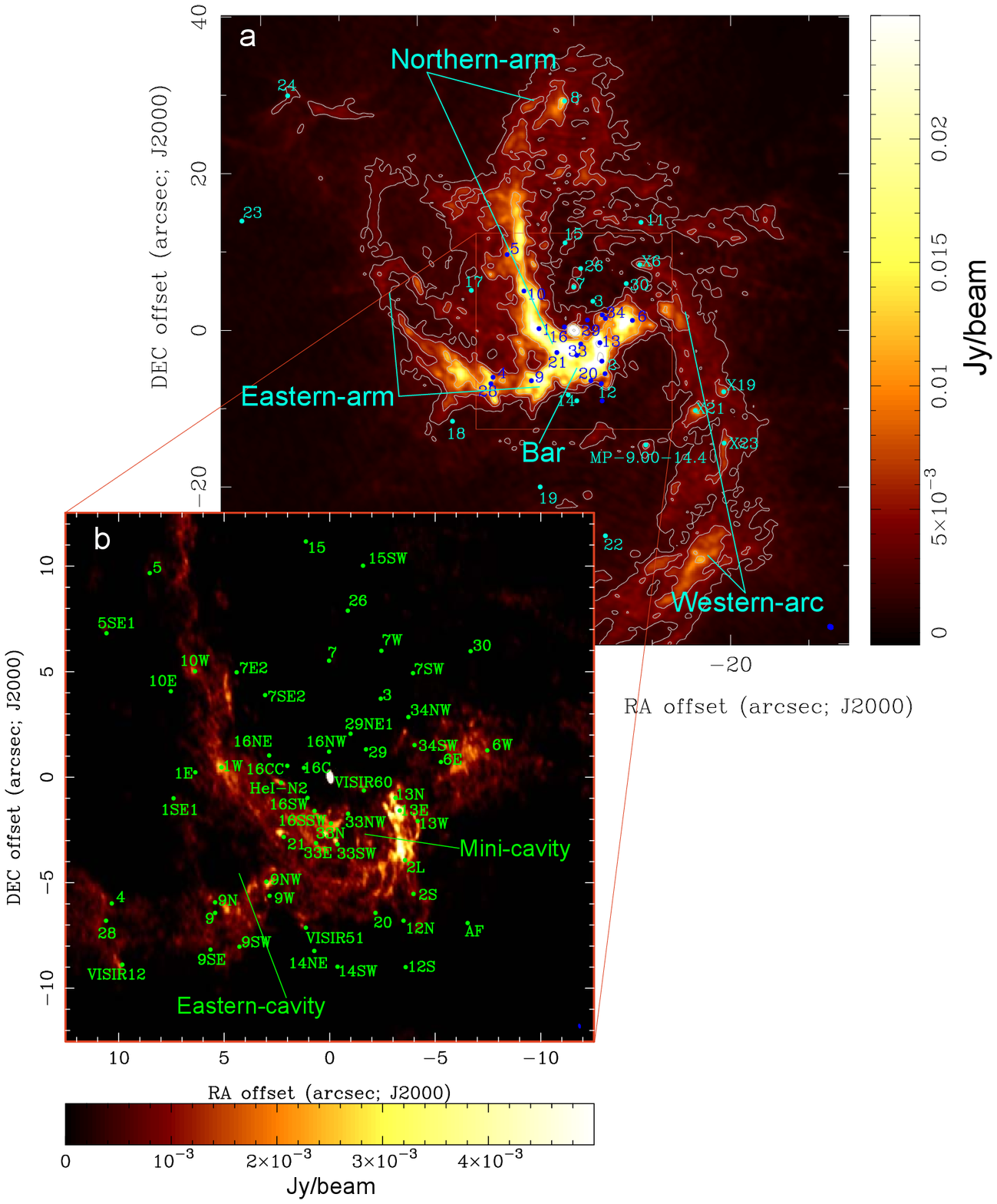}
\caption{The finding chart of IR sources at the Galactic center.
a. The VLA image (FWHM=0.70\arcsec$\times$0.57\arcsec, P.A.=57\arcdeg)
shows the radio emission at 3.6 cm from Sgr A West. Contours are
2.5 mJy beam$^{-1}\times$(1,2,4,16,256). b. The inset of 
the VLA image (FWHM=0.2\arcsec$\times$0.1\arcsec, P.A.=11\arcdeg)
shows the inner 20\arcsec~region of radio emission at 1.3 cm.
The dots mark the IRS positions
along with IRS ID numbers
\citep{blum96,genz00,geba06,paum04,paum06,vieh06}.
\label{fig6}}
\end{figure}
\clearpage
\begin{figure}[ht]
\figurenum{7}
\centering
\includegraphics[width=32pc,angle=-90]{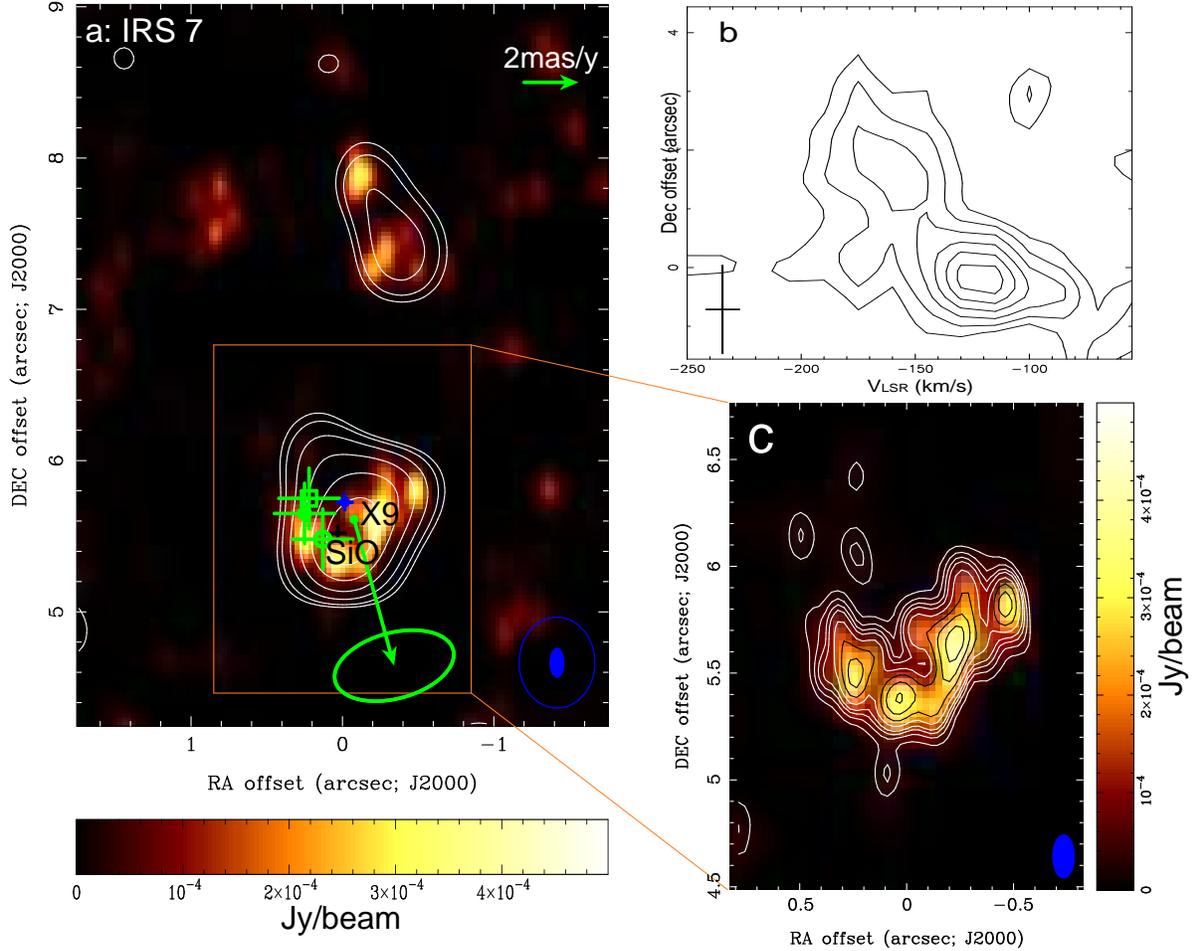}
\caption{a: Image of IRS 7 at 3.6 cm (contours, at epoch 2002X) and 
1.3 cm (color-scale, at the epoch 2005K). Contours are 0.15 mJy 
beam$^{-1}\times$(3,4,6,9,13,18). The FWHM beams are shown bottom-right 
for 3.6 cm (empty ellipse, 0.6\arcsec$\times$0.5\arcsec, P.A.=0\arcdeg) 
and 1.3 cm (solid ellipse, 0.2\arcsec$\times$0.1\arcsec, P.A.=0\arcdeg). 
The proper motion vector of X9 is shown with the scaling vector of 
2 mas y$^{-1}$ at top right. The 2$\sigma$ uncertainty in the proper 
motion is scaled with the open ellipses for magnitude (the semi-axes 
along the vector) and direction (the semi-axes perpendicular to the 
vector). The coordinates are the position offsets from Sgr A*. The 
green plus signs mark the IR positions obtained from various groups:
open circle from \citet{blum96}; open square from \citet{genz00};
dot from \citet{vieh06}. The blue plus-dot marks the radio position
from \citet{yusef89}  converting B1950 coordinates to J2000. The 
black plus marks the  position of the SiO maser \citep{reid07}. b: 
the velocity-versus-Dec diagram (H92$\alpha$) along the Declination 
cut across the position of X9 (IRS 7) given in Table 2. The plus
sign (bottom-left) indicates the FWHM beam size in Dec and the 
channel width in velocity. Contours are 0.15, 0.31, 0.46, 0.62, 
0.77 and 0.93 mJy beam$^{-1}$. The coordinate in Dec in the PV-diagram 
is the position offset from X9. c: inset of IRS 7 bow-shock at 1.3 cm. 
Contours are 44,66,88,132,198,286,396 $\mu$Jy beam$^{-1}$.
\label{fig7}}
\end{figure}
\clearpage
\begin{figure}[ht]
\figurenum{8}
\centering
\includegraphics[width=45pc,angle=-90]{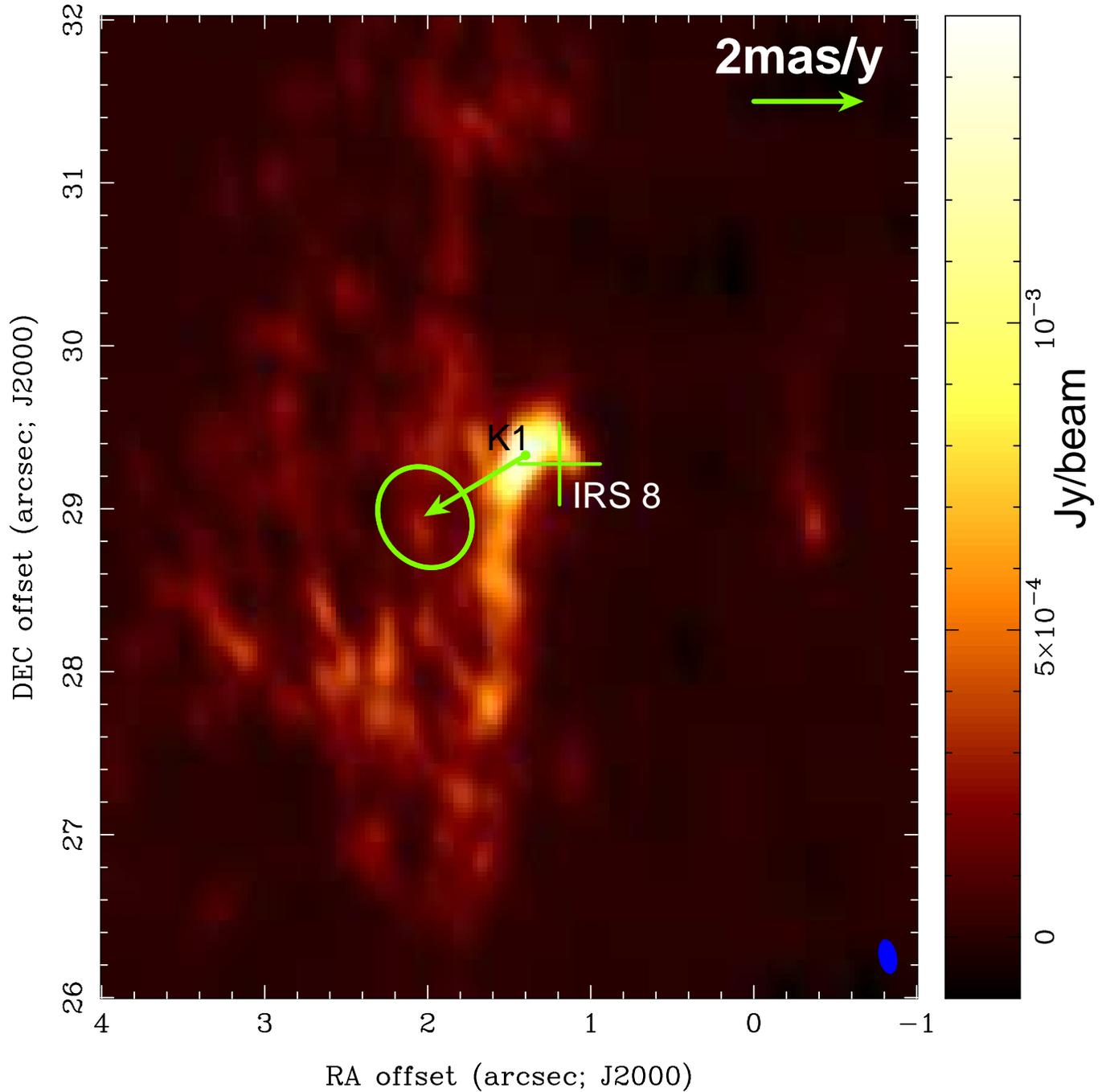}
\caption{Image of the IRS 8 bow-shock (K1) at 1.3 cm and the proper motion 
vector (scaled with the vector of 2 mas y$^{-1}$ at top-right ). The color 
wedge scales radio intensity in units of Jy beam$^{-1}$. The FWHM of the 
beam is 0.21\arcsec$\times$0.11\arcsec (P.A.=11\arcdeg). The semi-axes of the 
open ellipse scale the 2$\sigma$ uncertainty of the proper motion. The plus
sign is the star position of IRS 8  \citep{geba06}.
\label{fig8}}
\end{figure}
\clearpage
\begin{figure}[ht]
\figurenum{9}
\centering
\includegraphics[width=45pc,angle=-90]{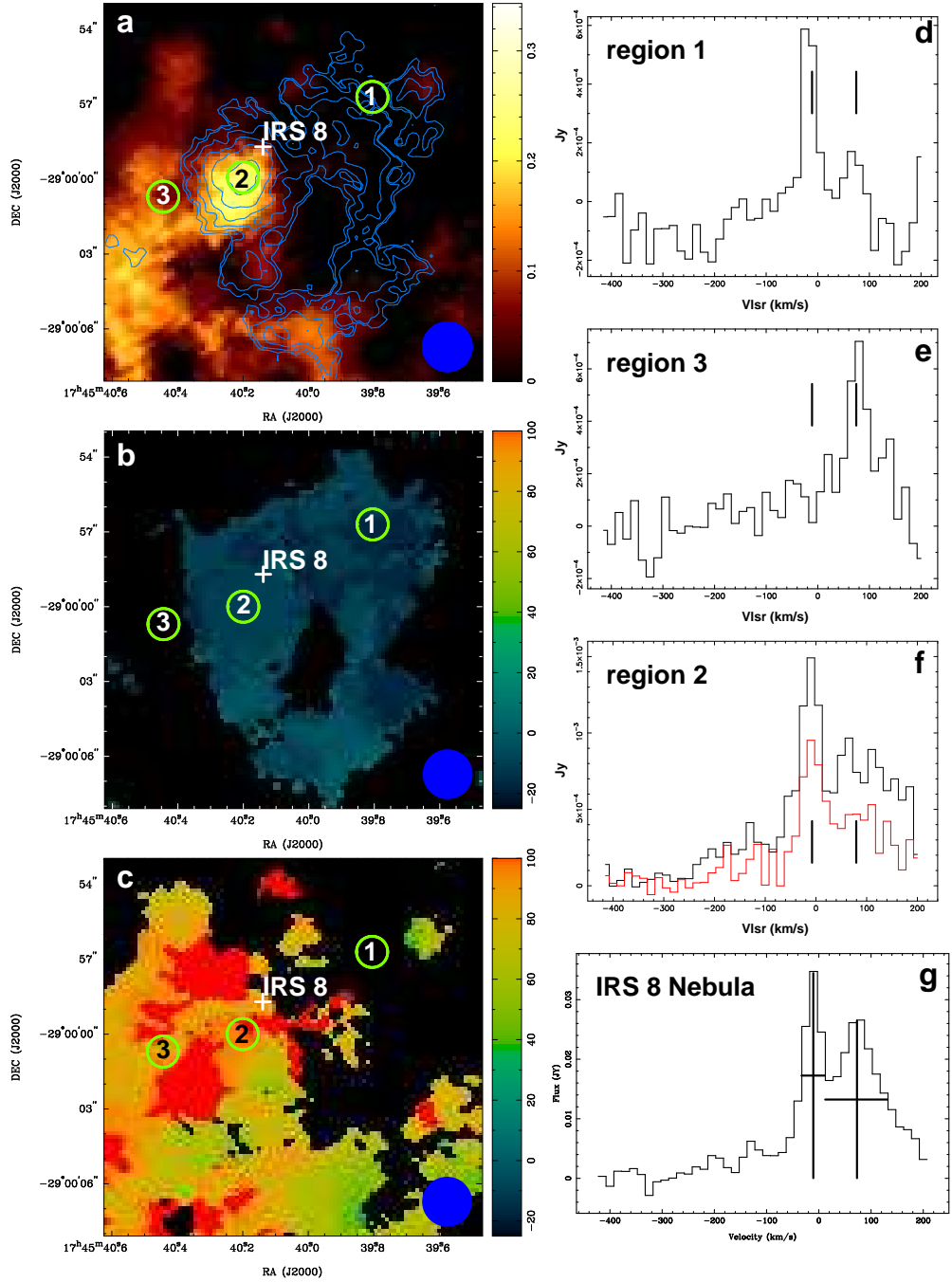}
\caption{Gas dynamics toward IRS 8: (see next page) \label{fig9}}
\end{figure}
\clearpage
\begin{figure}[ht]
\figurenum{9}
\caption{Gas dynamics toward IRS 8:
a: blue contours show the H92$\alpha$ recombination line intensity 
integrated from $-25$ to 5 \kms [contours: 5 mJy beam$^{-1}$ km s$^{-1}$
$\times$ (8,9,11,14,18,23)], with a circular beam of 2\arcsec, as 
shown in the bottom right corner, are overlaid on the color-scale 
image of intensity integrated between 35 and 170 \kms; this range 
corresponds to the Northern-Arm emission. The plus sign indicates 
the position of the stellar IR source. The circles indicate 
the three regions for which spectra are shown at the right: (1) a 
location at which there is emission only from the IRS 8 nebula, (2) 
a location corresponding to the bow shock, at which there is emission 
from both the Northern-Arm and the IRS 8 nebula, and (3) a location 
showing line emission only from the Northern Arm. b: plot of 
intensity-weighted velocity of gas in the range associated with the 
IRS8 nebula \citep{geba04}. The plus sign indicates the position of
the stellar IR source. c: Intensity-weighted velocity for the 
Northern Arm component, restricted to 35 to 170 \kms. For both the 
upper left and bottom left figures, the color wedge shows the velocity 
scale between $-25$ and +100 \kms. d, e, f: the spectra (black) on 
the right correspond to the three positions indicated, and are taken 
from the high-angular resolution (1.25\arcsec) line cube. The spectrum 
(red) in Fig. 9f is taken from the continuum peak (head of the bow shock) 
as compared with that (black) of region 2 (tail). g: the spectrum 
integrated from the region with a size of 6\arcsec$\times$5\arcsec 
(P.A.=0\arcdeg) centered at a position ($-$1.\arcsec2, 26.\arcsec3) 
offset from Sgr A*. The vertical bars indicate the peak velocities 
$-17$ \kms~and 76 \kms~for the nebula and bow-shock components, 
respectively.
\label{fig9}}
\end{figure}
\clearpage
\begin{figure}[ht]
\figurenum{10}
\centering
\includegraphics[width=37pc,angle=-90]{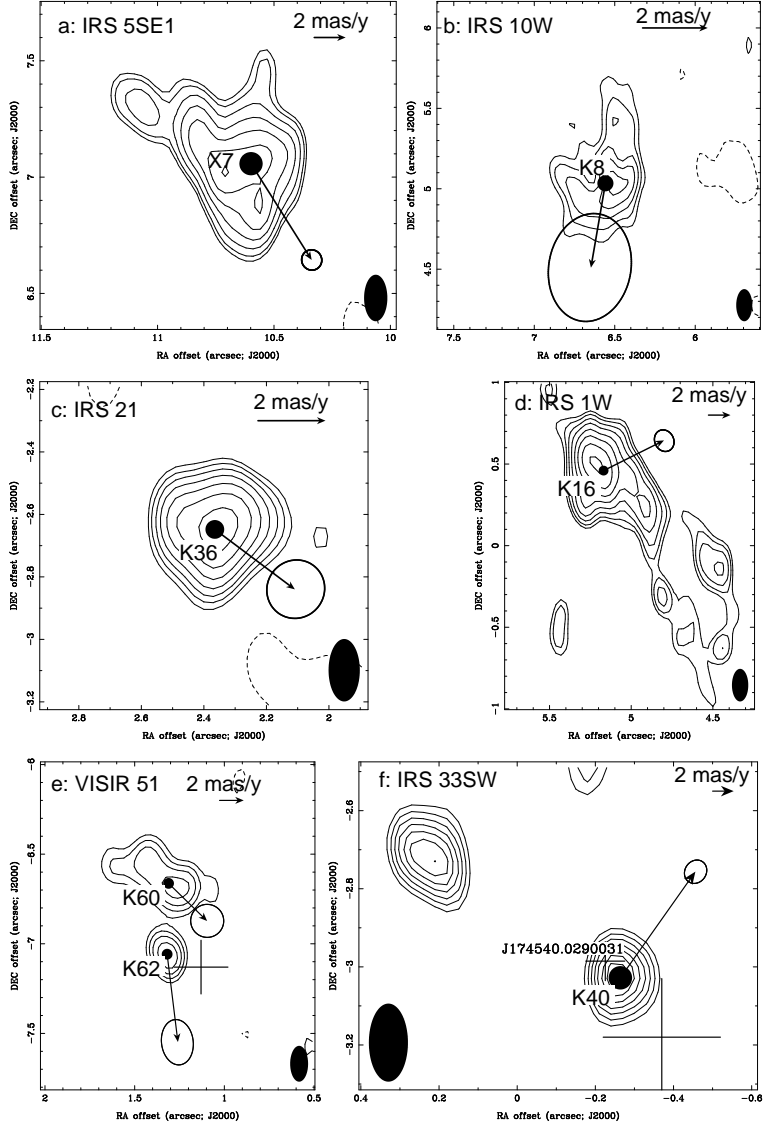}
\caption{Images of  compact radio sources at 1.3 cm in the Northern Arm, made 
from the 2005K epoch data with the long baseline data only ($>100{\rm k}\lambda$).
The vectors mark the proper motions, with the 2$\sigma$ uncertainty scaled with 
the open ellipses for magnitude (the semi-axes along the vectors) and direction 
(the semi-axes perpendicular to the vectors). The horizontal vectors on the
top-right of each panel represent a proper motion of 2 mas y$^{-1}$. a: radio 
emission (X7) of IRS 5SE1. The contours are 0.035 mJy beam$^{-1}\times$(--3, 3, 
4, 6, 9, 13, 18, 24 and 31). b: radio source  (K8) of IRS 10W with contours
of 0.035 mJy beam$^{-1}\times$ (--9, 9, 13, 18, 24 and 31). c: radio source (K36) 
of IRS 21 with contours of 0.035 mJy beam$^{-1}\times$ (--6,6,9,13,19,28,41 and 59). 
d: radio source (K16) of IRS 1W with contours of 0.035 mJy beam$^{-1}\times$ 
(--6,6,8,11,15,21,28,37 and 47). e: radio emission (K60 and K62) of VISIR 51 with
contours of 0.035 mJy beam$^{-1}\times$ (--8,8,12,17 and 24). f: radio emission 
(K40) of XJ174540.0290031/IRS 33SW with contours of 0.035 mJy beam$^{-1}\times$  
(--24,24,30,36,42,48 and 54). The FWHM beam shown in all the panels is 
0.2\arcsec$\times$0.1\arcsec (0\arcdeg). \label{fig10}}
\end{figure}
\clearpage
\begin{figure}[ht]
\figurenum{11}
\centering
\includegraphics[width=37pc,angle=-90]{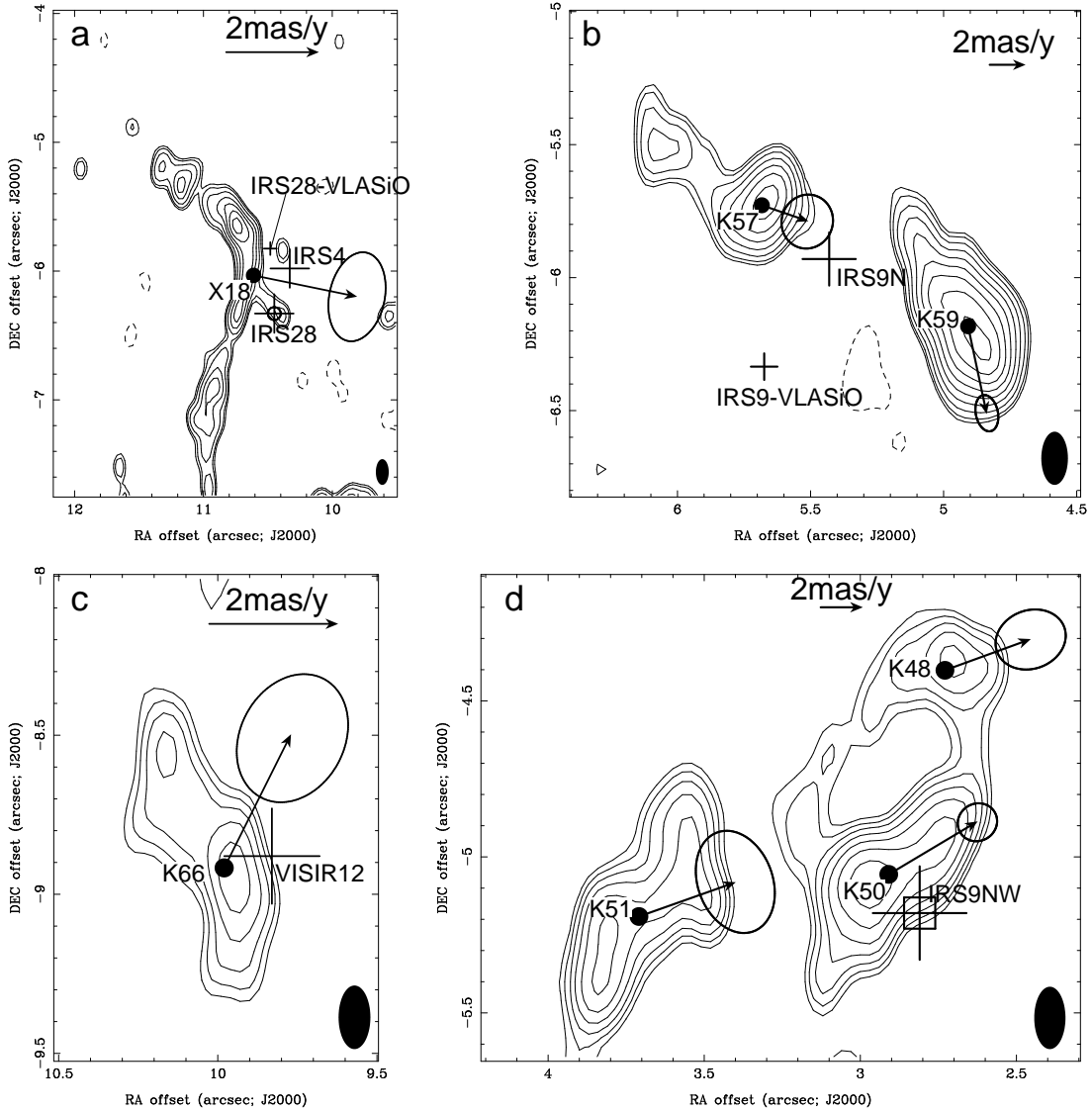}
\caption{Images of the compact radio continuum sources in the Eastern-Arm made 
from 2005 epoch data at 1.3 cm with the longer baselines ($>100k\lambda$).
a: radio emission (X18) of IRS 28/IRS 4/IRS 28-VLASiO. The contours are 0.035 mJy 
beam$^{-1}\times$(--4, 4, 5, 7, 10, 13, 17 and 22). b: radio emission (K57 and K59) 
of IRS 9N/IRS9-VLASiO. The contours are 0.035 mJy beam$^{-1}\times$
(--12, 4, 5, 7, 10, 14, 19, 25, 32 and 40). c:  radio emission of VISIR 12 (K66)
in contours of 0.035 mJy beam$^{-1}\times$(--8, 8, 12, 17, 23 and 30).
d: radio emission of IRS 9NW (K50). The contours are 0.035 mJy beam$^{-1}\times$
(--6, 6, 8, 11, 15, 21, 28, 36, 45 and 55). The FWHM beam is 0.\arcsec2$\times$0.\arcsec1 
(0\arcdeg). The vectors mark the proper motions of the radio sources with the 
open ellipses showing  the 2$\sigma$ errors in magnitude (the semi-axes along
the vector) and the direction (the semi-axes perpendicular
to the vector). The top-right vector on each panel gives the scale
for the proper motions (2 mas y$^{-1}$). \label{fig11}}
\end{figure}
\clearpage
\begin{figure}[ht]
\figurenum{12}
\centering
\includegraphics[width=38pc,angle=-90]{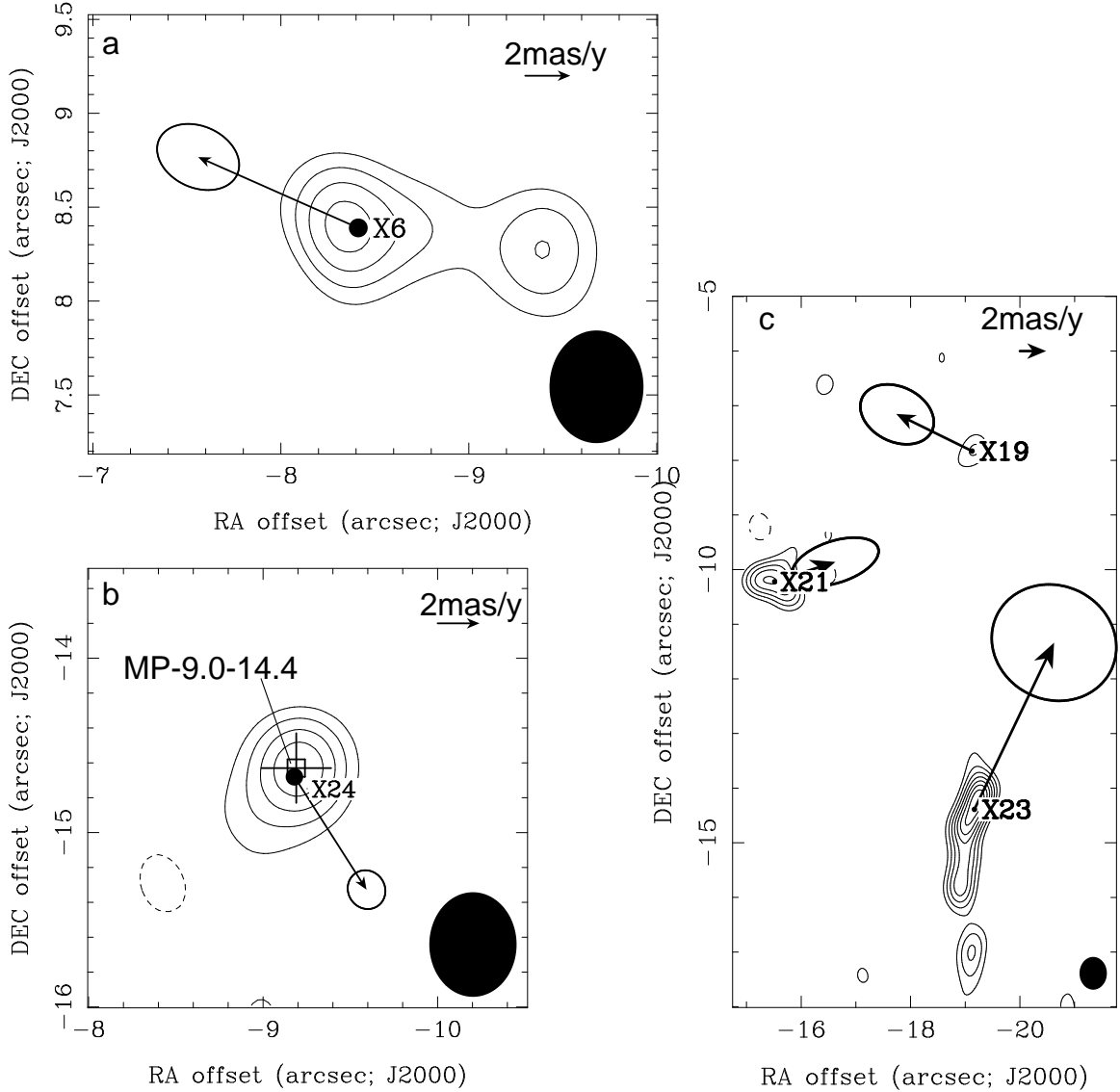}
\caption{Images of the compact radio continuum sources at 3.6 cm in the Western
Arc made from the 2002X epoch data with the longer baselines
($>100k\lambda$). 
a: the double radio emission (X6).
The contours are 0.4 mJy beam$^{-1}\times$(--2,2,3,4 and 5).
b: the radio emission (X24) of MP-09-14.4.
The contours are 0.4 mJy beam$^{-1}\times$(--2,2,3,4 and 5).
c: radio emission (X19, X21, X23)
in contours of  0.4 mJy beam$^{-1}\times$(--2,2,3,4,5,6,7 and 8).
The FWHM beam is 0.\arcsec6$\times$0.\arcsec5 (P.A.=0\arcdeg),
shown as a filled ellipse at bottom-right.
The vectors show the proper motions of the radio sources,
with the open ellipses
showing the 2$\sigma$ errors in magnitude (the semi-axes along
the vector) and  direction (the semi-axes perpendicular
to the vector). \label{fig12} }
\end{figure}
\clearpage
\begin{figure}[ht]
\figurenum{13}
\includegraphics[width=33pc,angle=-90]{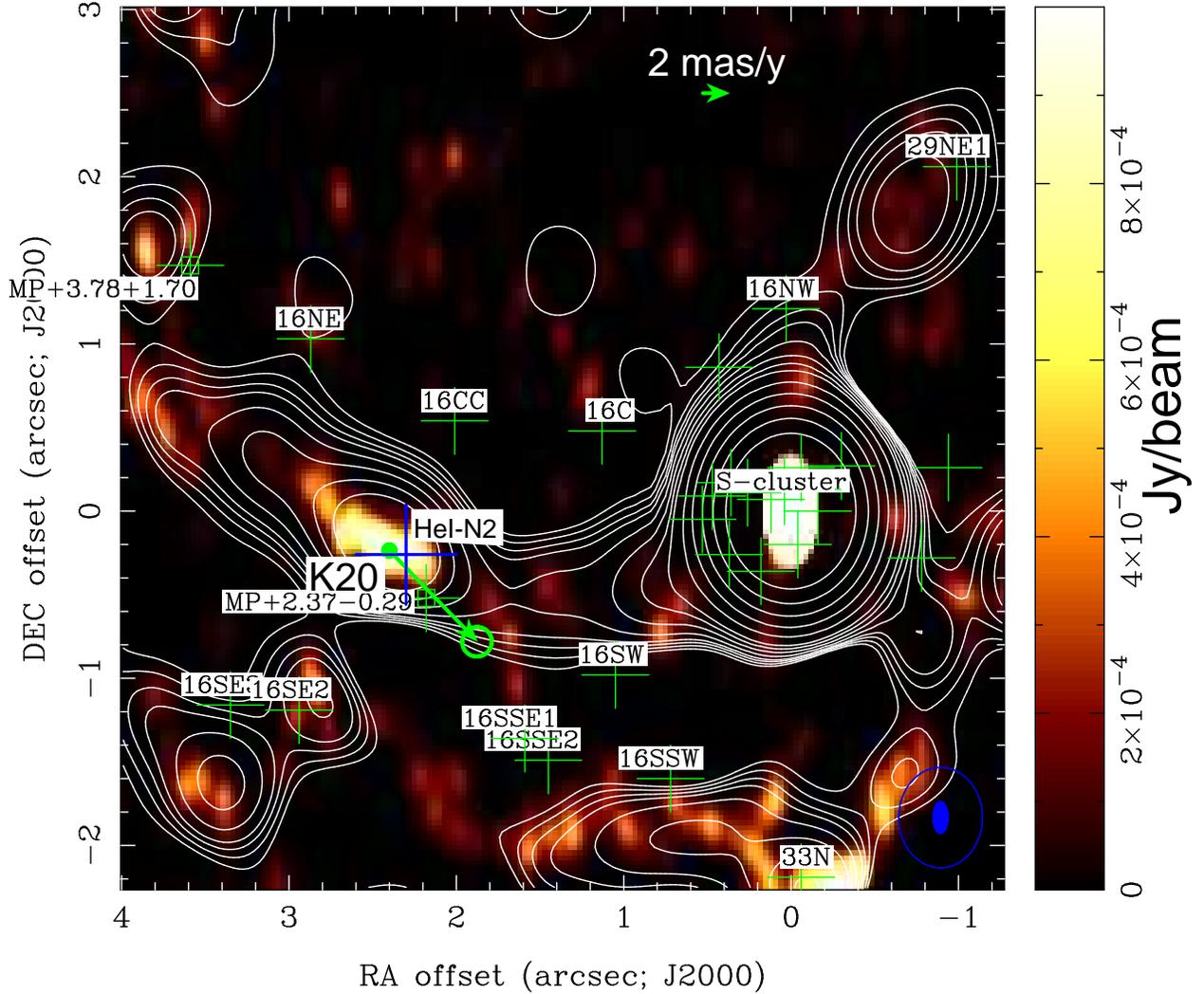}
\caption{Radio image of the compact components
in the region of IRS 16 constructed from
epoch 2005K  data at 1.3 cm (color) and
epoch 2002X  data at 3.6 cm (contours) using only
the longer baseline visibilities ($>100k\lambda$).
The FWHM beams are shown at bottom-right: the filled ellipse
(0.\arcsec2$\times$0.\arcsec1, 0\arcdeg)
and the open ellipse (0.\arcsec6$\times$0.\arcsec5, 0\arcdeg)
for images at 1.3 and 3.6 cm, respectively.
The contours are
0.25 mJy beam$^{-1}\times$(1, 2, 3, 4, 6, 9, 13, 18,
36, 72, 144, 288, 576).
The plus signs show the positions of the He I stars \citep{paum01}.
The plus-empty-squares are the positions of IR stars from
\citet{genz00}. 
The stars in the IRS 16 and S-cluster are marked with small plus
signs \citep{paum06}.
The vector marks the proper motion of
the source K20 (coincident with HeI-N2, marked with big plus) 
along with the open ellipse
showing the 2$\sigma$ uncertainty in magnitude
(the semi-axis along the vector) and direction
(the semi-axis perpendicular to the vector).
The horizontal vector at top-right shows a proper motion 2 mas y$^{-1}$. 
\label{fig13}}
\end{figure}
\clearpage
\begin{figure}
\figurenum{14}
\centering
\includegraphics[width=38pc]{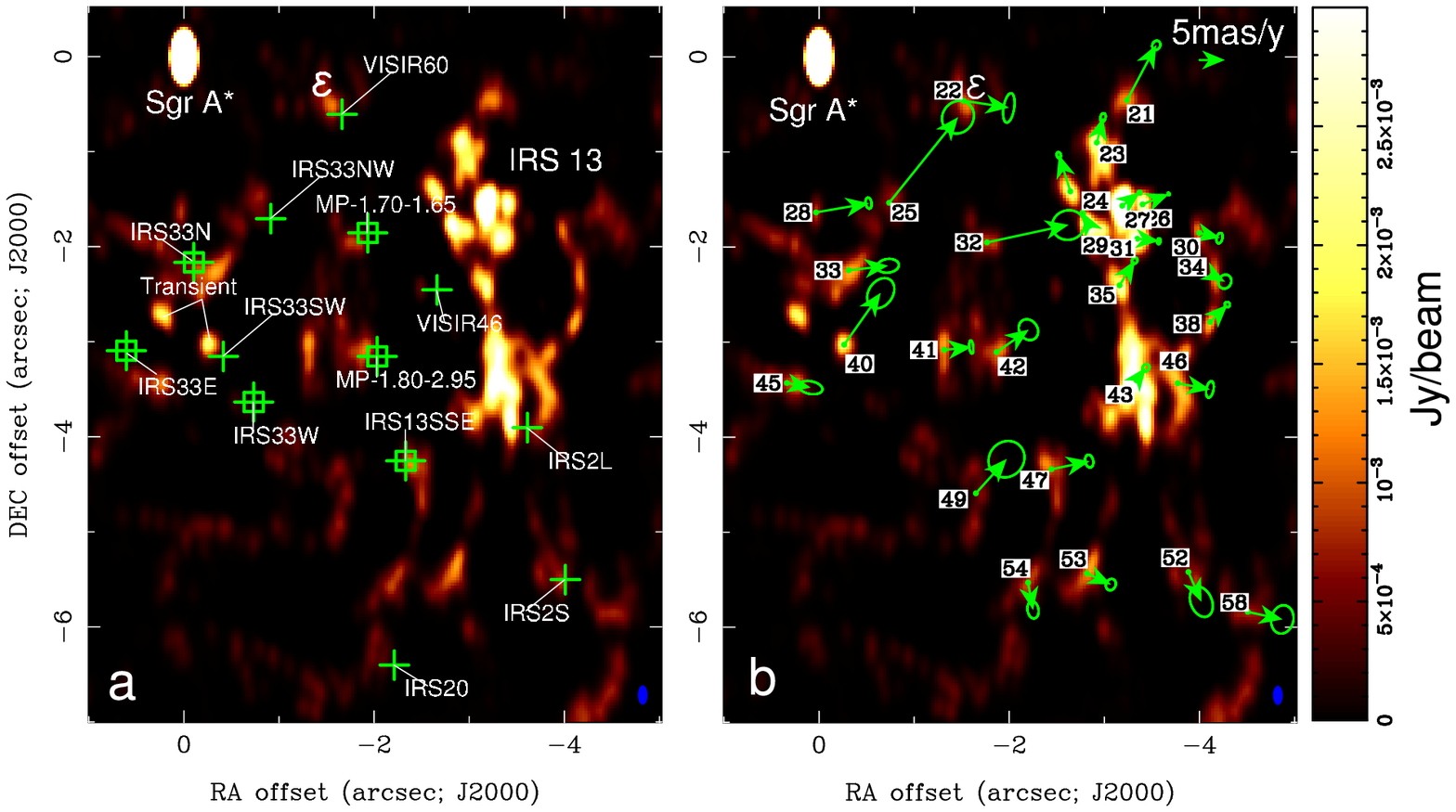}
\caption{The mini-cavity \citep{yusef89} observed
at 1.3 cm. a: Finding chart for IR sources: plus symbols from \cite{vieh06} 
and plus-open-square symbols from \cite{genz00} and \cite{paum06}. 
b: Proper motion vectors with 2$\sigma$ errors. 
The vector (5 mas y$^{-1}$) at top-right 
scales the proper motions. \label{fig14}}
\end{figure}
\begin{figure}
\centering
\figurenum{15}
\includegraphics[width=20.25pc,angle=-90]{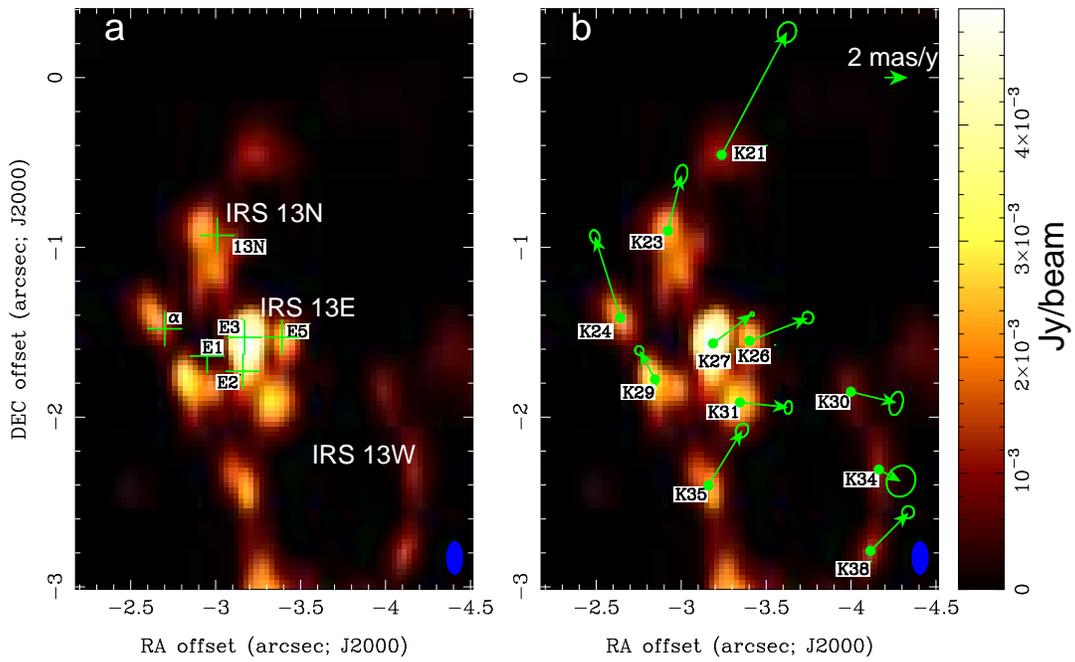}
\caption{IRS 13 observed at 1.3 cm.   
a: Finding chart. The plus signs
mark the IR positions of E1, E2, E3, E5, $\alpha$, and the averaged 13N components 
from \citet{muzi08}. b: Proper motion vectors with 
2$\sigma$ errors. The vector at top-right scales the proper motions.
\label{fig15}}
\end{figure}
\clearpage
\begin{figure}
\figurenum{16}
\centering
\includegraphics[width=25pc,angle=-90]{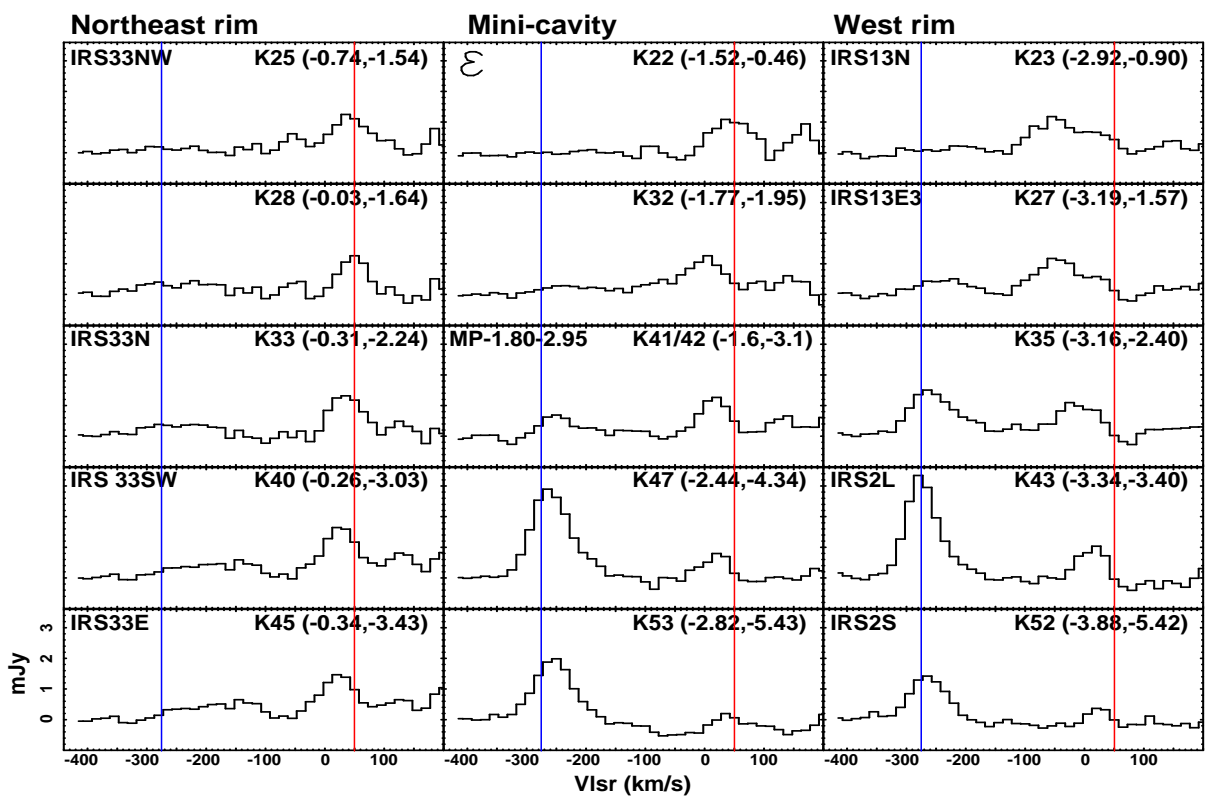}
\caption{The H92$\alpha$ line spectra
in the ``mini-cavity''
and the NE and West rim, taken from
the line image with  $\theta_{\rm FWHM}=1.25$\arcsec. 
The source ID and the position offsets
from SgrA* are given at top-right in each panel.
The commonly used names are at upper left. The
vertical blue and red lines indicate V$_{\rm LSR}=-275$ \kms and
V$_{\rm LSR}=+50$ \kms, respectively.
\label{fig16}    }
\end{figure}
\begin{figure}
\figurenum{17}
\centering
\includegraphics[width=22pc,angle=-90]{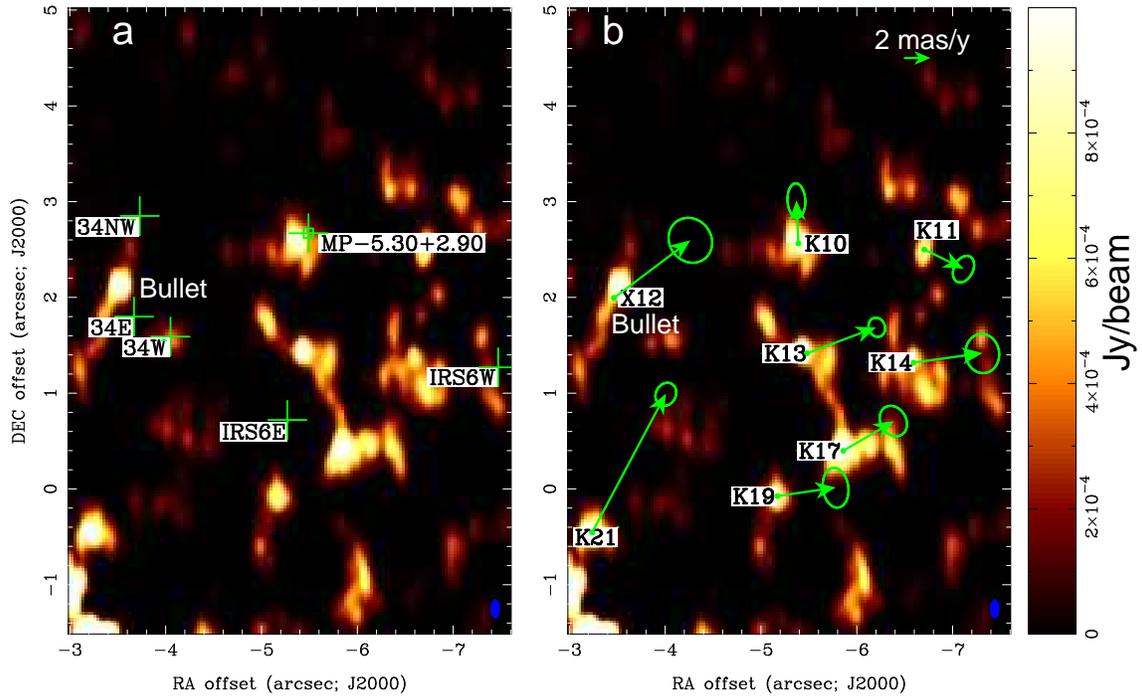}
\caption{The IRS 6/34 region observed
at 1.3 cm. a: finding chart. The IR sources 
are marked with plus signs \citep{paum06}
and plus-open-square \citep{genz00}.
b: proper motion vectors with 2$\sigma$ errors.
The vector (2 mas y$^{-1}$) at top-right scales 
proper motions.\label{fig17}
}
\end{figure}
\begin{figure}[ht]
\figurenum{18}
\centering
\includegraphics[width=42pc]{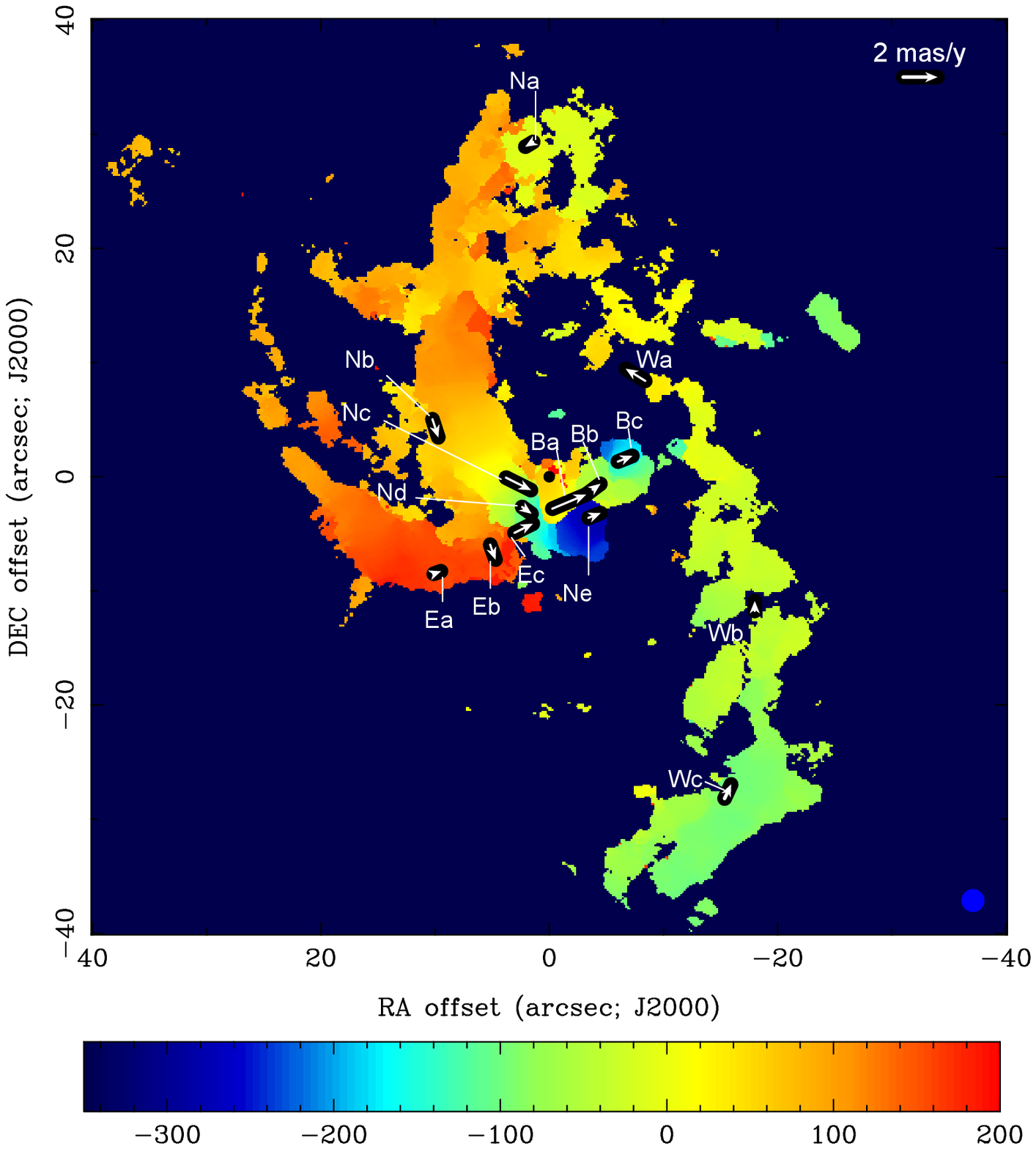}
\caption{Representation of the measured three-dimensional velocity field
in Sgr A West (the central 80 arcsec or 3 pc). The color map shows the radial
velocity ($V_z$) determined from fitting the peak velocity of H92$\alpha$ 
line spectra observed with the VLA. The color wedge shows the velocity 
scale from +200 to --350 \kms, red for redshift and blue for blueshift. 
The vectors show the mean transverse velocities ($V_x,V_y$) of the radio
components in regions summarized in Table 4. The scaling vector at top-right 
shows 300 \kms~in transverse velocity. The black dot indicates the position 
of Sgr A*. The $x$ and $y$ coordinates are the RA and DEC offsets from Sgr A*.
\label{fig18}}
\end{figure}
\begin{figure}[ht]
\figurenum{19}
\centering
\includegraphics[width=42pc,angle=-90]{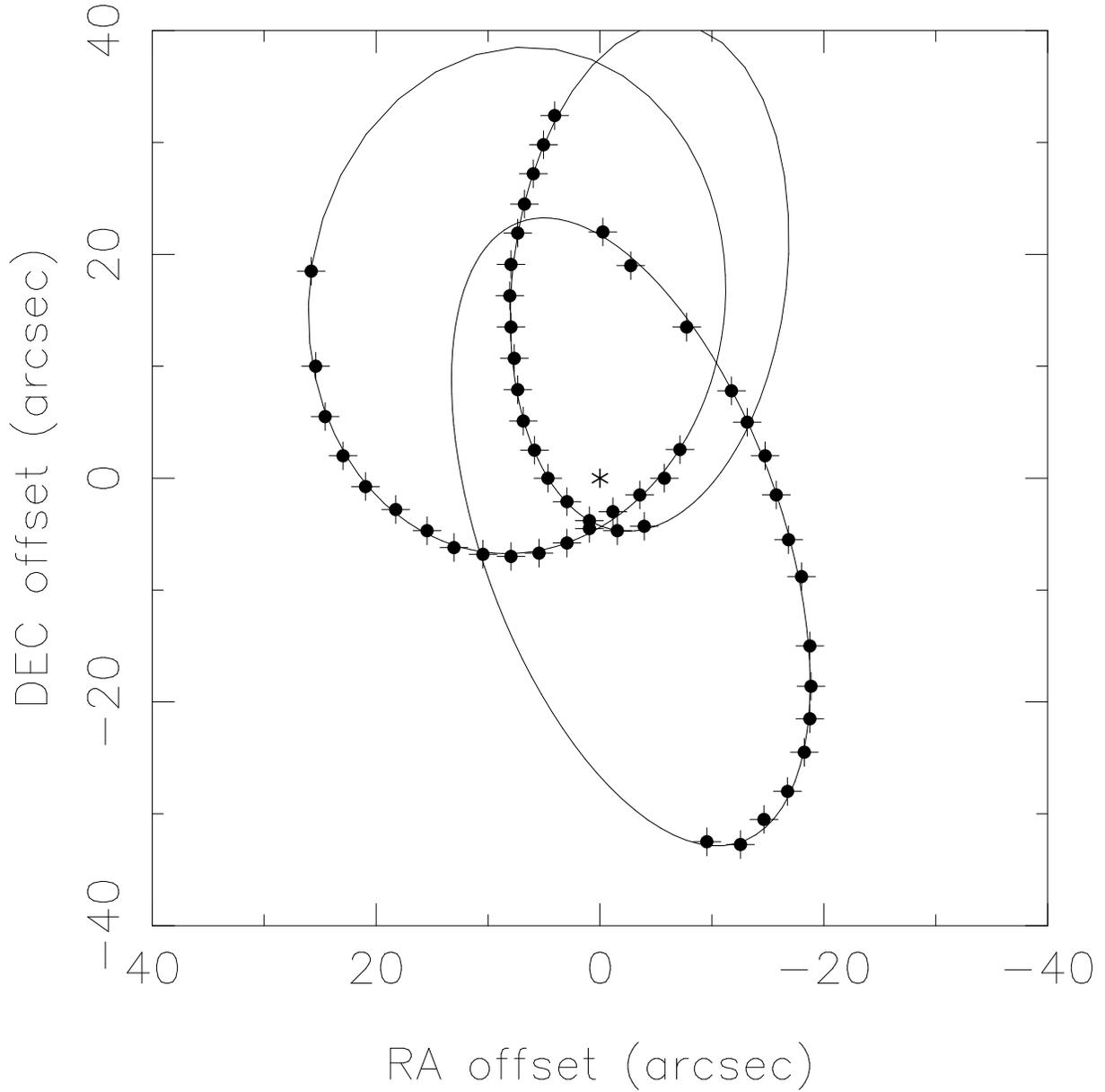}
\caption{The LSQ fitting (solid curves) to position data (dots)
 along the loci of the three streams observed in the H92$\alpha$
line emission from 
Sgr A West (the central 80 arcsec or 3 pc). The positions
are marked in Fig. 5a. The size of each cross represents the size
of the region over which the radial
spectra (Fig. 5b) were integrated.     
The star indicates the position of Sgr A*. The coordinates
are the RA and DEC offsets from Sgr A*, at the dynamic center.
\label{fig19}}
\end{figure}
\clearpage
\begin{figure}[ht]
\figurenum{20}
\centering
\includegraphics[width=42pc,angle=-90]{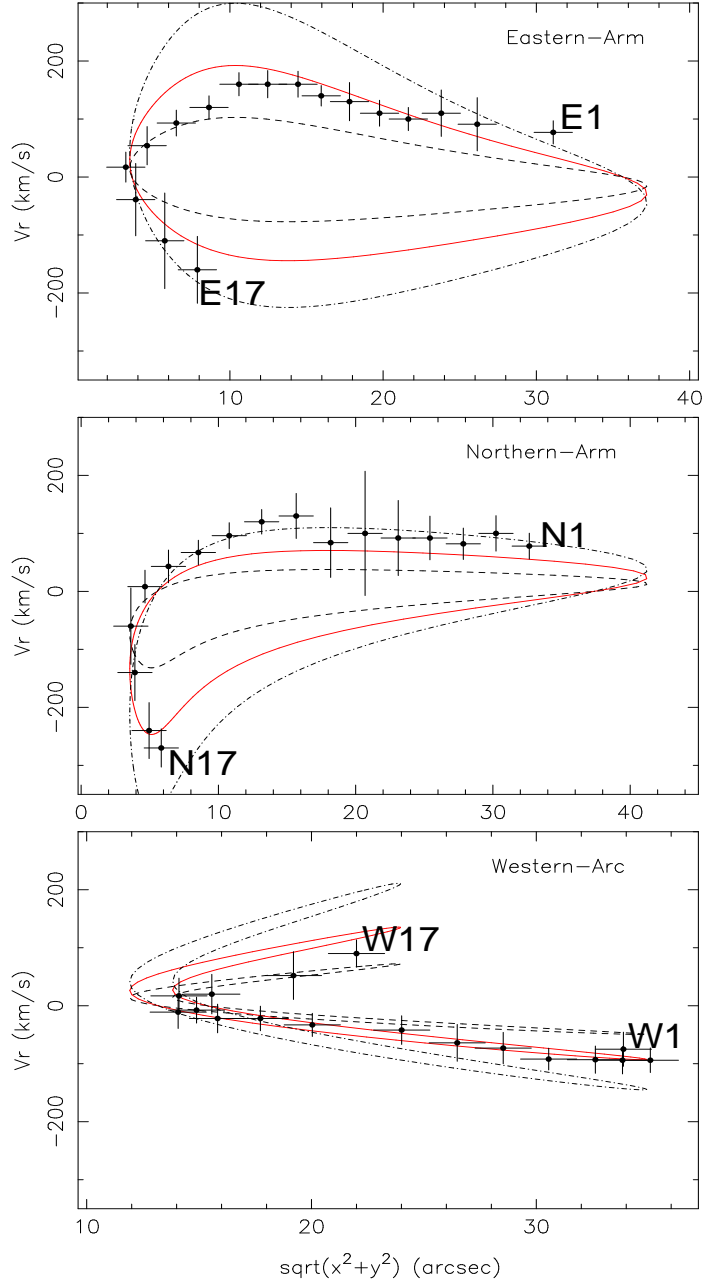}
\caption{Diagrams of radial velocity (V$_{z}$) vs. projected
distance ($\sqrt{x^2+y^2}$) with respect to the dynamical center for the three
streams associated with the Eastern Arm, Northern Arm and Western Arc.
Dots are the  measurements of radial velocities (V$_{\rm LSR}$).
The horizontal bars represent the radius of the region from which V$_{\rm LSR}$
was measured. The vertical bars are the FWHM of the velocity
component. The solid (red) lines  indicate
 the fits of the model (Eq. 7) with the best fitted orbital parameters
 ($a$,$e$,$\Omega$,$\omega$,$i$) (Table 5) and
a mass of $M_{\rm dyn}=4.2\times10^6$ M$_\odot$,
the mass of the SMBH, to the data. With the  orbital parameters
remaining fixed, 
the dashed lines indicate
the fits with a mass of 
$M_{\rm dyn}=1.2\times10^6$ M$_\odot$ while the dash-dot lines
correspond to the fits with a mass of 
$M_{\rm dyn}=10.2\times10^6$ M$_\odot$. \label{fig20}}
\end{figure}
\begin{figure}[ht]
\figurenum{21}
\plottwo{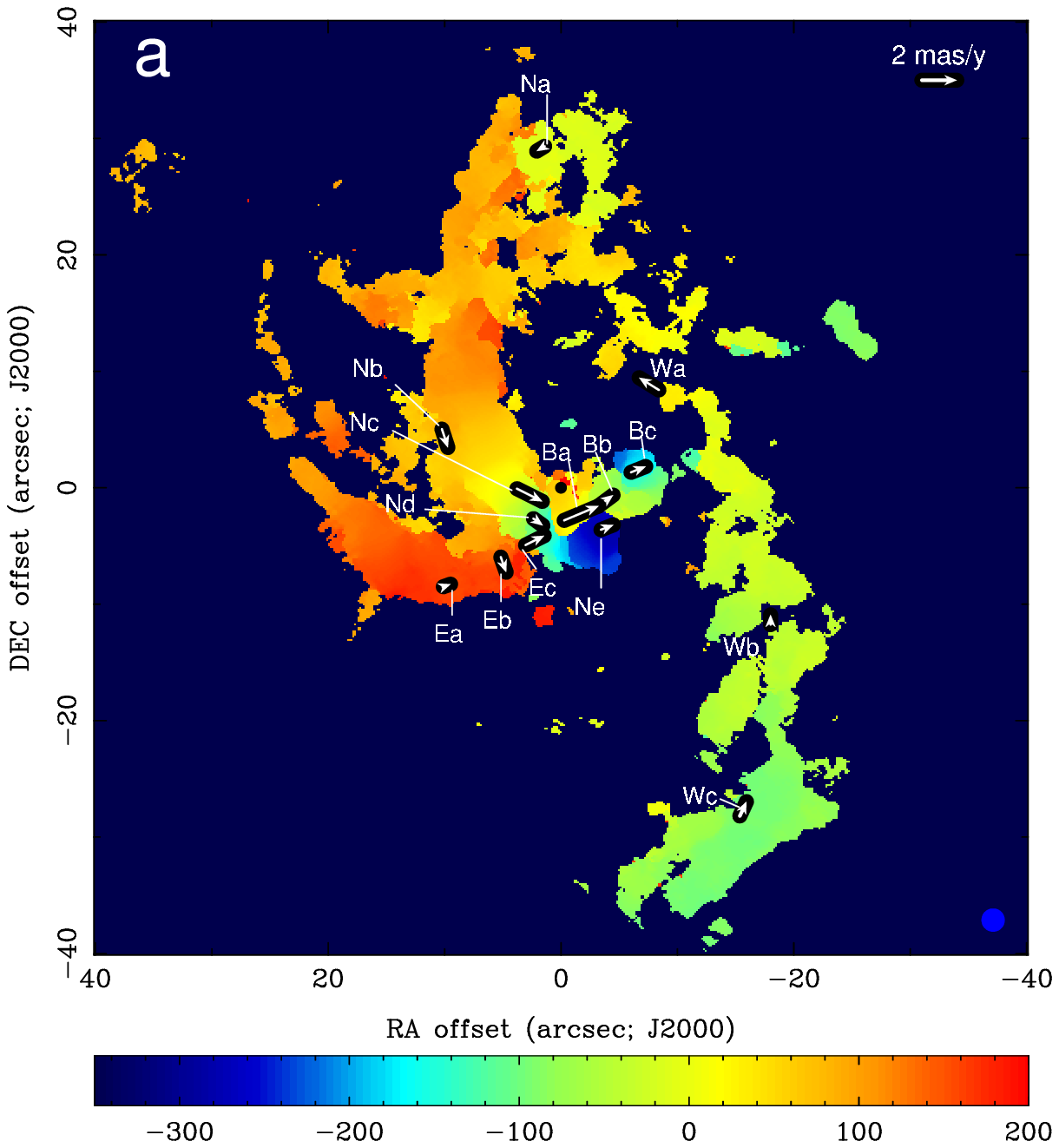}{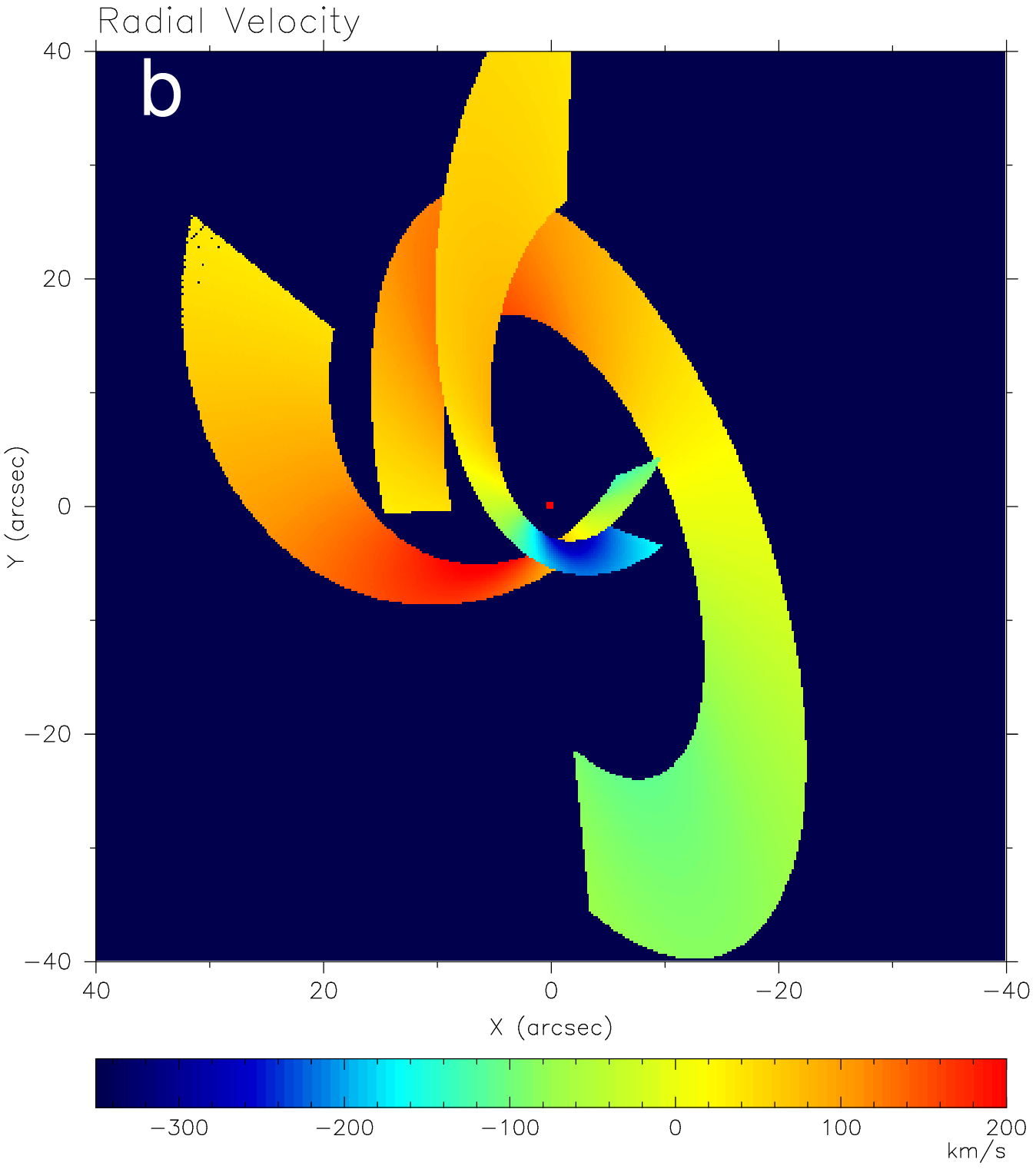}
\vfil
\plottwo{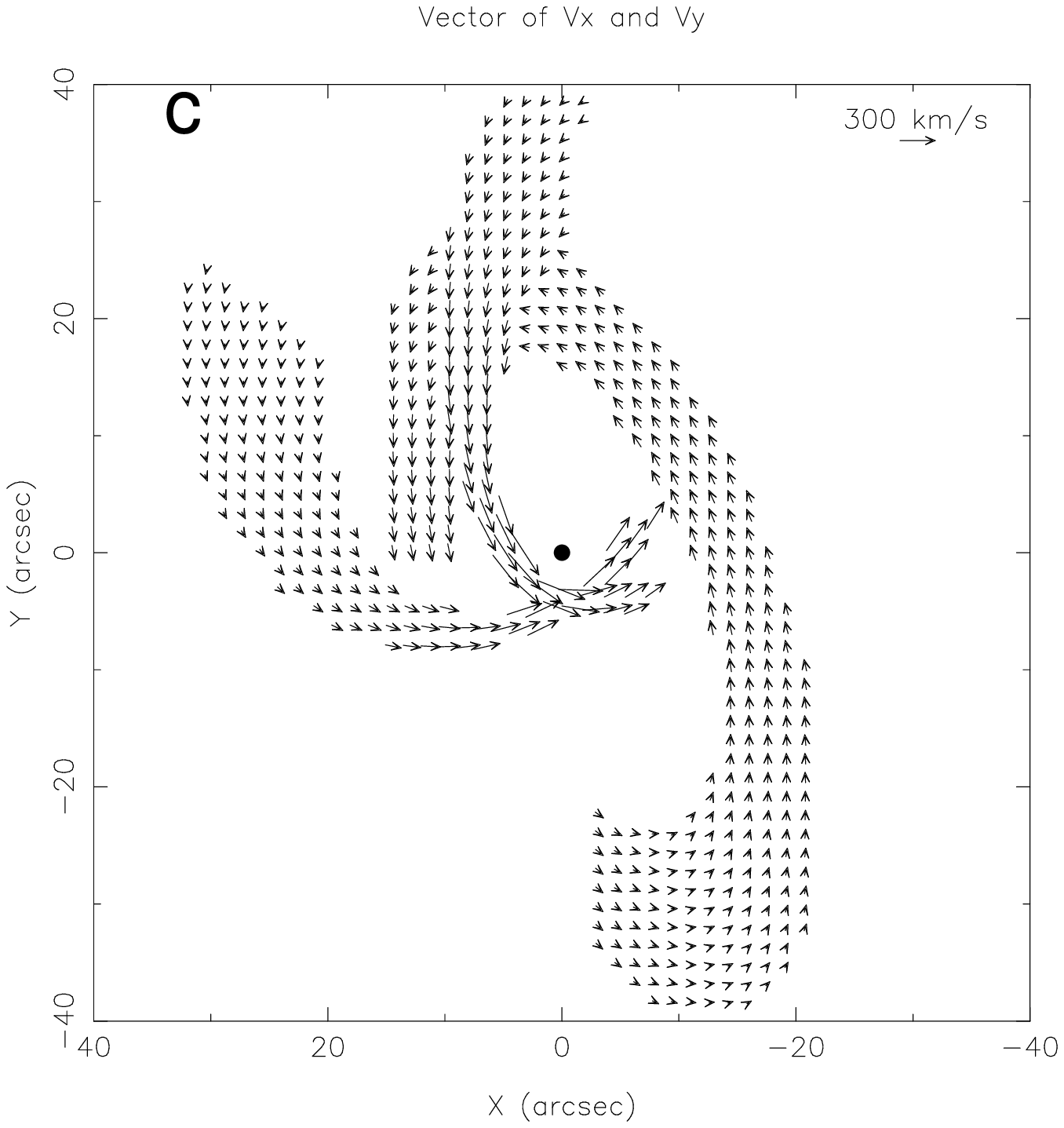}{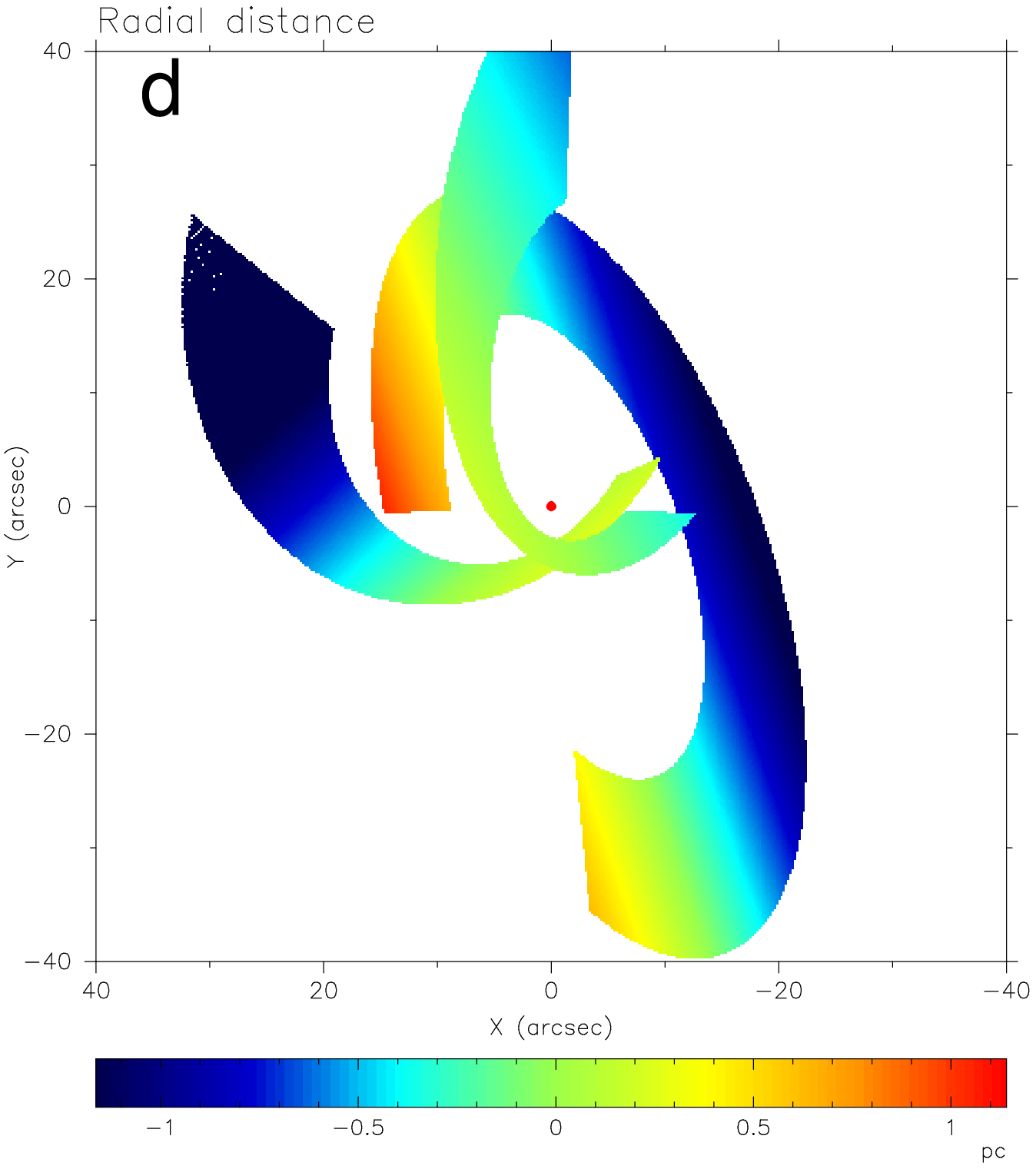}
\vfil
\plottwo{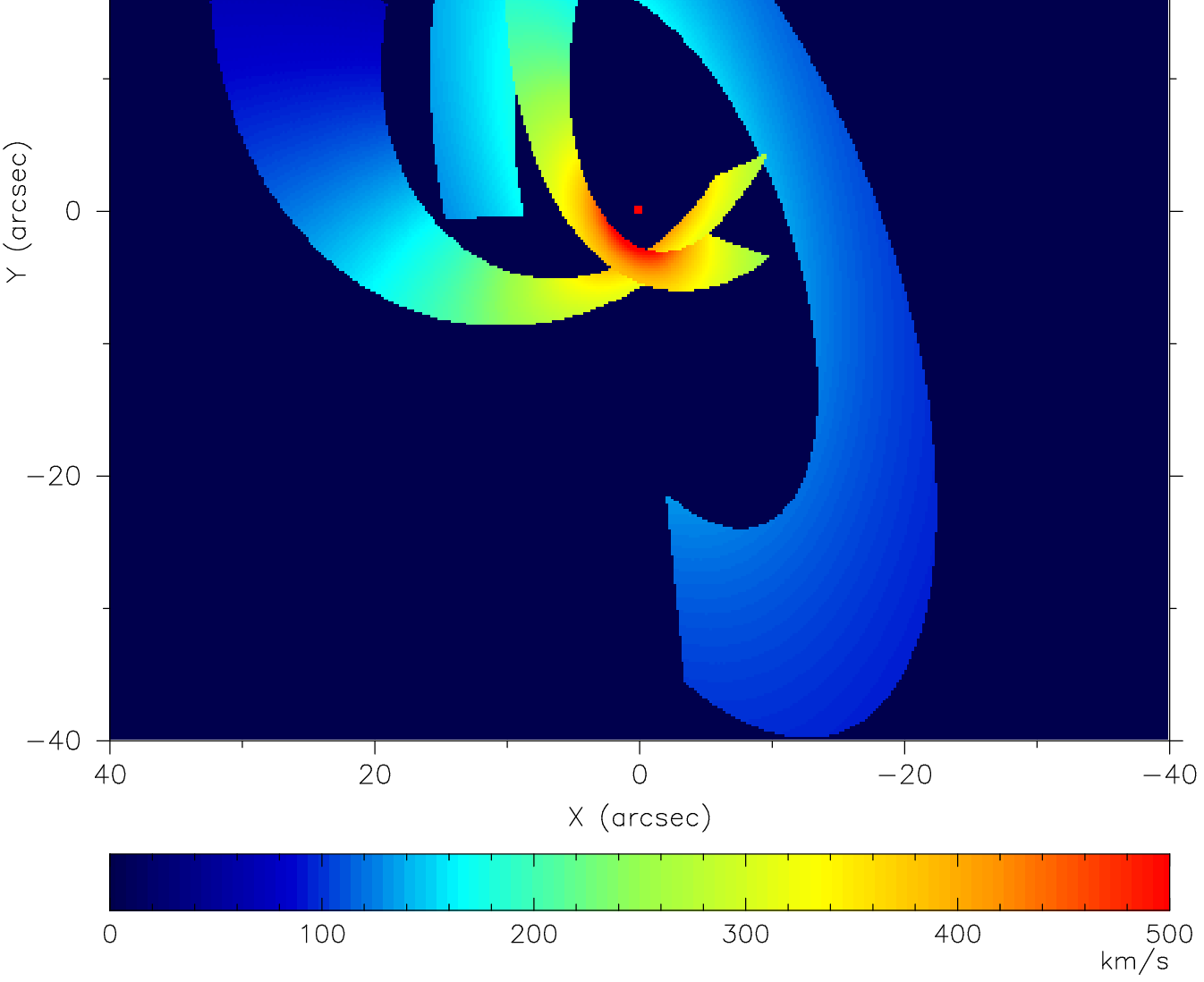}{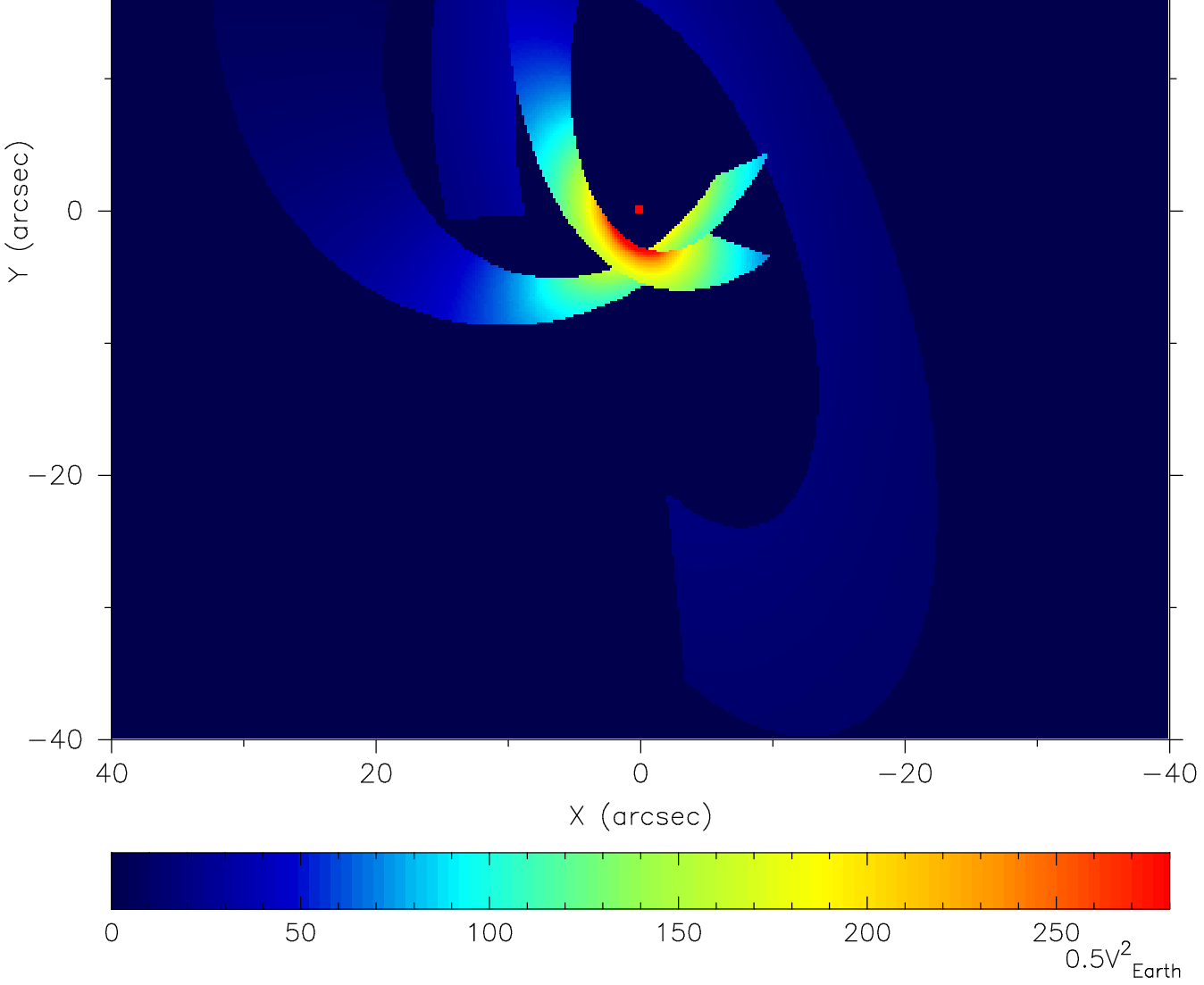}
\caption{}
\end{figure}
\clearpage
\begin{figure}
\figurenum{21}
\caption{The comparison between the observed kinematics of Sgr A
West and the results from the numerical calculation based on
the best fitted orbital model with the  
parameters summarized in Table 5. a) The observed radial velocity
(color) and transverse velocity determined
from proper motions (vectors). The  
axes are the position offsets from Sgr A*. 
b)  The radial velocity ($V_z$) of the three streams in Sgr A West,
calculated based on the best fitted orbital model. 
For comparison, the illustrated region is at the same scale as the observed
one in Fig. 21a. 
The wedge shows the same scale as that in Fig. 21a.
In Figs. 21b-f, the dot indicates the position of
the dynamical center, coincident with Sgr A*.
c) The calculated transverse velocity ($V_x, V_y$) of the three streams in
Sgr A West. The vector  at the top-right scales the transverse
velocity.
d) The three-dimensional geometry ($x,y,z$)  
of the three streams in Sgr A West with respect to
Sgr A*, calculated
using the fitted orbital model.
The color shows the line-of-sight distance (z) from Sgr A*   
between --1.16 to +1.16 pc.
The positive and negative values of z correspond to the regions 
behind and in front of the dynamical center, respectively.
e) The  total velocity ${\rm V = \sqrt{V_x^2 +V_y^2+V_z^2}}$ 
of the three streams in Sgr A West, 
calculated based on the best fitted results.
The  wedge of bottom scales V in units of \kms.
f) The kinetic energy per unit mass (${1\over2}V^2$) 
of the three streams in Sgr A West. 
The  wedge of bottom scales 
${1\over2}V^2$ in  units of ${1\over2}V^2_{Earth}$ where $V_{Earth}=29.8$ 
\kms~is the Earth's mean orbital velocity. 
In the model images (b, c, d, e, and f),
the X and Y coordinates are the RA and Dec offsets 
(arcsec) from the dynamical center.
\label{fig21}
}
\end{figure}

\begin{figure}
\figurenum{22}
\centering
\includegraphics[width=38pc]{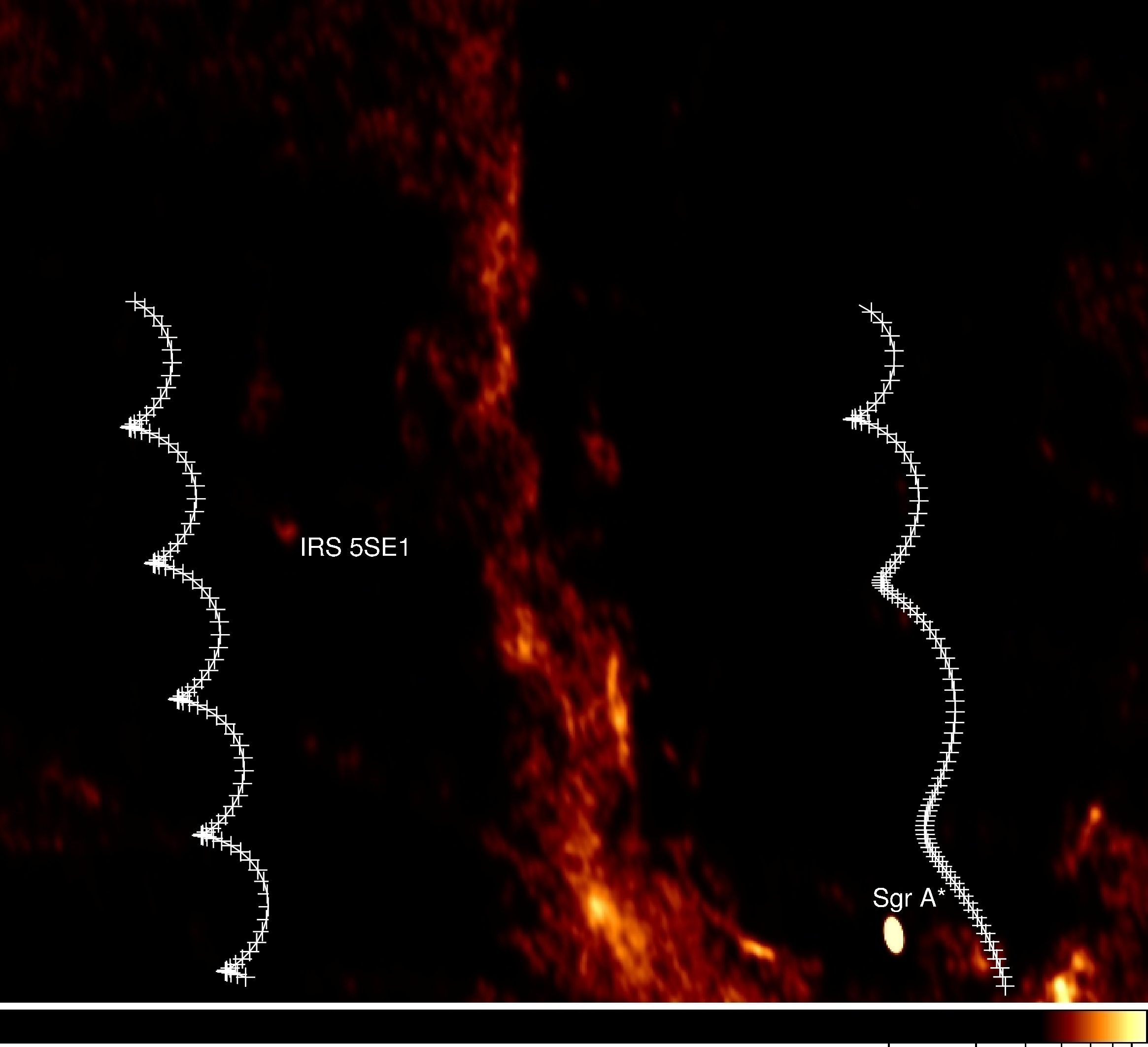}
\caption{The apparent helical structure of the ionized gas
in the Northern-Arm observed at 1.3 cm. 
The bright compact source to the bottom-right is Sgr A*.
The bow-shock source associated with IRS 5SE1 is also
labelled. Also shown are two projected model helices crafted to match 
the Northern Arm structure. Both are right-handed helices,
with the top tilted toward the observer. The model helix
at left is tilted by 45\arcdeg~(where 0\arcdeg~corresponds to
the helical axis in the plane of the sky) and has a pitch
angle of 49\arcdeg. The helix at the right has a wavelength that increases
linearly from top to bottom; it is titled by 50\arcdeg~and the initial
pitch angle (at top) is 45\arcdeg.
 \label{fig22}  
}
\end{figure}

\begin{deluxetable}{lcrrrrrrrrr}
\tabletypesize{\scriptsize}
\tablecaption{Log of Data and Observations}
\tablewidth{0pt}
\tablehead{
\colhead{Date}
&\colhead{RA}
&\colhead{DEC}
&\colhead{$\nu$}
&\colhead{$\nu_0$}
&\colhead{V$_{\rm lsr}$}
&\colhead{$\Delta\nu$}
&\colhead{N-chan}&
\colhead{Array Config} \\
\colhead{YYYY-MM-DD}
&\colhead{(B1950)}
&\colhead{(B1950)}
&\colhead{(GHz)}
&\colhead{(GHz)}
&\colhead{(km s$^{-1}$)}
&\colhead{(MHz)}
}
\startdata
\multicolumn{2}{c}{Continuum observations at 1.3 cm} \\
\cline{1-2} \\
2005-04-26 & 17:42:29.3180 &--28:59:18.390 & 22.485&\dots &\dots&50&2 &B \\
2004-12-23 & 17:42:29.3180 &--28:59:18.390 & 22.485&\dots &\dots&50&2 &A \\
1999-11-16 & 17:42:29.3180 &--28 59 18.390 & 22.485&\dots &\dots&50&2 &B \\
1999-08-19 & 17:42:29.3180 &--28 59 18.390 & 22.485&\dots &\dots&50&2 &A \\
1992-01-19 & 17:42:29.3180 &--28 59 18.390 & 22.235&\dots &\dots&50&2 &B \\
1991-06-22 & 17:45:40.0848\tablenotemark{a} & -29:00:27.779\tablenotemark{a} & 22.235&\dots &\dots&50&2 &A \\
\\
\multicolumn{2}{c}{H92$\alpha$ line observations at 3.5 cm} \\
\cline{1-2} \\
2002-05-22 & 17:42:29.3000 &--28:59:17.000 & 8.3156&8.3094&--200&0.3906&31&BnA\\
2002-05-21 & 17:42:29.3000 &--28:59:17.000 & 8.3156&8.3094&--200&0.3906&31&BnA\\
1999-10-18 & 17:42:29.3000 &--28:59:17.000 & 8.3145&8.3094&--200&0.3906&31&BnA\\1999-10-17 & 17:42:29.3000 &--28:59:17.000 & 8.3145&8.3094&--200&0.3906&31&BnA\\
1993-02-13 & 17:42:29.3000 &--28:59:17.000 & 8.3159&8.3094&--200&0.3906&31&BnA\\1993-02-12 & 17:42:29.3000 &--28:59:17.000 & 8.3159&8.3094&--200&0.3906&31&BnA\\
1991-02-19 & 17:42:29.3000 &--28:59:17.000 & 8.3104&8.3094&0&0.3906&31&D  \\
1990-10-02 & 17:42:29.3000 &--28:59:17.000 & 8.3089&8.3094&0&0.3906&31&CnB\\
1990-07-02 & 17:42:29.3000 &--28:59:17.000 & 8.3095&8.3094&0&0.3906&31&BnA\\
\enddata
\tablenotetext{a}{The J2000 equinox was used for the observations.}
\end{deluxetable}

\begin{deluxetable}{ccrrrrrrrr}
\tabletypesize{\scriptsize}
\tablecaption{Measurements of proper motion and radial velocity}
\tablewidth{0pt}
\tablehead{
\colhead{Source} &
\colhead{IR ID} &
\colhead{$\Delta\alpha$ (J2000)\tablenotemark{a} } &
\colhead{$\Delta\delta$ (J2000)\tablenotemark{a} } &
\colhead{$\theta_{maj}\times\theta_{min}$ (P.A.)\tablenotemark{b}}&
\colhead{$\mu_\alpha$\tablenotemark{c}} &
\colhead{$\mu_\delta$\tablenotemark{c}} 
 \\
&
&
\colhead{(asec)}&
\colhead{(asec)}&
\colhead{(asec$\times$asec(Deg))}&
\colhead{(mas y$^{-1}$)} &
\colhead{(mas y$^{-1}$)} \\ 
}
\startdata
\\
&&\multicolumn{5}{c}{Measurements from 1.3 cm observations} \\
\cline{3-7} \\

K1&IRS8 &   1.400$\pm$0.002&  29.328$\pm$0.002&0.44$\times$0.15 (142)&   1.8$\pm$ 0.4&--1.1$\pm$ 0.5\\
K2&&   6.956$\pm$0.004&  11.950$\pm$0.016&1.97$\times$0.32 (168)&   6.8$\pm$ 1.3&--28.0$\pm$ 5.4\\
K3&&   6.901$\pm$0.003&   9.873$\pm$0.008&1.27$\times$0.32 (168)&   1.6$\pm$ 0.6&--9.4$\pm$ 2.1\\
K8 &IRS10W&6.558$\pm$0.008&   5.034$\pm$0.004&0.56$\times$0.45 (164)&0.31$\pm$0.67&--1.9$\pm$ 0.8\\
K10&MP--5.30+2.90& --5.390$\pm$0.002&   2.564$\pm$0.003&0.35$\times$0.26 (22)&   0.2$\pm$ 0.3&3.3$\pm$0.7\\
K11\tablenotemark{d}&& --6.704$\pm$0.001&   2.500$\pm$0.002&0.22$\times$0.08 (166)& --3.0$\pm$0.3& --1.5$\pm$ 0.6\\
K13\tablenotemark{d}&& --5.482$\pm$0.002&1.417$\pm$0.002&0.35$\times$0.12 (64)&--5.4$\pm$0.3&   2.0$\pm$ 0.4\\
K14\tablenotemark{d}&& --6.592$\pm$0.003&   1.319$\pm$0.003&0.27$\times$0.17 (70)& --5.3$\pm$ 0.7&   0.7$\pm$ 0.8\\
K15&& --9.915$\pm$0.003&0.724$\pm$0.005&0.37$\times$0.21 (160)& --2.8$\pm$ 0.5& --0.5$\pm$ 1.0\\
K16&IRS1W&5.169$\pm$0.002&0.460$\pm$0.002&0.49$\times$0.28  (39)& --5.5$\pm$0.4&2.7$\pm$0.5\\
K17\tablenotemark{d}&& --5.858$\pm$0.002&0.398$\pm$0.002&0.37$\times$0.22 (156)& --3.9$\pm$0.4&   2.3$\pm$ 0.7\\
K19\tablenotemark{d}&& --5.171$\pm$0.002& --0.075$\pm$0.002&0.21$\times$0.17 (177)& --4.5$\pm$ 0.5& 0.6$\pm$ 0.9\\
K20&HeI-N2/IRS16&   2.400$\pm$0.002& --0.236$\pm$0.001&0.54$\times$0.10 (64)& --6.5$\pm$ 0.6& --6.7$\pm$0.4\\
K21\tablenotemark{e}&&--3.238$\pm$0.002&--0.455$\pm$0.002&0.28$\times$0.21 (125)&--5.7$\pm$0.4&10.7$\pm$ 0.5\\
K22 &VISIR60& --1.519$\pm$0.002& --0.461$\pm$0.003&0.36$\times$0.17 (26)& --8.8$\pm$ 0.5& --1.4$\pm$ 1.5\\
&$Source-\varepsilon$&&&&&\\
K23\tablenotemark{e}&&--2.922$\pm$0.001& --0.902$\pm$0.002&0.27$\times$0.14 (48)& --1.2$\pm$0.2& 4.9$\pm$ 0.4\\
K24 &IRS13N-$\alpha$& --2.642$\pm$0.001& --1.415$\pm$0.001&0.30$\times$0.07 (36)&2.2$\pm$0.2&7.1$\pm$ 0.3\\
K25 &IRS33NW&--0.737$\pm$0.014&--1.535$\pm$0.018&0.36$\times$0.21 (144)&
--13.4$\pm$ 1.2&  16.7$\pm$ 1.9\\
K26&IRS13E5&--3.402$\pm$0.001&--1.551$\pm$0.001&0.13$\times$0.12 (23)& --5.1$\pm$ 0.2& 2.0$\pm$ 0.3\\
K27&IRS13E3& --3.188$\pm$0.001& --1.566$\pm$0.001&0.21$\times$0.11 (174)& --3.4$\pm$ 0.1&   2.6$\pm$ 0.1\\
K28 & &0.032$\pm$0.001& --1.636$\pm$0.002&0.14$\times$0.01 (45)&--10.3$\pm$ 0.3&   1.8$\pm$ 0.6\\
K29\tablenotemark{f}&&--2.846$\pm$0.001&  --1.778$\pm$0.001&0.27$\times$0.14 (48)& 1.4$\pm$0.2&   2.5$\pm$0.2\\
K30&IRS13W& --3.999$\pm$0.001& --1.851$\pm$0.002&0.21$\times$0.05 (19)& --3.9$\pm$ 0.3& --1.0$\pm$ 0.5 \\
K31\tablenotemark{f}&&--3.346$\pm$0.001&--1.913$\pm$0.001&0.27$\times$0.22  (37)&--4.2$\pm$0.2&--0.5$\pm$0.3 \\
K32&MP-1.70-1.65&--1.766$\pm$0.007&--1.951$\pm$0.003&0.25$\times$0.14 (72)&--15.9$\pm$ 1.6& 3.4$\pm$ 1.4 \\
&$Source-\zeta$&&&&&\\
K33 &IRS33N&--0.312$\pm$0.005&--2.245$\pm$0.003&0.60$\times$0.22 (113)&--7.7$\pm$1.1&   0.9$\pm$ 0.7 \\
K34 &&--4.164$\pm$0.001&  -2.309$\pm$0.003&0.49$\times$0.05 (4)&--1.9$\pm$0.6&  --1.0$\pm$0.7\\
K35 && --3.160$\pm$0.001& --2.403$\pm$0.001&0.34$\times$0.12  (22)& --2.9$\pm$ 0.2& 4.8$\pm$ 0.4\\
K36 &IRS21&2.365$\pm$0.001& --2.648$\pm$0.001&0.24$\times$0.15 (98)& --2.4$\pm$0.4& --1.8$\pm$0.5\\
K37 &&   6.536$\pm$0.005& --2.781$\pm$0.007&0.46$\times$0.20 (29)&0.8$\pm$0.6&   9.0$\pm$ 0.9 \\
K38 && --4.114$\pm$0.001& --2.788$\pm$0.001&0.23$\times$0.01 (150)& --3.3$\pm$ 0.2& 3.4$\pm$ 0.3 \\
K39 && --5.458$\pm$0.004& --2.844$\pm$0.004&0.37$\times$0.17 (131)& --3.2$\pm$ 0.8&1.6$\pm$1.0 \\
K40 &XJ174540.0290031&--0.264$\pm$0.001&--3.028$\pm$0.002&0.12$\times$0.01 
(92)& --7.1$\pm$ 0.4&  10.1$\pm$ 1.9 \\
K41 && --1.317$\pm$0.001& --3.083$\pm$0.003&0.50$\times$0.09 (178)& --5.3$\pm$ 0.2&   0.6$\pm$ 0.6 \\
K42 &MP-1.80-2.95&--1.867$\pm$0.001&--3.106$\pm$0.002&0.33$\times$0.20 (167)&  
--6.2$\pm$0.7&4.4$\pm$1.2\\
&$Source-\eta$&&&&&\\
K43\tablenotemark{g}&&--3.337$\pm$0.001&  -3.402$\pm$0.002&0.99$\times$0.23 (3)&  --1.9$\pm$ 0.2&   2.4$\pm$ 0.5 \\
K45 &IRS33E&0.338$\pm$0.004& --3.430$\pm$0.002&0.48$\times$0.13  (77)& --4.6$\pm$1.2& --0.9$\pm$0.6 \\
K46\tablenotemark{g}& & --3.771$\pm$0.001&--3.433$\pm$0.004&0.83$\times$0.19 (9)& --6.3$\pm$0.3& --1.2$\pm$0.8 \\
K47&IRS13SSE& --2.443$\pm$0.002& --4.338$\pm$0.003&0.43$\times$0.17  (22)&--7.4$\pm$ 0.4&1.6$\pm$ 0.6 \\
K48&&2.729$\pm$0.003& --4.402$\pm$0.003&0.30$\times$0.21  (88)& --4.1$\pm$ 0.9&   1.5$\pm$ 0.7 \\
K49&& --1.649$\pm$0.003& --4.591$\pm$0.016&1.11$\times$0.16 (176)& --6.0$\pm$ 0.6&   6.7$\pm$ 2.4 \\
K50 &IRS9NW&   2.909$\pm$0.002& --5.056$\pm$0.002&0.49$\times$0.16 (129)& --4.2$\pm$ 0.5&   2.5$\pm$ 0.4 \\
K51 &&   3.709$\pm$0.003& --5.191$\pm$0.004&0.67$\times$0.20 (145)& --4.6$\pm$ 0.8&   1.7$\pm$ 1.3 \\
K52 &IRS2S& --3.884$\pm$0.004& --5.419$\pm$0.005&0.39$\times$0.31 (10)& --2.5$\pm$ 0.9& --5.8$\pm$ 1.6\\
K53& & --2.817$\pm$0.002& --5.434$\pm$0.003&0.45$\times$0.14 (149)& --4.7$\pm$ 0.4& --2.2$\pm$ 0.6 \\
K54 && --2.198$\pm$0.003& --5.530$\pm$0.003&0.44$\times$0.13 (148)& --1.0$\pm$ 0.5& --5.3$\pm$ 0.9 \\
K55 &&  14.575$\pm$0.003& --5.629$\pm$0.007&0.68$\times$0.27 (10)&   0.3$\pm$ 1.2& --8.7$\pm$ 2.3 \\
K57 &IRS9N&5.683$\pm$0.002& --5.728$\pm$0.002&0.24$\times$0.15 (122)& --2.5$\pm$0.7& --0.9$\pm$ 0.8 \\
K58&& --4.508$\pm$0.004& --5.838$\pm$0.003&0.40$\times$0.21 (93)& --6.8$\pm$ 1.1& --1.5$\pm$ 1.4 \\
K59 &&   4.908$\pm$0.001& --6.183$\pm$0.002&0.43$\times$0.19 (29)& --1.0$\pm$ 0.3& --4.8$\pm$ 0.5 \\
K60\tablenotemark{h}& & 1.312$\pm$0.004&--6.663$\pm$0.003&0.48$\times$0.17 (65)&--3.2$\pm$ 0.7&--3.1$\pm$ 0.6 \\
K62\tablenotemark{h}&&1.319$\pm$0.001&--7.058$\pm$0.001&0.14$\times$0.08 (60)& --0.8$\pm$ 0.7& --7.3$\pm$ 0.9\\
K63 &&   5.789$\pm$0.006& --7.736$\pm$0.005&0.47$\times$0.28 (83)&   1.6$\pm$ 1.5& --5.7$\pm$1.2 \\
K64 &&   7.155$\pm$0.003& --7.763$\pm$0.003&0.25$\times$0.19 (119)& --3.0$\pm$ 0.5&   0.8$\pm$ 0.7 \\
K66 &VISIR12& 9.980$\pm$0.001&--8.917$\pm$0.001&0.73$\times$0.19  (23)&--1.1$\pm$ 0.4&   2.0$\pm$ 0.6 \\
\\
&&\multicolumn{5}{c}{Measurements from 3.5 cm observations} \\
\cline{3-7} \\
X4 &&   7.321$\pm$0.003&  15.249$\pm$0.006&1.26$\times$0.43 (163)&   2.0$\pm$ 0.4& --1.0$\pm$ 0.7 \\
X6 &MP-8.20+7.80& --8.412$\pm$0.004&   8.390$\pm$0.005 & unresolved& 7.2$\pm$ 1.0&   3.2$\pm$ 0.6 \\
X7 &IRS5SE1& 10.599$\pm$0.002& 7.056$\pm$0.003&0.50$\times$0.30  (53)& --4.4$\pm$ 0.9& --7.0$\pm$ 0.8\\
X8 &&  19.847$\pm$0.005&   5.650$\pm$0.013&2.8$\times$0.3  (20)& --3.0$\pm$ 0.7& --3.1$\pm$ 1.7 \\
X9&IRS7   & -0.077$\pm$0.004&   5.611$\pm$0.004&0.6$\times$0.5 (120)&--1.5$\pm$1.2&--5.4$\pm$0.6 \\
X11 && 4.928$\pm$0.002&4.229$\pm$0.004&2.2$\times$0.7 (158)& --4.8$\pm$0.8& --4.8$\pm$ 1.0 \\
X12&IRS34NW&--3.465$\pm$0.005&1.995$\pm$0.010&unresolved&--6.7$\pm$0.8& 5.1$\pm$ 1.2 \\
&(Bullet)&&&&& \\
X13&MP+5.70+2.40&5.498$\pm$0.002&1.714$\pm$0.002&unresolved& --4.7$\pm$ 0.2&--14.0$\pm$ 0.2 \\
X17&&  16.231$\pm$0.004& --3.306$\pm$0.004&unresolved & --2.3$\pm$ 0.6& --7.3$\pm$1.4 \\
X18 &IRS28/&  10.611$\pm$0.002&  -6.034$\pm$0.003&1.5$\times$0.8 (2)&--2.3$\pm$ 0.3&  --0.5$\pm$ 0.5 \\
&IRS4&&&&& \\
X19 &&--19.124$\pm$0.007& --7.830$\pm$0.008&unresolved&   5.8$\pm$1.8&2.8$\pm$0.9 \\
X20 &&  10.086$\pm$0.001& --8.940$\pm$0.001&unresolved& --0.06$\pm$ 0.14& --0.8$\pm$ 0.2 \\
X21 &&--15.516$\pm$0.003&--10.221$\pm$0.003&unresolved& --4.7$\pm$ 1.2&   1.5$\pm$ 0.6 \\
X23 &&--19.169$\pm$0.003&--14.389$\pm$0.004&unresolved& --6.1$\pm$ 1.1&  12.8$\pm$ 2.1 \\
X24 &MP-09-14.4& --9.180$\pm$0.005&--14.680$\pm$0.010&unresolved& --3.5$\pm$0.6&  --5.4$\pm$1.3 \\
X25 &&--15.366$\pm$0.004&--28.188$\pm$0.004&unresolved& --1.7$\pm$ 0.4&  3.7$\pm$ 0.5 \\
\enddata
\tablenotetext{a}{Position offsets with respect to Sgr A*.
~~$^b$Deconvolved source size.
~~$^c$Proper motions.
~~$^d$The radio component is located in the IRS 6/34 region.
~~$^e$The radio component is located in the IRS 13N region.
~~$^f$The radio component is located in the IRS 13E region.
~~$^g$The radio component is located in the IRS 2L region.
~~$^h$The radio component is located in the VISIR 51 region.}
\end{deluxetable}

\begin{deluxetable}{rrrrrrrrr}
\tablenum{3}
\tabletypesize{\scriptsize}
\rotate
\tablecaption{The radial velocities from three mini-spiral arms}
\tablewidth{0pt}
\tablehead{
\colhead{ID\tablenotemark{a}} &
\colhead{$ {\Delta\alpha} $,
${\Delta\delta}$\tablenotemark{b} } &
\colhead{$V_{\rm LSR}$,
$\Delta V_{\rm FWHM}$\tablenotemark{c}}&
\colhead{ID\tablenotemark{a}} &
\colhead{$ {\Delta\alpha} $,
${\Delta\delta} $\tablenotemark{b}} &
\colhead{$V_{\rm LSR}$,
$\Delta V_{\rm FWHM}$\tablenotemark{c}}&
\colhead{ID\tablenotemark{a}} &
\colhead{$ {\Delta\alpha} $,
$ {\Delta\delta}$\tablenotemark{b} } &
\colhead{$V_{\rm LSR}$,
$\Delta V_{\rm FWHM}$\tablenotemark{c}} \\
\colhead{}&
\colhead{(arcsec)}&
\colhead{(km s$^{-1}$)}&
\colhead{}&
\colhead{(arcsec)}&
\colhead{(km s$^{-1}$)}&
\colhead{}&
\colhead{(arcsec)}&
\colhead{(km s$^{-1}$)} \\ 
}
\startdata
\\
\multicolumn{3}{c}{Northern Arm} & \multicolumn{3}{c}{Eastern Arm} & \multicolumn{3}{c}{Western Arm} \\
N1&4.05, 32.40&78,   45~~~~~~~~&E1&25.00,   18.5&77,  40~~~~~~~~&W1  &  --9.55, --32.50  &--75, 59~~~~~~~~\\
N2&5.05, 29.80&100,  61~~~~~~~~&E2&24.15,   10.0&91,  92~~~~~~~~&W2  & --12.55, --32.75  &--94, 42~~~~~~~~\\
N3&5.95, 27.20&82,   54~~~~~~~~&E3&23.15,   5.50&110, 80~~~~~~~~&W3  & --14.65, --30.50  &--94, 47~~~~~~~~\\
N4&6.75, 24.50&92,   75~~~~~~~~&E4&21.55,   2.00&100, 40~~~~~~~~&W4  & --16.75, --28.00  &--93, 47~~~~~~~~\\
N5&7.35, 21.90&92,  129~~~~~~~~&E5&19.75, --0.75&110, 45~~~~~~~~&W5  & --18.25, --24.50  &--92, 38~~~~~~~~\\
N6&7.95, 19.10&100, 214~~~~~~~~&E6&17.55, --3.0 &130, 66~~~~~~~~&W6  & --18.75, --21.50  &--73, 54~~~~~~~~\\
N7&8.05, 16.30&84,  120~~~~~~~~&E7&15.25, --4.7 &140, 35~~~~~~~~&W7  & --18.85, --18.60  &--64, 64~~~~~~~~\\
N8&7.95, 13.50&130,  78~~~~~~~~&E8&13.05, --6.2 &160, 45~~~~~~~~&W8  & --18.75, --15.00  &--42, 49~~~~~~~~\\
N9&7.65, 10.70&120,  42~~~~~~~~&E9&10.45, --6.80&160, 47~~~~~~~~&W9  & --18.00, --8.80   &--33, 40~~~~~~~~\\
N10&7.35, 7.90&96,   45~~~~~~~~&E10&7.95, --7.00&160, 40~~~~~~~~&W10 & --16.85, --5.50  &--22, 42~~~~~~~~\\
N11&6.85, 5.10&67,   42~~~~~~~~&E11&5.45, --6.70&120, 40~~~~~~~~&W11 & --15.75, --1.50  &--22, 49~~~~~~~~\\
N12&5.85, 2.50&43,   57~~~~~~~~&E12&2.95, --5.80&93,  45~~~~~~~~&W12 & --14.75,  2.00  &--8, 45~~~~~~~~\\
N13&4.65, 0.00&8,    57~~~~~~~~&E13&0.95, --4.50&54,  66~~~~~~~~&W13 & --13.15,  5.00  &--11, 57~~~~~~~~\\
N14&2.95, --2.10&-60, 131~~~~~~~~&E14&--1.15, --3.00&--17, 52~~~~~~~~&W14&--11.75,  7.80  &17, 61~~~~~~~~\\
N15&0.95, --3.80&--140, 97~~~~~~~~&E15&--3.55, --1.50&--39, 125~~~~~~~~&W15&--7.75, 13.50  &20, 68~~~~~~~~\\
N16&--1.55, --4.70&--240, 97~~~~~~~~&E16&--5.75, --0.00&--110, 165~~~~~~~~&W16&--2.75, 19.00  &52, 82~~~~~~~~\\
N17&--3.95, --4.30&--270, 66~~~~~~~~&E17&--7.55, 2.25&--160, 115~~~~~~~~&W17&--0.25, 22.00  &90, 47~~~~~~~~\\
\enddata
\tablenotetext{a}{IDs of locations in the mini-spiral arms.~~~$^{b}$Position offsets from
Sgr A*.~~~$^{c}$The LSR velocity and FWHM velocity width of the spectral features.}
\end{deluxetable}

\begin{deluxetable}{lllrrrrrr}
\tablenum{4}
\tabletypesize{\scriptsize}
\tablecaption{The three-dimensional velocities in Sgr A West}
\tablewidth{0pt}
\tablehead{
\colhead{Region} &
\colhead{IDs of radio knots} &
\colhead{IR sources} &
\colhead{$\overline {\Delta \alpha} $\tablenotemark{a}} &
\colhead{$\overline {\Delta \delta} $\tablenotemark{a}} &
\colhead{$\overline {V_x}$\tablenotemark{b}} &
\colhead{$\overline {V_y}$\tablenotemark{b}} &
\colhead{$\overline {V_{\rm LSR}}$\tablenotemark{b}}
 \\
\colhead{}&
\colhead{}&
\colhead{}&
\colhead{(asec)}&
\colhead{(asec)}&
\colhead{(km s$^{-1}$)} &
\colhead{(km s$^{-1}$)} &
\colhead{(km s$^{-1}$)}
\\ 
}
\startdata
\\
\multicolumn{2}{l}{Northern Arm} &&&& & &\\
Na&K1     &IRS8          &1.40$\pm$ 0.01&29.25$\pm$ 0.01
                      &69$\pm$14&--42$\pm$19&--12$\pm$6\\
    &   &&&& & &                              93$\pm$28 \\
Nb&K8, X7 &IRS5, IRS10   &10.22$\pm$0.01&5.03$\pm$ 0.01
                      &--54$\pm$20     & --168$\pm$20 & 67$\pm$2\\
Nc&K16,K20&IRS1, IRS16   &3.78$\pm$0.01&--0.08$\pm$0.01
                      &--218$\pm$13&--110$\pm$13& 15$\pm$1\\
Nd&K36& IRS21            &2.37$\pm$0.01&--2.65$\pm$0.01
                      &--91$\pm$15&--67$\pm$17&--92$\pm$4\\
Ne&K43,K46,K52&IRS2      &--3.43$\pm$0.01&--3.61$\pm$ 0.01
                      &--107$\pm$6&42$\pm$15&--274$\pm$1\\
 &&&&& & & \\
\multicolumn{2}{l}{Eastern Arm} &&&& & & \\
Ea&K66,X18&IRS4,IRS28 
                      &10.12$\pm$0.01&--8.50$\pm$0.01
                      &--67$\pm$8&24$\pm$14&158$\pm$3\\
Eb&K57,K59&IRS9          &5.17$\pm$0.01&--5.96$\pm$0.01
                      &--48$\pm$10&--130$\pm$16&139$\pm$3\\
Ec&K48,K50,K51&IRS9NW    &3.05$\pm$0.01&--4.92$\pm$0.01
                      &--161$\pm$14&81$\pm$13&165$\pm$13\\
&&&&& & & \\
\multicolumn{2}{l}{Western Arc}  &&&& & & \\
Wa&X6&\dots              &--8.41$\pm$0.01&8.37$\pm$0.01
                      &193$\pm$38&121$\pm$23&31$\pm$11\\
Wb&X19,X21,X23&\dots     &--18.00$\pm$0.01&--11.75$\pm$0.01  
                      &--137$\pm$28&96$\pm$18&--67$\pm$5\\ 
Wc&X25&\dots         &--15.37$\pm$0.01&--3.52$\pm$0.01
                      &--64$\pm$  15&140$\pm$19&--90$\pm$ 4\\
&&&& & & \\
\multicolumn{2}{l}{Bar} &&&& & &  \\
Ba& K25,K33,K40 & IRS33 &--0.27$\pm$0.01&--2.81$\pm$0.01
                                     &--296$\pm$14&124$\pm$22&--85$\pm$ 6\\
&        &&&& & &                     34$\pm$ 3\\            
&        &&&& & &                     161$\pm$4\\
Bb&K21,K23,K24,K26& IRS13   &--3.18$\pm$0.01&--1.58$\pm$0.01
                     &--110$\pm$2&95$\pm$   3 &--37$\pm$ 2\\ 
&K27,K29,K31,K35&&&& & &       --247$\pm$ 4\\   
Bc&K10,K11,K13,& IRS6,IRS34& --6.01$\pm$0.01&1.33$\pm$0.01
                           &--128$\pm$6& 51$\pm$9&--180$\pm$ 4\\
&K14,K17,K19&&&& & & \\
\enddata
\tablenotetext{a}{The mean position offsets derived by averaging the
position offsets of the radio knots listed in column 2 of this table.}
\tablenotetext{b}{The mean velocities derived by averaging the
velocity components of the radio knots listed in column 2 of this table.}
\end{deluxetable}

\begin{deluxetable}{llllll}
\tablenum{5}
\tabletypesize{\scriptsize}
\tablecaption{Orbital parameters of the three streams in Sgr A West}
\tablewidth{0pt}
\tablehead{
\colhead{Orbital parameters} &(Units)&
\colhead{Northern Arm} &
\colhead{Eastern Arm} &
\colhead{Western Arc} &
 \\
}
\startdata
\\
Distance (${\rm D}$)&(kpc)&8&8&8\\  
Mass (${\rm M_{dyn}}$)&(10$^6$ M$_\odot$)&4.2$^{+6}_{-3}$&4.2$^{+6}_{-3}$&4.2$^{+6}_{-3}$ \\
Eccentricity (e)   &   &0.83$\pm0.10$&0.82$\pm$0.05&0.20$\pm$0.15\\
Semi major axis ($a\pm\Delta a$)&(kAU)&205$\pm$91&289$\pm$140&236$\pm$12\\
                   & (pc)&0.99$\pm$0.44&1.40$\pm$0.68&1.11$\pm$0.06\\
Longitude of ascending node ($\Omega\pm\Delta\Omega$)\tablenotemark{a}&(deg)&64$\pm$28&--42$\pm$11&71$\pm$6\\
Argument of perifocus   ($\omega\pm\Delta\omega$)&(deg)&132$\pm$40&--280$\pm$8&22$\pm$48\\
Inclination             (i$\pm\Delta$i)\tablenotemark{a}   &(deg) &139$\pm$10&122$\pm$5&117$\pm$3\\
Perifocal  distance (q) & (kAU) & 34&52&184\\
                        &  (pc) & 0.17 &0.25 &0.89\\
Period (T)              & (10$^3$ y) & 45 & 76&54\\
\enddata
\tablenotetext{a}{Following the convention of astrodynamics \citep{boul91}, 
$\Omega$ is the angle between the +x-axis (East) and a line 
from the dynamic center to the point where the object crosses 
the sky plane from negative to positive 
side of the z-axis measured counterclockwise as viewed from
the +z direction (pointing away from the Earth). The
inclination i is the angle between the +z-axis and the angular
momentum vector {\bf h}, which is perpendicular to the orbit
plane, measured from 0\arcdeg to 180\arcdeg. If i$>$90\arcdeg,
the orbital motion is counterclockwise as viewed from the
Earth. If i$<$90\arcdeg, the orbital motion is clockwise as viewed from the
Earth.  
}
\end{deluxetable}


\begin{thebibliography}{}
\bibitem[Aitken et al. (1991)]{aitk91}
Aitken, D.K., Gezari, D., Smith, C.H., McCaughrean, M. \& Roche, P.F. 1991, 
\apj, 380, 419
\bibitem[Aitken et al. (1998)]{aitk98}
Aitken, D.K., Smith, C.H., Moore, T.J.T. \& Roche, P.F. 1998, \mnras, 299, 743
\bibitem[An et al. (2005)]{ant05}
An, T., Goss, W. M., Zhao, J.-H., Hong, X. Y., Roy, S., Rao, A. P.,
 Shen, Z.-Q., 2005, \apjl, 634, L49
\bibitem[Becklin \& Neugebauer (1975)]{beck75} Becklin, E. E. 
\& Neugebauer, G., 1975, \apjl,  200, L71
\bibitem[Blum, Sellgren, \& Depoy (1996)]{blum96} Blum, R. D., 
Sellgren, K., \& Depoy, D. L., 1996, \apj, 470, 864
\bibitem[Boulet (1991)]{boul91} Boulet, D. L. 1991, Methods of
Orbit Determination (Virginia: Willmann-Bell, Inc.)
\bibitem[Brandt (1999)]{bran99} Brandt, S. 1999, Data Analysis:
Statistical and Computational Methods for Scientists and Engineers
(Springer-Verlag New York Inc.)
\bibitem[Bower et al. (2005)]{bow05} Bower, G. C.,
Roberts, D. A., Yusef-Zadeh, F., Backer, D. C.,
Cotton, W. D., Goss, W. M., Lang, C. C.,
\& Lithwick, Y., 2005, \apj, 633, 218
\bibitem[Brandt (1999)]{bran99} Brandt, S. 1999, Data Analysis:
Statistical and Computational Methods for Scientists and Engineers
(Springer-Verlag New York Inc.)
\bibitem[Chandrasekhar \& Fermi (1953)]{chan53}
Chandrasekhar, S. \& Fermi, E. 1953, \apj, 118, 113 
\bibitem[Doeleman et al. (2008)]{doel08}
Doeleman, S. S., Weintroub, J., Rogers, A. E. E. et al., 2008, Nature, 455, 78
\bibitem[Ekers et al. (1983)]{eker83} 
Ekers, R. D., van Gorkom, J. H., Schwarz, U. J., \& Goss, W. M.,
1983, \aap, 122, 143
\bibitem[Geballe et al. (1987)]{geba87}
Geballe, T. R., Wade, R., Krisciunas, K., Gatley, I., \& Bird, M. C., 1987,
\apj, 320, 570
\bibitem[Geballe et al. (2004)]{geba04} Geballe, T. R., 
Rigaut, F., Roy, J.-R., \& Draine, B. T., 2004, \apj, 602, 770
\bibitem[Geballe et al. (2006)]{geba06} Geballe, T. R., Najarro, F., 
Rigaut, F., \& Roy, J.-R., 2006, \apj, 652, 370
\bibitem[Genzel et al. (2000)]{genz00} Genzel, R., Pichon, C., Eckart, A., Gerhard, O. E., \& Ott, T., 2000, \mnras, 317, 348
\bibitem[Ghez et al. (2003)]{ghez03} Ghez, A. M., et al. 2003, \apjl, 586, L127
\bibitem[Ghez et al. (2005)]{ghez05} Ghez, A. M., et al. 2005, \apjl, 620, 744
\bibitem[Ghez et al. (2008)]{ghez08} Ghez, A. M., et al. 2008, ApJ, 689, 1044
\bibitem[Gillessen et al. (2009)]{gill09}
Gillessen, S., Eisenhauer, F., Trippe, S., Alexander, T.,
Genzel, R., Martins, F., Ott, T., 2009, \apj, 692, 1075
\bibitem[Glasse, Aitken \& Roche (2003)]{glas03}
Glasse, A.C.H., Aitken, D.K. \& Roche, P.F. 2003, Astron. Nachr. 324, No. S1,  563
\bibitem[Herbst et al (1993)]{herb93}
Herbst, T. M., Beckwith, S. V. W., Forrest, W. J., \& Pipher, J. L.,
1993, \aj, 105, 956
\bibitem[Krabbe et al. (1991)]{krab91}
Krabbe, A., Genzel, R., Drapatz, S., \& Rotaciuc, V.,
1991, \apjl, 382, L19
\bibitem[Lacy et al. (1980)]{lacy80}
Lacy, J. H., Townes, C. H., Geballe, T. R., \& Hollenbach, D. J.,
1980, \apjl, 241, L132
\bibitem[Lacy, Achtermann, \& Serabyn (1991)]{lacy91}
Lacy, J.H., Achtermann, J. M. \& Serabyn, E., 1991, \apjl, 380, L71
\bibitem[Lebofsky, Rieke, \& Tokunaga (1982)]{lebo82} Lebofsky, M. J., Rieke, G. H., \& Tokunaga, A. T., 1982, \apj, 263, 736
\bibitem[Liszt (2003)]{liszt03} Liszt, H. S., 2003, A\&A, 408, 1009
\bibitem[Lo \& Claussen (1983)]{lo83} Lo, K. Y. \& Claussen, M. J.,
1983, \nat, 306, 647
\bibitem[Menten et al. (1997)]{ment97} Menten, K. M., Reid, M. J.,
 Eckart, A., \& Genzel, R. 1997, \apj, 475, 111
\bibitem[Morris \& Yusef-Zadeh (1987)]{morr87} Morris, M. \& Yusef-Zadeh, F. 1987, 
in ``The Galactic Center,'' ed: D.C. Backer (New York: AIP), p. 127
\bibitem[Muno et al. (2005)]{muno05} Muno, M. P., Pfahl, E.,
Baganoff, F. K., Brandt, W. N., Ghez, A., Lu, J., \&  
Morris, M. R. 2005, \apjl, 622, L113 
\bibitem[Mu\v{z}i\'{c} et al. (2007)]{muzi07} Mu\v{z}i\'{c}, K., Kckart, A. , Sch\"odel, R.,
Meyer, L. \& Zensus, A., 2007, A\&A, 469, 993
\bibitem[Mu\v{z}i\'{c} et al. (2008)]{muzi08} Mu\v{z}i\'{c}, K., Sch\"odel, R., Eckart, A., Meyer, L., \& Zensus, A., 2008, \aap, 482, 173
\bibitem[Panagia (1973)]{pana73}
Panagia, N., 1973, \apj, 78, 929
\bibitem[Paumard et al. (2001)]{paum01} Paumard, T., Maillard, J. P., 
Morris, M., \& Rigaut, F., 2001, \aap, 366, 466
\bibitem[Paumard et al. (2004)]{paum04} Paumard, T., Maillard, J. P.,
\& Morris, M., 2004, \aap, 426, 81
\bibitem[Paumard et al. (2006)]{paum06}
Paumard, T., Genzel, R., Martins, F., Nayakshin, S.,
Beloborodov, A. M., Levin, Y., Trippe, S., Eisenhauer, F.,
Ott, T., Gillessen, S., Abuter, R., Cuadra, J., Alexander, T., 
Sternberg, A., 2006, ApJ, 643, 101 
\bibitem[Perger et al. (2008)]{perg08} Perger, M., 
Moultaka, J., Eckart, A., Viehmann, T., Sch\"odel, R., 
\& Mu\v{z}i\'{c}, K., 2008, \aap, 478, 127
\bibitem[Porquet et al. (2005)]{porq05}
Porquet, D., Grosso, N., Belanger, G., Goldwurm, A., 
Yusef-Zadeh, F., Warwick, R. S., \& Predehl, P., 2005, \aap, 443, 571
\bibitem[Reid et al. (2003)]{reid03} Reid, M. J., Menten, K. M., 
Genzel, R., Ott, T., Sch\"odel, R., Eckart, A., 2003, \apj, 587, 203
\bibitem[Reid et al. (2007)]{reid07} Reid, M. J.,
Menten, K. M., Trippe, S., Ott, T., \&  Genzel, R., 2007, \apj,
659, 378
\bibitem[Roberts \& Goss (1993)]{rob93} Roberts, D. A. \&
Goss, W. M., 1993, \apjs, 86, 133
\bibitem[Roberts, Yusef-Zadeh \& Goss (1996)]{rob96}
Roberts, D. A., Yusef-Zadeh, F., \& Goss, W. M., 1996, \apj, 459, 627
\bibitem[Sch\"odel et al. (2002)]{scho02} Sch\"odel, R., Ott, T., Genzel, R.,
\& et al., 2002, \nat, 419, 694 
\bibitem[Serabyn \& Lacy (1985)]{sera85}
Serabyn, E. \&  Lacy, J. H., 1985, \apj, 293, 445
\bibitem[Serabyn et al. (1988)]{sera88}
Serabyn, E., Lacy, J. H., Townes, C. H., \& Bharat, R., 1988,
\apj, 326, 171
\bibitem[Serabyn, Lacy \& Achtermann (1991)]{sera91}
Serabyn, E., Lacy, J. H., \& Achtermann, J. M., 1991, \apj, 378, 557
\bibitem[Schwarz, Bregman \& van Gorkom (1989)]{schw89}
Schwarz, U. J., Bregman, J. D. \& van Gorkom, J. H., 1989, A\&A, 215, 33
\bibitem[Tanner et al. (2002)]{tann02}  Tanner, A., Ghez, A. M., Morris, M.,
Becklin, E. E., Cotera, A., Ressler, M., Werner, M., \& Wizinowich, P., 2002,
\apj, 575, 860
\bibitem[Tanner et al. (2005)]{tann05}
Tanner, A., Ghez, A. M., Morris, M. R., \& Christou, J. C., 2005, \apj, 620, 744
\bibitem[Viehmann et al. (2006)]{vieh06} Viehmann, T., Eckart, A.,
Sch\"odel, R., Pott, J.-U., \& Moultaka, J., 2005, \apj, 642, 861
\bibitem[Vollmer \& Duschl (2000)]{voll00}
Vollmer, B. \& Duschl, W.J. 2000, New Astronomy, 4, 581
\bibitem[Wardle \& Yusef-Zadeh (1992)]{ward92}
Wardle, M. \& Yusef-Zadeh, F., 1992, Nature, 357, 308
\bibitem[Wollman, Smith, \& Larson (1982)]{woll82} Wollman, E. R., Smith, H. A., \& Larson, H. P., 1982, \apj, 258, 506
\bibitem[Yusef-Zadeh, Morris, \& Ekers (1989)]{yusef89}
Yuzef-Zadeh, F., Morris, M., \& Ekers, R., 1989, in IAU Symp. 136,
The Center of The Galaxy, ed. Mark Morris (Norwell: Kluwer), 443
\bibitem[Yusef-Zadeh, Morris \& Ekers (1990)]{yusef90}
Yusef-Zadeh, F., Morris, M., \& Ekers, R. D., 1990, Nature, 348, 45
\bibitem[Yusef-Zadeh \& Morris (1991)]{yusef91}
Yusef-Zadeh, F., \& Morris, M., 1991, \apjl, 371, L59
\bibitem[Yusef-Zadeh \& Melia (1992)]{yusef92} 
Yusef-Zadeh, F., \&  Melia, F., 1992,
\apjl, 385, 41
\bibitem[Yusef-Zadeh \&  Wardle (1993)]{yusef93}
Yusef-Zadeh, F., Wardle, M., 1993, \apj, 405, 584
\bibitem[Yusef-Zadeh, Roberts \& Biretta (1998)]{yusef98}
Yusef-Zadeh, F., Roberts, D. A., \& Biretta, J., 1998, \apjl, 503, L191
\bibitem[Zhao et al. (1991)]{zhao91}
Zhao, J.-H., \&
Goss, W. M, Lo, K. Y., Ekers, R. D., 1991, Nature, 354, 46
\bibitem[Zhao \& Goss (1998)]{zhao98} Zhao, J.-H., \&  
Goss, W. M, 1998, \apjl, 499, L163
\bibitem[Zhao \& Goss (1999)]{zhao99} Zhao, J.-H., \&
Goss, W. M, 1999, in The Central Parsecs of the Galaxy, ASP Conference Series, Vol. 186. Edited by Heino Falcke, Angela Cotera, Wolfgang J. Duschl, Fulvio Melia, and Marcia J. Rieke, 224
\end{thebibliography}
\end{document}